\documentclass[10pt,twocolumn,letterpaper]{article}

\usepackage[pagenumbers]{cvpr} 

\usepackage[export]{adjustbox}
\usepackage{array}
\usepackage{amsmath}
\usepackage{amssymb}
\usepackage{booktabs}
\usepackage{capt-of}
\usepackage{color}
\usepackage{contour}
\usepackage{float}
\usepackage{fancyhdr}
\usepackage{fouridx}
\usepackage{graphicx}
\usepackage{listings}
\usepackage{tikz}
\usepackage{multirow}
\usepackage{mycomments}
\usepackage{nicefrac}
\usepackage{silence}
\usepackage{ulem}
\usepackage{soul}
\usepackage{lipsum}

\usetikzlibrary{spy, calc, tikzmark}
\def\commenton{1}  
\def\editon{1}     

\definecolor{cvprblue}{rgb}{0.21,0.49,0.74}
\usepackage[pagebackref,breaklinks,colorlinks,allcolors=cvprblue]{hyperref}

\usepackage[capitalize]{cleveref}
\crefname{section}{Sec.}{Secs.}
\Crefname{section}{Section}{Sections}
\Crefname{table}{Table}{Tables}
\crefname{table}{Tab.}{Tabs.}

\def\paperID{4} 
\def\confName{CVPR}
\def\confYear{2026}
    
\mathchardef\mhyphen="2D
\newcommand{\tref}{{t_{\text{ref}}}}
\newcommand{\origevents}{{\mathcal{E}}}
\newcommand{\warpedevents}{{\mathcal{E}_{\tref}'}}
\newcommand{\boxscaler}{0.97}
\newcommand{\labelscaler}{0.18}

\DeclareMathOperator*{\argmax}{arg\,max}

\contourlength{0.8pt}
\newcommand{\myul}[1]{%
  \uline{\phantom{#1}}%
  \llap{\contour{white}{#1}}
}

\lstdefinestyle{mystyle}{
    backgroundcolor=\color{backcolour},   
    commentstyle=\color{codegreen},
    keywordstyle=\color{magenta},
    numberstyle=\tiny\color{codegray},
    stringstyle=\color{codepurple},
    basicstyle=\ttfamily\footnotesize,
    breakatwhitespace=false,         
    breaklines=true,                 
    captionpos=b,                    
    keepspaces=true,                 
    numbers=left,                    
    numbersep=5pt,                  
    showspaces=false,                
    showstringspaces=false,
    showtabs=false,                  
    tabsize=2
}
\definecolor{PERSIAN_BLUE}{RGB}{51,51,178}
\definecolor{PERSIAN_BLUE_COMP_MUD_BROWN}{RGB}{178,114,51}
\definecolor{PERSIAN_BLUE_MONOCHROME}{RGB}{74,74,152}
\definecolor{PERSIAN_BLUE_ANALOG_1}{RGB}{114,51,178}
\definecolor{PERSIAN_BLUE_ANALOG_2}{RGB}{51,178,178}
\definecolor{PERSIAN_BLUE_SPLIT_COMP_1}{RGB}{178,146,51}
\definecolor{PERSIAN_BLUE_SPLIT_COMP_2}{RGB}{178,83,51}
\definecolor{PERSIAN_BLUE_TRIADIC_1}{RGB}{51,178,51}
\definecolor{PERSIAN_BLUE_TRIADIC_2}{RGB}{178,51,51}
\definecolor{PERSIAN_BLUE_TETRADIC_1}{RGB}{178,114,51}
\definecolor{PERSIAN_BLUE_TETRADIC_2}{RGB}{178,178,51}
\definecolor{PERSIAN_BLUE_TETRADIC_3}{RGB}{178,51,178}
\definecolor{RAISIN_BLACK}{RGB}{42,45,52}
\definecolor{RUSSIAN_GREEN}{RGB}{92,148,110}
\definecolor{BLIZZARD_BLUE}{RGB}{160,221,230}
\definecolor{DARK_ORANGE}{RGB}{255,143,41}
\definecolor{PICOTEE_BLUE}{RGB}{38,38,134}
\definecolor{BISTRE}{RGB}{55,44,37}
\definecolor{GRAY_WEB}{RGB}{131,131,131}
\definecolor{DARK_SKY_BLUE}{RGB}{147,183,190}
\definecolor{MINT_CREAM_1}{RGB}{241,255,250}
\definecolor{MINT_CREAM_2}{RGB}{247,255,247}
\definecolor{PALE_SILVER}{RGB}{213,199,188}
\definecolor{OPAL}{RGB}{148,191,190}
\definecolor{LASER_LEMON}{RGB}{252,252,98}
\definecolor{MINDARO}{RGB}{218,255,125}
\definecolor{GRANNY_SMITH_APPLE}{RGB}{178,239,155}
\definecolor{RHYTHM}{RGB}{140,134,170}
\definecolor{SILVER}{RGB}{197,197,197}
\definecolor{SPANISH_GRAY_1}{RGB}{130,145,145}
\definecolor{SPANISH_GRAY_2}{RGB}{148,155,150}
\definecolor{BRICK_RED}{RGB}{195,60,84}
\definecolor{SKY_BLUE_CRAYOLA}{RGB}{142,227,239}
\definecolor{CELESTE}{RGB}{174,243,231}
\definecolor{HOOKERS_GREEN}{RGB}{88,125,113}
\definecolor{GREEN_PANTONE}{RGB}{77,170,87}
\definecolor{NAPLES_YELLOW}{RGB}{255,130,109}
\definecolor{AIR_SUPERIORITY_BLUE}{RGB}{108,166,193}
\definecolor{FIRE_BRICK}{RGB}{179,0,27}
\definecolor{ICE_BERG}{RGB}{126,163,204}
\definecolor{TAN}{RGB}{204,173,143}
\definecolor{MIDDLE_BLUE}{RGB}{247,255,247}
\definecolor{LAVENDER_BLUSH}{RGB}{238,229,233}
\definecolor{CINNABAR}{RGB}{214,73,51}
\definecolor{BABY_POWDER}{RGB}{247,247,242}
\definecolor{AQUA_MARINE}{RGB}{133,255,199}
\definecolor{CORAL}{RGB}{255,133,82}
\definecolor{PLATINUM}{RGB}{230,230,230}
\definecolor{BROWN}{RGB}{126,89,32}
\definecolor{FULVOUS}{RGB}{220,133,31}
\definecolor{YELLOW_ORANGE}{RGB}{255,167,55}
\definecolor{BITTER_SWEET}{RGB}{255,111,89}
\definecolor{ZOMP}{RGB}{67,170,139}
\definecolor{AMARANTH_RED}{RGB}{215,29,45}
\definecolor{SPACE_CADET}{RGB}{49,61,90}
\definecolor{CELADON_BLUE}{RGB}{66,129,164}
\definecolor{ELECTRIC_BLUE}{RGB}{142, 237, 247}
\definecolor{BABY_BLUE_EYES}{RGB}{161, 205, 241}
\definecolor{STAR_COMMAND}{RGB}{34, 116, 165}
\definecolor{DENIM_BLUE}{RGB}{57, 67, 183}
\definecolor{BLUE_JEANS}{RGB}{72, 172, 240}
\definecolor{UPSDELL_RED}{RGB}{180, 24, 37}
\definecolor{RED_MUNSELL}{RGB}{228, 58, 72}
\definecolor{BLUE_MUNSELL}{RGB}{27, 154, 170}
\definecolor{LIBERTY}{RGB}{76, 76, 157}
\definecolor{BABY_BLUE}{RGB}{108, 207, 246}
\definecolor{CAROLINA_BLUE}{RGB}{27, 152, 224}
\definecolor{MINION_YELLOW}{RGB}{242, 220, 93}
\definecolor{SANDY_BROWN}{RGB}{242, 163, 89}
\definecolor{FLAME}{RGB}{215, 78, 9}
\definecolor{BLOOD_RED}{RGB}{110, 14, 10}
\definecolor{CLARET}{RGB}{137, 4, 61}
\definecolor{EMERALD}{RGB}{68, 207, 108}
\definecolor{ILLUMINATING_EMERALD}{RGB}{50, 147, 111}
\definecolor{MAROON_X_11}{RGB}{184, 19, 101}
\definecolor{CARRIBEAN_GREEN}{RGB}{29, 211, 176}
\definecolor{RED_CRAYOLA}{RGB}{239, 45, 86}
\definecolor{EGG_SHELL}{RGB}{244, 236, 214}
\definecolor{AMAZON}{RGB}{74, 120, 86}
\definecolor{GOOGLE_G}{RGB}{60, 186, 84}
\definecolor{GOOGLE_Y}{RGB}{244, 194, 13}
\definecolor{GOOGLE_R}{RGB}{219, 50, 54}
\definecolor{GOOGLE_B}{RGB}{72, 133, 237}
\definecolor{PRINCETON_ORANGE}{RGB}{245, 128, 38}
\definecolor{YALE_BLUE_2}{RGB}{0, 68, 128}
\definecolor{MIKADO_YELLOW}{RGB}{255, 193, 0}
\definecolor{SELECTIVE_YELLOW}{RGB}{255, 186, 8}
\definecolor{MISTY_ROSE}{RGB}{255, 227, 220}
\definecolor{HONEY_YELLOW}{RGB}{247, 179, 43}
\definecolor{RUBY_RED}{RGB}{154, 3, 30}
\definecolor{BRONZE}{RGB}{203, 121, 58}
\definecolor{CELADON}{RGB}{172,247,193}
\definecolor{GREEN_SHEEN}{RGB}{128,194,175}
\definecolor{GREEN_SHEEN_2}{RGB}{136, 183, 181}
\definecolor{CITRINE}{RGB}{224, 202, 60}
\definecolor{MELLOW_APRICOT}{RGB}{255, 184, 111}
\definecolor{GLOSSY_GRAPE}{RGB}{171,146,191}
\definecolor{PINK_LAVENDER}{RGB}{212, 178, 216}
\definecolor{MAGIC_MINT}{RGB}{139, 232, 203}
\definecolor{MANATEE}{RGB}{141, 153, 174}
\definecolor{ALICE_BLUE}{RGB}{237, 242, 244}
\definecolor{SPANISH_BISTRE}{RGB}{111, 115, 47}
\definecolor{CAMEL}{RGB}{179, 138, 88}
\definecolor{PORTLAND_ORANGE}{RGB}{244, 96, 54}
\definecolor{PERSIAN_GREEN}{RGB}{27, 153, 139}
\definecolor{BEIGE}{RGB}{235, 235, 211}
\definecolor{CERISE}{RGB}{218, 65, 103}
\definecolor{LAVENDER_BLUE}{RGB}{183, 195, 243}
\definecolor{CYCLAMEN}{RGB}{221, 117, 150}
\definecolor{POPSTAR}{RGB}{175, 93, 99}
\definecolor{FERN_GREEN}{RGB}{91, 117, 83}
\definecolor{MIDNIGHT_BLUE}{RGB}{25, 25, 89}
\definecolor{DAVYS_GRAY}{RGB}{87, 87, 97}
\definecolor{COOL_GRAY}{RGB}{132, 143, 165}
\definecolor{BATTLESHIP_GRAY}{RGB}{133, 135, 134}
\definecolor{GREEN_BLUE}{RGB}{0, 100, 177}
\definecolor{SEAL_BROWN}{RGB}{82, 43, 24}
\definecolor{CHESTNUT}{RGB}{139, 84, 61}
\definecolor{AVOCADO}{RGB}{67, 125, 3}
\definecolor{SHEEN_GREEN}{RGB}{126, 203, 1}
\definecolor{MALACHITE}{RGB}{47, 212, 82}
\definecolor{STEEL_BLUE}{RGB}{51, 123, 178}

\definecolor{PRUSSIAN_BLUE}{RGB}{0, 51, 96}
\definecolor{OXFORD_BLUE}{RGB}{0, 34, 64}
\definecolor{AUBURN}{RGB}{164, 44, 45}
\definecolor{CADMIUM_ORANGE}{RGB}{237, 140, 65}
\definecolor{CASTLETON_GREEN}{RGB}{40, 89, 67}
\definecolor{BLACK_CHOCOLATE}{RGB}{35, 35, 26}
\definecolor{BITTER_LIME}{RGB}{188, 237, 9}
\definecolor{DARK_SIENNA}{RGB}{45, 6, 5}
\definecolor{TYRIAN_PURPLE}{RGB}{76, 8, 39}
\definecolor{GOLD_FUSION}{RGB}{117, 112, 78}
\definecolor{RUBY}{RGB}{216, 17, 89}
\definecolor{QUINACRIDONE_MAGENTA}{RGB}{143, 45, 86}
\definecolor{PUCE}{RGB}{195, 125, 146}
\definecolor{DEEP_TAUPE}{RGB}{132, 98, 103}
\definecolor{MOUNTBATTEN_PINK}{RGB}{147, 116, 138}
\definecolor{RED_VIOLET_CRAYOLA}{RGB}{175, 77, 152}
\definecolor{WILD_ORCHID}{RGB}{214, 107, 160}
\definecolor{MEDIUM_SLATE_BLUE}{RGB}{125, 131, 255}
\definecolor{BLUE_JEANS}{RGB}{0, 166, 251}
\definecolor{ANTIQUE_BRONZE}{RGB}{100, 101, 54}
\definecolor{OLIVE_DRAB_7}{RGB}{72, 61, 3}
\definecolor{BRIGHT_MAROON}{RGB}{179, 57, 81}
\definecolor{BOTTLE_GREEN}{RGB}{34, 111, 84}
\definecolor{WINTER_GREEN_DREAM}{RGB}{75, 143, 140}
\definecolor{LIVER}{RGB}{101, 66, 54}
\definecolor{FOREST_GREEN_CRAYOLA}{RGB}{124, 169, 130}
\definecolor{PURPLE_MOUNTAIN_MAJESTY}{RGB}{159, 134, 192}
\definecolor{LILAC}{RGB}{190, 149, 196}
\definecolor{CERULEAN_CRAYOLA}{RGB}{0, 167, 225}
\definecolor{MEDIUM_PURPLE}{RGB}{147, 129, 255}
\definecolor{PARADISE_PINK}{RGB}{236, 64, 103}
\definecolor{KOMBU_GREEN}{RGB}{36, 49, 25}
\definecolor{OPERA_MAUVE}{RGB}{200, 132, 166}
\definecolor{TITANIUM_YELLOW}{RGB}{240, 225, 0}
\definecolor{DARK_LAVA}{RGB}{82, 70, 50}
\definecolor{SPANISH_CARMINE}{RGB}{209, 17, 73}
\definecolor{SADDLE_BROWN}{RGB}{128, 47, 0}
\definecolor{INDIGO_DYE}{RGB}{21, 65, 103}
\definecolor{SEA_GREEN}{RGB}{0, 128, 60}
\definecolor{NAVY_BLUE}{RGB}{30, 0, 128}
\definecolor{BROWN}{RGB}{128, 79, 0}
\definecolor{BARN_RED}{RGB}{128, 15, 0}
\definecolor{SPANISH_BISTRE}{RGB}{128, 111, 0}
\definecolor{BURGUNDY}{RGB}{128, 0, 34}
\definecolor{INDIGO}{RGB}{94, 0, 128}
\definecolor{MAGENTA_DYE}{RGB}{198, 15, 123}
\definecolor{MAGENTA_PROCESS}{RGB}{249, 0, 147}
\definecolor{ASPARAGUS}{RGB}{136, 171, 117}
\definecolor{MAXIMUM_GREEN}{RGB}{103, 148, 54}
\definecolor{BLUE_MUNSELL}{RGB}{45, 147, 173}
\definecolor{RIFLE_GREEN}{RGB}{78, 83, 64}
\definecolor{NICKEL}{RGB}{105, 114, 104}
\definecolor{SPANISH_VIRIDIAN}{RGB}{12, 124, 89}
\definecolor{PACIFIC_BLUE}{RGB}{42, 183, 202}
\definecolor{CHARTREUSE_TRADITIONAL}{RGB}{228, 255, 26}
\definecolor{FIERY_ROSE}{RGB}{255, 83, 118}
\definecolor{LIGHT_SEA_GREEN}{RGB}{0, 166, 166}
\definecolor{YELLOW_GREEN}{RGB}{138, 201, 38}
\definecolor{FRENCH_RASPBERRY}{RGB}{195, 49, 73}
\definecolor{RADICAL_RED}{RGB}{255, 56, 100}
\definecolor{FIREBRICK}{RGB}{179, 0, 27}
\definecolor{STEEL_TEAL}{RGB}{80, 132, 132}
\definecolor{CRIMSON_UA}{RGB}{166, 38, 57}
\definecolor{ROSE}{RGB}{255, 114, 118}
\definecolor{INDIGO_2}{RGB}{99, 93, 198}
\definecolor{HONEY_DEW}{RGB}{213, 242, 227}
\definecolor{STEEL_BLUE_2}{RGB}{89, 133, 170}
\definecolor{CAMBRIDGE_BLUE}{RGB}{167, 202, 177}
\definecolor{ORANGE_WEB}{RGB}{255, 164, 0}
\definecolor{CAROLINA_BLUE_2}{RGB}{0, 159, 253}

\definecolor{YALE_BLUE}{RGB}{0, 68, 124}
\definecolor{YALE_BLUE_DARK}{RGB}{0, 34, 62}

\definecolor{VIRIDIS_START_PURPLE}{RGB}{68, 1, 84}
\definecolor{VIRIDIS_START_OD1}{RGB}{181, 222, 43}
\definecolor{VIRIDIS_END_OD1}{RGB}{229, 228, 25}
\definecolor{VIRIDIS_END_YELLOW}{RGB}{253, 231, 37}

\title{Inertia-Informed Orientation Priors for Event-Based Optical Flow Estimation}

\author{
Pritam P. Karmokar and William J. Beksi\\
The University of Texas at Arlington\\
Arlington, TX, USA\\
{\tt\small pritam.karmokar@mavs.uta.edu, william.beksi@uta.edu}
}

\begin{document}

\maketitle

\begin{abstract}
Event cameras, by virtue of their working principle, directly encode motion
within a scene. Many learning-based and model-based methods exist that estimate
event-based optical flow, however the temporally dense yet spatially sparse
nature of events poses significant challenges. To address these issues,
contrast maximization (CM) is a prominent model-based optimization methodology
that estimates the motion trajectories of events within an event volume by
optimally warping them. Since its introduction, the CM framework has undergone
a series of refinements by the computer vision community. Nonetheless, it
remains a highly non-convex optimization problem. In this paper, we introduce a
novel biologically-inspired hybrid CM method for event-based optical flow
estimation that couples visual and inertial motion cues. Concretely, we propose
the use of orientation maps, derived from camera 3D velocities, as priors to
guide the CM process. The orientation maps provide directional guidance and
constrain the space of estimated motion trajectories. We show that this
orientation-guided formulation leads to improved robustness and convergence in
event-based optical flow estimation. The evaluation of our approach on the
MVSEC, DSEC, and ECD datasets yields superior accuracy scores over the
state of the art.
\end{abstract}
    
\section{Introduction}
\label{sec:introduction}
\begin{figure}
\centering
\includegraphics[width=\linewidth]{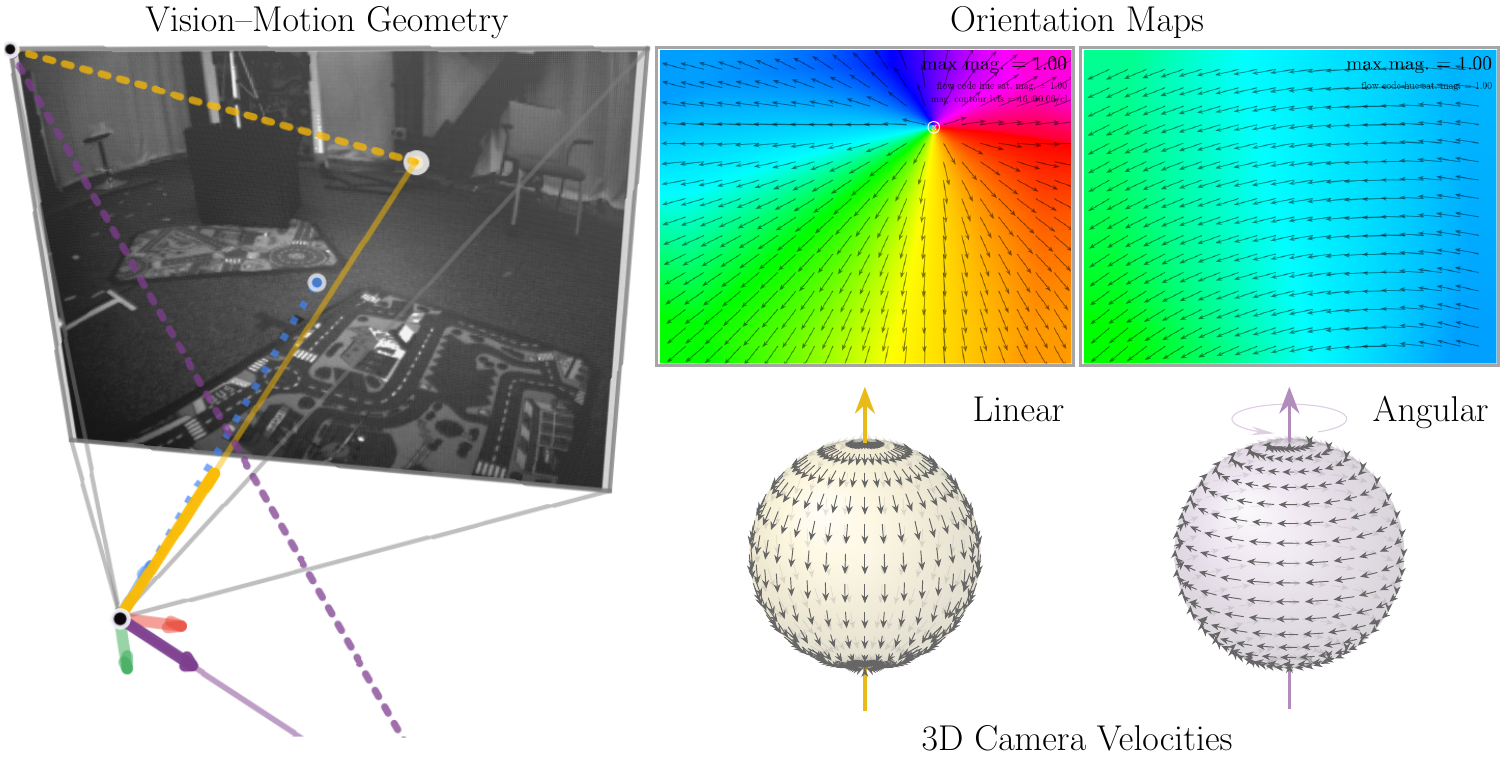}
\caption{The camera motion and image plane geometry (left) and induced
orientation maps and flow patterns (right) due to camera linear (yellow) and
angular (purple) velocities. Structured inertia-informed orientation priors
generated in the form of orientation maps from camera motion cues are used to
provide guidance to the CM process for event-based optical flow estimation.}
\label{fig:intro}
\end{figure}


Event cameras are visual sensing devices with a bioinspired working principle.
While conventional imaging sensors in frame-based cameras synchronously record
photo-intensities across a grid of pixels by integrating incident photons over
an interval of time, event-based imaging sensors capture the logarithmic change
in the photo-intensities at each pixel and rapidly fire spikes encoded as
\textit{events} asynchronously. This paradigm of operation endows event cameras
with appealing advantages over conventional cameras such as high temporal
resolution and high dynamic range at low latencies, low data bandwidth, and low
power. 

Since motion in a scene induces changes in the incident photo-intensities on
the image plane, event cameras make ideal imaging sensors to directly encode
this motion. As a result, past works have proposed approaches to estimate
event-based optical flow. These can be broadly classified into learning-based
and model-based optimization methods, where learning-based techniques can be
further categorized on the basis of their supervisory signals into (i)
supervised learning (SL)~\cite{gehrig2019evflownetest,
stoffregen2020evflownetplus,gehrig2021eraft} using ground-truth flow, (ii)
semi-supervised learning
(SSL)~\cite{zhu2018evflownet,lee2020spikeflownet,ding2022steflownet} using
conventional frames, and (iii) unsupervised learning (USL) using none.

Among model-based methods~\cite{gallego2018unifying,nagata2021matchtimesurf,
akolkar2020real,brebion2021realtimeflow}, contrast maximization
(CM)~\cite{gallego2018unifying} is the most notable approach.  In CM, the
parameters of the motion trajectories of events are optimized to maximize a
\textit{contrast} objective. This objective quantifies the sharpness of the
image of warped events (IWE), which is constructed by spatially accumulating
warped events over the image plane.  Therefore, metrics that can quantify
contrast have been abundantly studied
(\eg,~\cite{gallego2019focus,stoffregen2019analysis}). Nevertheless, CM remains
highly non-convex. It requires good initialization and regularization,
especially in challenging high degree-of-freedom motion scenarios, to yield
robust motion trajectories. 

In biology, acquiring and processing information from the visual and vestibular
sensory systems is highly intertwined. Inspired by this connection, we
introduce a technique to incorporate motion cues in the form of orientation
maps that directly benefit the CM optimization process. Specifically, the
orientation maps provide directional guidance and assist in constraining the
space of estimated motion trajectories, Fig.~\ref{fig:intro}. \textit{To the
best of our knowledge, we are the first to introduce a methodology that
generates visual cues in the form of orientation maps from 3D linear and
angular camera velocities to directly support event-based optical flow
estimation.} Our contributions are summed up as follows.
\begin{enumerate}
  \item We introduce a new biologically-inspired hybrid CM framework for
  event-based optical flow estimation that couples visual and inertial motion
  cues.
  \item We propose the use of orientation maps, derived from 3D camera 
  velocities, as priors to guide the CM process.
  \item We show that our orientation-guided formulation leads to improved
  robustness and convergence in event-based optical flow estimation.
  \item We preserve the modularity of CM by leveraging 3D camera velocities
  when they are available, and fallback to the standard framework when they are
  not.
\end{enumerate}
The source code for this project is publicly available at \cite{opcm}.

\section{Related Work}
\label{sec:related_work}
\subsection{Event-Based Optical Flow Estimation}
\label{subsec:event_optical_flow}
The majority of the advancements in event-based optical flow estimation can be
attributed to adaptations from preexisting frame-based techniques. For example,
EKLT~\cite{gehrig2020eklt} estimated spatially sparse temporally smooth optical
flow using event data to track image features in the blind time between frames.
Inspired by EKLT and~\cite{zhu2017event},~\cite{seok2020robust} performed
spatially sparse and continuous-time feature tracking on local event image
patches by regressing B{\'e}zier curves. A learning-based optical flow
computation paradigm for frames was introduced in RAFT~\cite{teed2020raft},
which was modified for events by E-RAFT~\cite{gehrig2021eraft} through an SL
technique. Motivated by~\cite{seok2020robust}, E-RAFT was further adapted
in~\cite{gehrig2024dense} to build multiple correlation volumes in time to
estimate B{\'e}zier curve parameters via asynchronous events and infer dense
continuous-time nonlinear trajectories. 

Among the first to adopt model-based approaches, Gallego
\etal~\cite{gallego2018unifying} introduced CM to estimate event-based optical
flow. Thereafter, CM was refined and further developed by several works
(\eg,~\cite{gallego2019focus,shiba2022eventcollapse,shiba2022secrets,hamann2024motion,shiba2025simultaneous}).
The traditionally unimodal events-only CM framework was extended to operate
bimodally through assimilation of frames into the optimization
framework~\cite{wang2020jointfiltering,karmokar2025secrets}. Influenced
by~\cite{zhu2019unsupervised,ye2020unsupervised}, CM was also employed
in~\cite{shiba2024secrets} to estimate the depth and egomotion of a camera.
Nonetheless, the incorporation of additional sensing modalities to achieve
robust event-based optical flow remains underexplored. In contrast to these
prior works, we focus on leveraging the first-order kinematics of camera motion
(\ie, velocities) and generate orientation priors to constrain the CM
optimization process to converge to valid flow fields.

\subsection{Camera Velocity Estimation}
\label{subsec:camera_velocity_estimation}
Camera egomotion, optical flow field, and depth of the scene are known to be
tightly coupled to each other~\cite{srinivasan2013high}. For the past 40 years,
visual motion fields~\cite{trucco1998introductory} have been utilized for
estimating egomotion~\cite{raudies2012review} from inertial data. Velocity
estimation is an integral part of egomotion estimation. With the help of
artificial neural networks (\eg, bidirectional recurrent neural networks) and
Kalman filters, single~\cite{feigl2020rnn} and
multiple~\cite{yuan2014localization} inertial measurement units (IMUs) have
been successfully used to estimate accurate, drift-free velocity from purely
inertial data. 
Gated recurrent units have also been employed to estimate
velocities from a combination of IMU data and the global navigation satellite
system~\cite{xiao2025inertial}. 
Utilizing similar architectures, velocities can
be reliably estimated in real-time when paired with additional visual data from
frames~\cite{deng2018visual,buchanan2022deep} or
events~\cite{lu2023event,xu2023tight}. Although there exists a plethora of
algorithms that purely estimate 3D camera velocities, the use of such
velocities in multimodal optical flow estimation approaches are relatively
scarce or non-existent. In our work, we make direct use of 3D camera velocities
to generate orientation maps that aid CM in estimating event-based optical
flow. 

\subsection{Visual-Vestibular Coupling}
\label{subsec:visual-vestibular_coupling}
In biology, the coupling between the visual and vestibular~\footnote{The
vestibular system, situated in the inner ear, is essential for balance, spatial
awareness, and coordinating movements. It informs the brain about motion, head
position, and spatial orientation, aiding posture and stable vision.} sensory
systems, along with their processing, is notably
significant~\cite{gu2008neural,fetsch2010visual,butler2010bayesian}. Despite
the conflicting cues, which are known to be the major cause of motion sickness
in humans~\cite{cohen2003critical,bertolini2016moving}, a forced fusion of
information from both these systems is still observed~\cite{winkel2015forced}.
During integration, an optimal balance between the two cues is maintained
through reliable weighting~\cite{campos2012multimodal,horst2015reliability}.
Nonetheless, a higher bias towards vestibular information has been
observed~\cite{fetsch2009dynamic,campos2014contributions}. We are inspired by
the pervasive and resilient utility of the visual-vestibular coupling found in
nature. Specifically, our motivation to employ camera egomotion is rooted in
understanding this particular phenomenon.

\section{Preliminaries}
\label{sec:preliminaries}
\begin{figure}
    \centering
    \includegraphics[width=\linewidth]{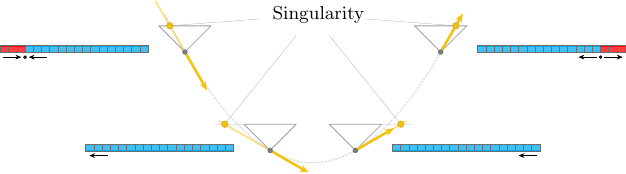}
    \caption{%
        Singularity points due to camera velocity. The linear motion of a 1D
        camera is depicted along a parabola with instantaneous velocity vectors
        (yellow). This velocity ray intersects with the image plane (in- or
        out-of-sensor) at the singularity point at which zero optical flow is
        observed. The corresponding color coded 1D flow orientations are
        illustrated for each pose, where cyan and red colors indicate left and
        right flow orientations, respectively. The top-left and top-right 1D
        arrays demonstrate the FOC and FOE, respectively.
    }
    \label{fig:singularity}
\end{figure}

Trucco and Verri~\cite[ch.~8, p.~183]{trucco1998introductory} define a motion
field as \textit{the 2D vector field of velocities of the image points, induced
by the relative motion between the viewing camera and the observed scene}. When
perceived by a moving camera in a static environment, this motion field
exhibits predictable patterns. For example, under pure translational motion,
the generated motion field\footnote{Note that the motion field integrated over
a fixed delta time yields the flow displacement field (\ie, optical flow
field).  In~\cite{trucco1998introductory}, the terms optical flow field and
motion field are used interchangeably throughout the text.} appears to either
(i) converge toward a unique point with zero motion or (ii) diverge away from
such a point, called the focus of expansion (FOE) and focus of contraction
(FOC), respectively. For consistency in terminology, we refer to this point as
the ``singularity'' point $\boldsymbol{s}_{\text{lin}}$, and likewise
$\boldsymbol{s}_{\text{ang}}$ for the rotational case (\cref{fig:singularity}).
Correspondingly, the motion field will appear to converge in or diverge out of
$\boldsymbol{s}_{\text{lin}}$, and likewise curl clockwise or counter-clockwise
around $\boldsymbol{s}_{\text{ang}}$.

This strong association between self-motion/egomotion and global optical flow
patterns has been known and thoroughly studied in both the computer
vision~\cite{gibson1950perception,gibson1966senses,
verri1989mathematical,gluckman1998ego} and
neuroscience~\cite{warren1988direction,lappe1999perception,duffy2000optic,
sunkara2016joint,henning2022populations} communities. For instance, in robotic
systems ``motion cues'' are sensed using an IMU. Furthermore, estimating
velocity from an IMU is a highly-investigated research area
(\cref{subsec:camera_velocity_estimation}). Similarly, in biology the
processing of cues from both the visual and vestibular sensory systems is
highly coupled (\cref{subsec:visual-vestibular_coupling}). In this context,
prior works have focused on either (i) inferring motion from visual cues or
(ii) fusing information after separately inferring motion from each system.
However, leveraging camera motion to aid visual processing is an uninvestigated
field of research. 

\section{Method}
\label{sec:method}
\subsection{Contrast Maximization}
\label{subsec:contrast_maximization}
Within the context of CM, we process all the data from an event stream in parts
based on time intervals. More formally, we work on a single ($i$-th) event set
$\mathcal{E}^{(i)} \doteq \{e^{(i)}_k\}^{N^{(i)}_{e}}_{k=1}$ at a time, which
consists of $N^{(i)}_{e}$ individual events. For conciseness, we exclude the
superscript ($\cdot^{(i)}$). Each event $e_k \doteq (\mathbf{x}_k, t_k, p_k)$ is
recorded as a 4-tuple consisting of the $x\mhyphen y$ coordinates of the
location of the triggered pixel $\mathbf{x}_k = (x_k, y_k)$ along with the
timestamp $t_k$ and the polarity $p_k$ of the event. 

Let $\mathbf{X}_{\origevents} = \{\mathbf{x}_k\}^{N_e}_{k=1}$ and
$\mathbf{X}_{\warpedevents} = \{\mathbf{x}_k'\}^{N_e}_{k=1}$ be the set of event
coordinates that belong to the original event set $\origevents$ and the warped
event set $\warpedevents$ at the reference time $\tref$, respectively. We assume
that all local motions are linear within a small time window. A warping function
$\mathbf{W}(\mathbf{x}_{k}, t_k; \boldsymbol{\theta}_k, \tref)$ defines a
mapping $\origevents \mapsto \warpedevents$ that transports $\mathbf{x}_k \in
\mathbf{X}_{\origevents}$ to $\mathbf{x}_k' \in \mathbf{X}_{\warpedevents}$ at
the reference time $\tref$ according to 
\begin{equation}
\begin{aligned}
  \mathbf{x}_k' &\doteq \mathbf{W}(\mathbf{x}_{k}, t_k; \boldsymbol{\theta}_k, \tref) \\
                &= \mathbf{x}_k + \boldsymbol{\theta}_k(\tref - t_k), \quad 1 \leq k \leq N_e.
  \label{eq:warp_function}
\end{aligned}
\end{equation}
This occurs along trajectories modeled by motion parameters $\boldsymbol{\Theta}
= \{\boldsymbol{\theta}_k\}^{N_e}_{k=1}$, where $\boldsymbol{\theta}_k =
\mathbf{v}(\mathbf{x}_k)$ is the velocity vector at $\mathbf{x}_k$. We employ 
these warped events $\warpedevents$ to create the IWE by organizing them by
their coordinates $\mathbf{x}_k' \in \mathbf{X}_{\warpedevents}$ and collecting
them on the image plane via 
\begin{equation}
  I_{\text{events}}(\mathbf{x}; \boldsymbol{\Theta}, \tref) \doteq \sum^{N_e}_{k=1} \delta(\mathbf{x} - \mathbf{x}_k'), 
  \label{eq:iwe}
\end{equation}
where $\delta(\cdot)$ is the Dirac delta function. When implementing CM,
$\delta$ is substituted with a smooth approximation such as a Gaussian, \ie,
$\delta_{\sigma}(\mathbf{x} - \boldsymbol{\mu}) \doteq \left.
\mathcal{N}(\mathbf{x}; \boldsymbol{\mu}, \sigma^2 \texttt{Id})
\right\vert_{\sigma=1 \text{\,pixel}}$. We ignore event polarities when
constructing the IWE, but one could choose to construct a two-channel IWE or a
summation of the two polarities. The \emph{contrast} of this IWE is then defined
by a CM objective function $f(\boldsymbol{\Theta})$ such as the variance
function,
\begin{equation}
  \begin{aligned}
  & \text{Var}\bigl(I_{\text{events}}(\boldsymbol{\Theta}; \tref)\bigr) \\
  & \doteq \frac{1}{\vert \Omega \vert} \int_{\Omega} \bigl( I_{\text{events}}(\mathbf{x}; \boldsymbol{\Theta}, \tref) - \mu_{I_{\text{events}}(\boldsymbol{\Theta}; \tref)} \bigr)^2 d\mathbf{x},
  \end{aligned}
  \label{eq:variance}
\end{equation}
with mean $\mu_{I_{\text{events}}(\boldsymbol{\Theta}; \tref)} \doteq
\tfrac{1}{\vert \Omega \vert} \int_{\Omega} I_{\text{events}}(\mathbf{x};
\boldsymbol{\Theta}, \tref)\, d\mathbf{x}$, where $\Omega$ denotes the sensor
coordinate subspace. The intention of the CM objective function is to implicitly
measure how well the motion parameters $\boldsymbol{\Theta}$ model the original
motion of the intensity gradients that generated the original events. A higher
measure of \emph{contrast} is correlated with improved alignment of the warped
events as well as more accurate motion parameters. Finally, CM is the task of
maximizing this \emph{contrast} with respect to $\boldsymbol{\Theta}$ in order
to find the optimal motion parameters $\boldsymbol{\Theta}^{\ast} =
\argmax_{\boldsymbol{\Theta}} f(\boldsymbol{\Theta})$.

We utilize the mean squared magnitude of the IWE gradient,
\begin{equation}
  G(\boldsymbol{\Theta}; \tref) \doteq \frac{1}{\vert \Omega \vert} \int_{\Omega} \lVert \nabla I_{\text{events}}(\mathbf{x}; \boldsymbol{\Theta}, \tref) \rVert^2 d\mathbf{x},
  \label{eq:gradient_magnitude}
\end{equation}
as our CM objective function~\cite{gallego2019focus}. We select $G(\cdot)$ since
(i) in contrast to the zeroth-order with $\text{Var}(\cdot)$, it encodes a
first-order spatial constraint by its definition, and (ii) it empirically
provides improved and quicker convergence in the experiments. 
The \emph{relative contrast} used in our \emph{contrast} objective corresponding
to $t_{\text{ref}}=t_0$ is computed as
\begin{equation}
  f_{\text{rel}}(\boldsymbol{\Theta}, \tref) = \frac{G(\boldsymbol{\Theta}; \tref)}{G(\mathbf{0}_{\boldsymbol{\Theta}};-)}.
  \label{eq:rel_contrast}
\end{equation}
Since computing this objective at multiple reference times has been shown to
offer superior performance~\cite{shiba2022secrets,karmokar2025secrets}, we
perform the computation at each $\tref \in t_{\text{multiple}} = \{t_0,
t_{0.25}, t_{0.5}, t_{0.75}, t_1\}$. Lastly, a Gaussian-weighted mean of the
\textit{relative contrast} is computed to be used in our final objective,
\begin{equation}
  f_{\text{rel}}(\boldsymbol{\Theta}) = \frac{\sum_{t \in t_{\text{multiple}}} \mathcal{G}_w(t) f_{\text{rel}}(\boldsymbol{\Theta}, t)}{\sum_{t \in t_{\text{multiple}}} \mathcal{G}_w(t)},
\end{equation}
where $\mathcal{G}_w(t) = \mathcal{N}(t; 0.5, 1)$.

\subsection{Orientation Priors from 3D Camera Velocities}
\label{subsec:orientation_prior_contrast_maximization}
Traditionally, global optical flow patterns can be generated using the relative
motion field equations~\cite{trucco1998introductory} of the form
\begin{equation}
  \begin{aligned}
      \dot{x} &= \frac{v_Z x - v_X f}{Z} - \omega_Y f + \omega_Z y + \omega_X \frac{x y}{f} - \omega_Y \frac{x^2}{f}, \\
      \dot{y} &= \frac{v_Z y - v_Y f}{Z} + \omega_X f - \omega_Z x - \omega_Y \frac{x y}{f} - \omega_X \frac{y^2}{f},
      \label{eq:motion_field_equation}
  \end{aligned}
\end{equation}
where $\boldsymbol{x} = [x, y]^{\top}$ is an image point, $\dot{\boldsymbol{x}}
= [\dot{x}, \dot{y}]^{\top}$ is the image point velocity, and $\boldsymbol{v} =
[v_X, v_Y, v_Z]^{\top}$, $\boldsymbol{\omega} = [\omega_X, \omega_Y,
\omega_Z]^{\top}$, and $Z$ are linear velocity, angular velocity, and depth of
a moving scene point, $\boldsymbol{X} = [X, Y, Z]^{\top}$, relative to the
camera, respectively. In our orientation priors for CM (OPCM) formulation, we
focus on orientation and not on the magnitude of the flow field generated due
to the moving camera. 

\subsubsection{Generating Distorted Orientation Maps}
\label{subsubsec:generating_dist_o_maps}
\begin{figure}[]
    \centering
    \subfloat[Linear velocity.]{
    \includegraphics[height=0.3\linewidth, trim={0 2.5cm 0 0}, clip, cfbox=gray 0pt 0pt]{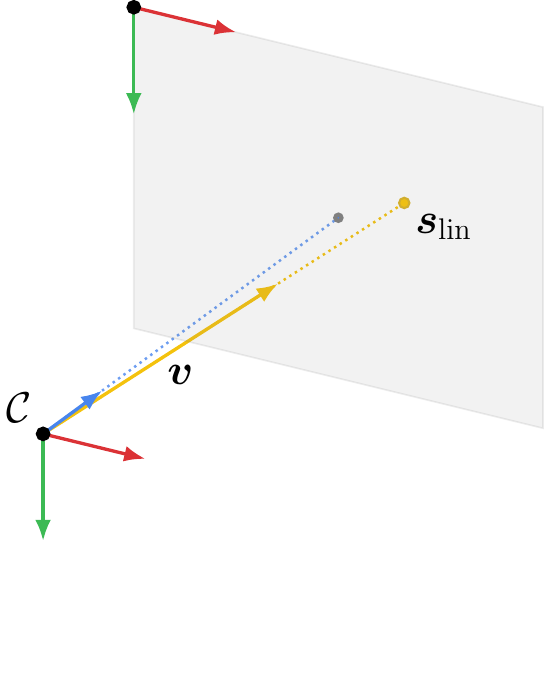} 
    } \hspace{0.1\linewidth}
    \subfloat[Angular velocity.]{
    \includegraphics[height=0.3\linewidth, trim={0 2.5cm 0 0}, clip, cfbox=gray 0pt 0pt]{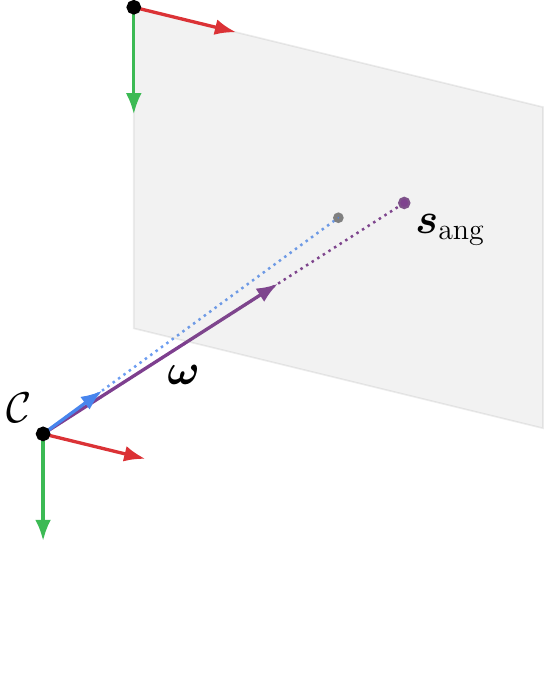} 
    }\\
    \subfloat[Linear orientation map.]{
    \includegraphics[width=0.4\linewidth, cfbox=gray!75!white 0.5pt -0.2pt]{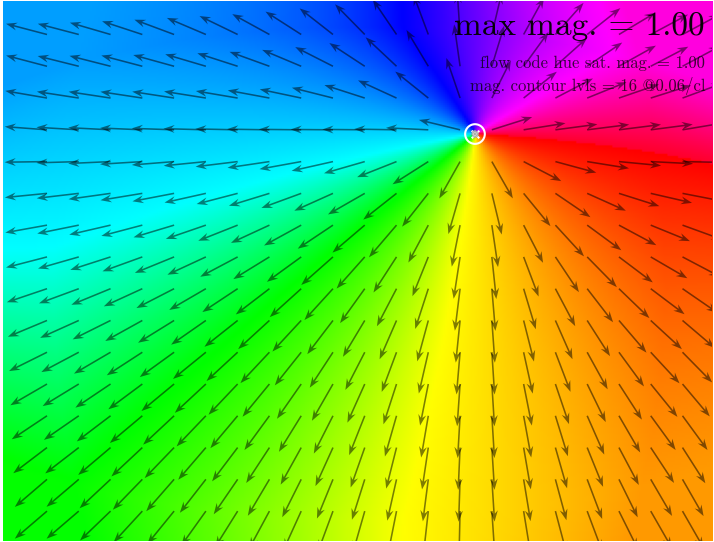} 
    }
    \subfloat[Angular orientation map.]{
    \includegraphics[width=0.4\linewidth, cfbox=gray!75!white 0.5pt -0.2pt]{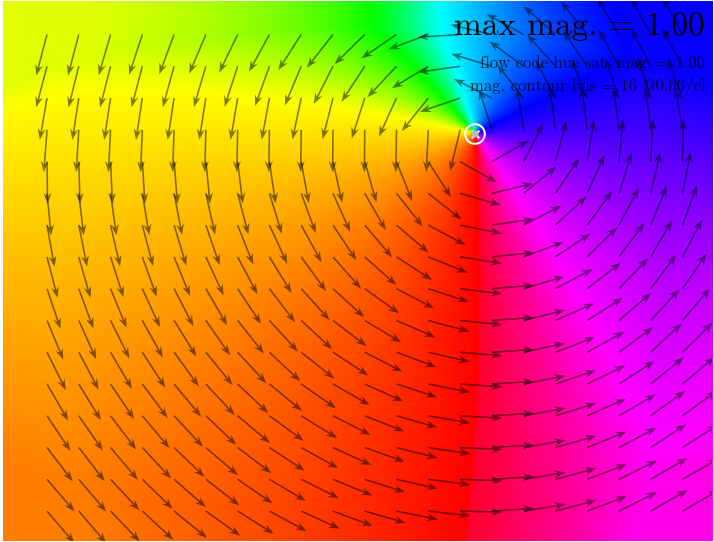} 
    }
    \caption{Generating distorted orientation maps. (a) and (b) depict the
    camera linear and angular velocities along with their corresponding
    intersections $\boldsymbol{s}_{\text{lin}}$ and
    $\boldsymbol{s}_{\text{ang}}$ with the image plane, respectively. (c) and
    (d) demonstrate the final generated linear and angular orientation maps
    corresponding to (a) and (b), respectively.}
    \label{fig:o_maps}
\end{figure}

The distorted orientation maps are extracted in three steps: (i) compute
singularities $\boldsymbol{S} = \{\boldsymbol{s}_{\text{lin}},
\boldsymbol{s}_{\text{ang}}\}$, (ii) generate orientation maps
$\boldsymbol{\mathcal{O}} = \{\mathcal{O}_{\text{lin}},
\mathcal{O}_{\text{ang}}\}$, and (iii) apply lens distortion to yield the
distorted orientation maps
$\fourIdx{\text{dist}}{}{}{}{\boldsymbol{\mathcal{O}}}$. The outcome of this
process is visualized in Fig.~\ref{fig:o_maps}.

\vspace{0.5em}
\noindent\myul{Compute Singularities:}
To obtain the singularities $\boldsymbol{S}$, we project the 3D velocities
expressed in the camera reference frame, $\boldsymbol{v} = [v_X, v_Y,
v_Z]^{\top}$, onto the image plane at $Z=1$ and map to pixel coordinate units
using the camera intrinsic matrix $K$ as follows:
\begin{equation}
    \tilde{\boldsymbol{s}}_{\text{lin}} = 
    \begin{bmatrix}
        {s_x}_{\text{lin}}\\
        {s_y}_{\text{lin}} \\
        1        \\
    \end{bmatrix} = 
    K \boldsymbol{v} = 
    \begin{bmatrix}
        f_x & 0   & c_x \\
        0   & f_y & c_y \\
        0   & 0   & 1   \\
    \end{bmatrix}
    \begin{bmatrix}
        \nicefrac{v_X}{v_Z} \\
        \nicefrac{v_Y}{v_Z} \\
        1                   \\
    \end{bmatrix},
    \label{eq:intrinsic_projection}
\end{equation}
where $\tilde{\boldsymbol{s}}_{\text{lin}}$ is the ideal undistorted
homogeneous pixel coordinate of the intersection of the 3D linear velocity with
the image plane. Similarly, we can also obtain
$\tilde{\boldsymbol{s}}_{\text{ang}} = [{s_x}_{\text{ang}}, {s_y}_{\text{ang}},
1]^{\top} = K\boldsymbol{\omega}$ for the 3D angular velocity.

\vspace{0.5em}
\noindent\myul{Generate Orientation Maps:}
The linear orientation maps are created by
\begin{equation}
    \mathcal{O}_{\text{lin}}(\boldsymbol{x}) = 
    \begin{cases}
        \boldsymbol{x} - \boldsymbol{s}_{\text{lin}}, & v_Z > 0 \\
        \boldsymbol{s}_{\text{lin}} - \boldsymbol{x}, & v_Z < 0 \\
        [v_X, v_Y]^{\top}, & v_Z = 0.
    \end{cases} 
\end{equation}
Likewise, the angular orientation maps are generated by  
\begin{equation}
    \mathcal{O}_{\text{ang}}(\boldsymbol{x}) = 
    \begin{cases}
        e^{-\iota \frac{\pi}{2}} \cdot (\boldsymbol{x} - \boldsymbol{s}_{\text{ang}}), & \omega_Z > 0 \\
        e^{-\iota \frac{\pi}{2}} \cdot (\boldsymbol{s}_{\text{ang}} - \boldsymbol{x}), & \omega_Z < 0 \\
        e^{-\iota \frac{\pi}{2}} \cdot [\omega_X, \omega_Y]^{\top}, & \omega_Z = 0,
    \end{cases}
\end{equation}
where $e^{-\iota \frac{\pi}{2}}$ represents a $90^{\circ}$ clockwise rotation or
the matrix transformation $\begin{bmatrix} 0 & 1 \\ -1 & 0 \end{bmatrix}$. 

\begin{figure}
\centering
\begin{adjustbox}{max width=\linewidth}
\subfloat[Undistorted.]{
  \includegraphics[width=0.25\linewidth,cfbox=SILVER 0.1pt 0pt]{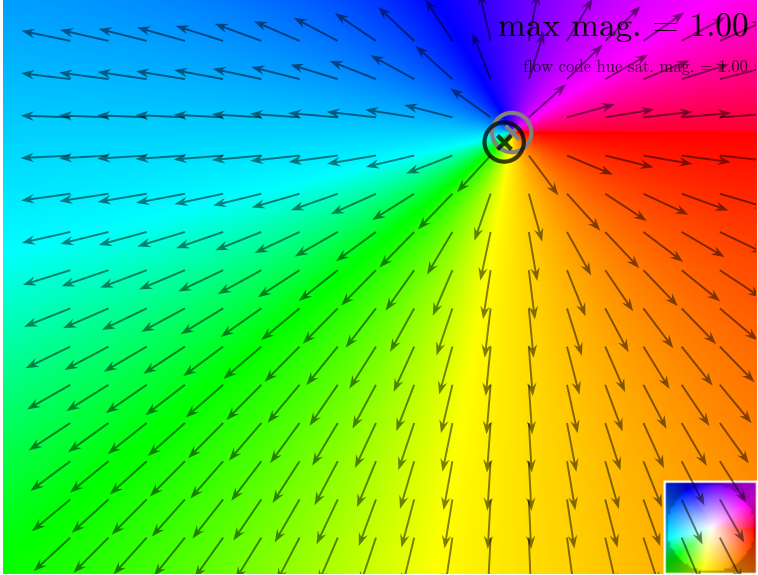}    
}
\subfloat[Distorted.]{
  \includegraphics[width=0.25\linewidth,cfbox=SILVER 0.1pt 0pt]{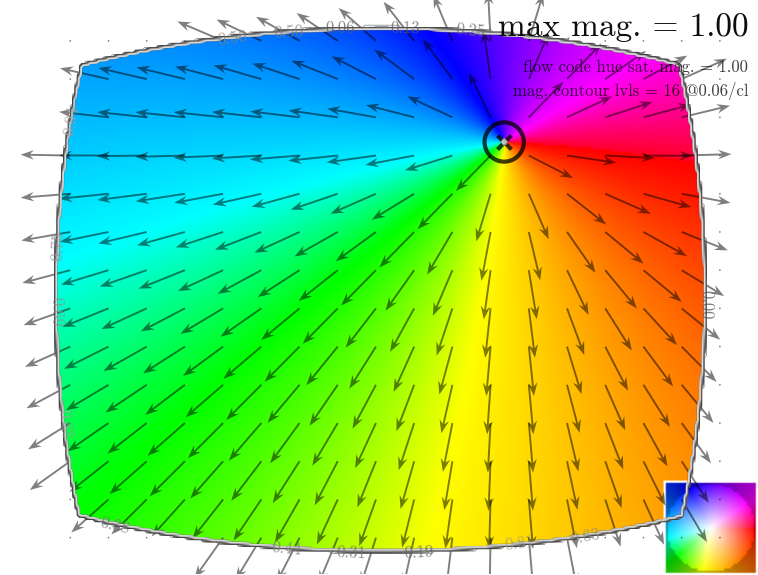}\label{fig:omap_dist_b}
}
\subfloat[Filled.]{
  \includegraphics[width=0.25\linewidth,cfbox=SILVER 0.1pt 0pt]{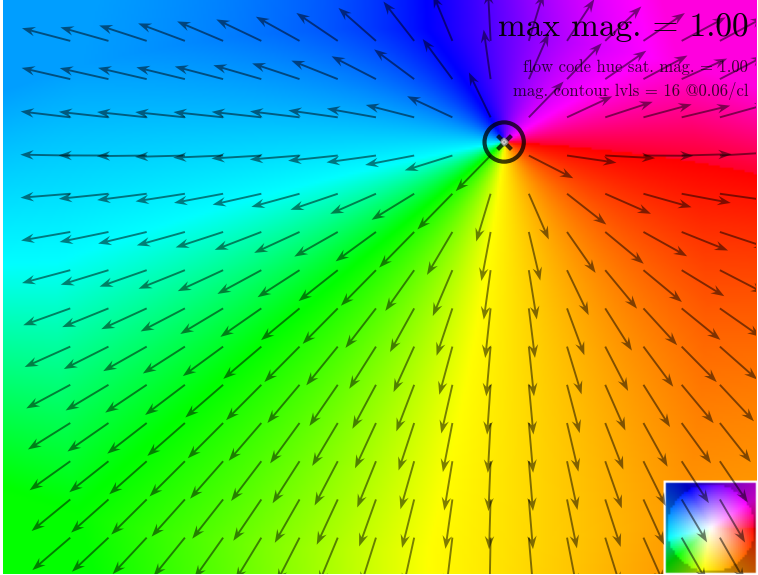}    
}
\subfloat[Difference.]{
  \includegraphics[width=0.25\linewidth,cfbox=SILVER 0.1pt 0pt]{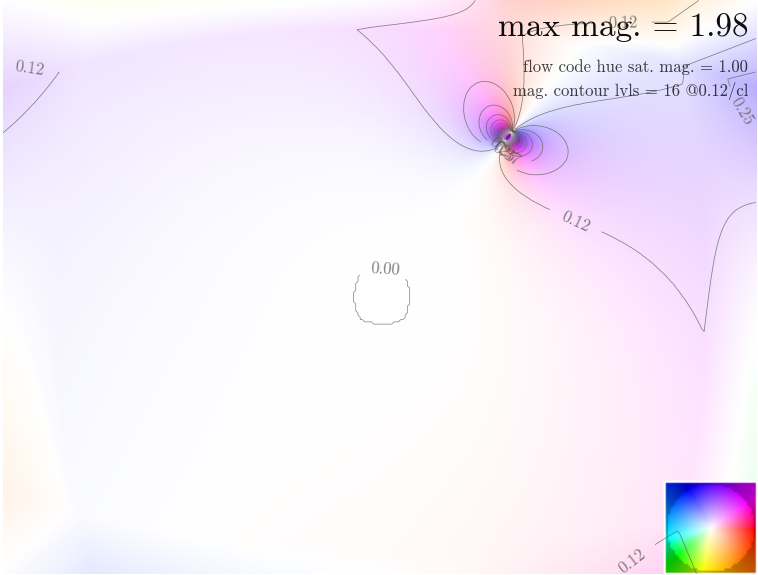}    
}
\end{adjustbox}
\caption{Applying lens distortion to the orientation maps. (a) shows an example
of an undistorted orientation map. (b) illustrates the distorted map with empty
regions as a result of applying lens distortion. (c) and (d) depict the filled
map and its difference with the undistorted map, respectively.} 
\label{fig:omap_dist}
\end{figure}

\vspace{0.5em}
\noindent\myul{Apply Lens Distortion:}
The currently derived orientation maps have not incorporated corrections for 
lens distortion, a factor that can significantly influence the guidance.
To enable lens distortion, we have two options: (i) apply distortion to the
singularity points $\boldsymbol{S} = \{\boldsymbol{s}_{\text{lin}},
\boldsymbol{s}_{\text{ang}}\}$ to obtain
\fourIdx{\text{dist}}{}{}{}{\boldsymbol{S}}, then generate orientation maps
$\fourIdx{\text{dist}}{}{}{}{\boldsymbol{\mathcal{O}}} =
\{\fourIdx{\text{dist}}{}{}{}{\mathcal{O}_{\text{lin}}},
\fourIdx{\text{dist}}{}{}{}{\mathcal{O}_{\text{ang}}}\}$; (ii) generate
orientation maps $\boldsymbol{\mathcal{O}}$ and apply distortion on the map to
obtain $\fourIdx{\text{dist}}{}{}{}{\boldsymbol{\mathcal{O}}}$. Since the lens
distortion parameters obtained from the camera's internal calibration are only
reliable inside the field of view, the first option is not only unreliable, but
also inaccurate. Therefore, we use the second option. Furthermore, applying lens 
distortion empties out the boundary regions of the orientation map, as depicted 
in \cref{fig:omap_dist}. Therefore, we additionally fill these regions utilizing 
the OpenCV library. For more details, please see the supplementary material.

\subsubsection{Contrast Maximization with Orientation Priors}
\label{subsubsec:opcm_objective}
Note that both orientation maps are normalized to have unit magnitude to yield pure
orientation maps $\hat{\mathcal{O}}_{\cdot} =
\nicefrac{\mathcal{O}_{\cdot}}{\vert \mathcal{O}_{\cdot} \vert}$. Subsequently, the
correlation between these maps and the orientation of the motion parameters
($\hat{\boldsymbol{\Theta}} = \nicefrac{\boldsymbol{\Theta}}{\vert
\boldsymbol{\Theta} \vert}$) is quantified. We compute this correlation using 
mean squared error (MSE) scores: 
\begin{equation}
\begin{aligned}
    g_{\text{lin}}(\boldsymbol{\Theta}) &= \frac{1}{\vert \Omega \vert} \int_{\Omega} \left\Vert \bigl( \hat{\boldsymbol{\Theta}}(\boldsymbol{x}) - \hat{\mathcal{O}}_{\text{lin}}(\boldsymbol{x}) \bigr) \right\Vert^{2} d\mathbf{x}, \\
    g_{\text{ang}}(\boldsymbol{\Theta}) &= \frac{1}{\vert \Omega \vert} \int_{\Omega} \left\Vert \bigl( \hat{\boldsymbol{\Theta}}(\boldsymbol{x}) - \hat{\mathcal{O}}_{\text{ang}}(\boldsymbol{x}) \bigr) \right\Vert^{2} d\mathbf{x}.
\end{aligned}
\end{equation}
It is worthwhile to note that since orientation maps are unit vector fields,
other correlation metrics such as the negative cosine similarity or the dot
product are identical to the MSE. For more details, please see the
supplementary material.  

\begin{figure}
    \centering
    \includegraphics[width=\linewidth]{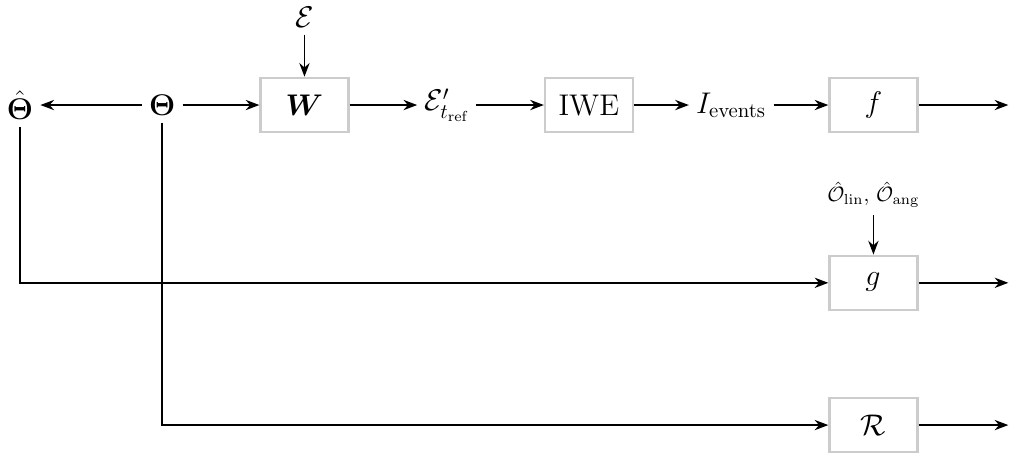}
    \caption[Orientation Prior Contrast Maximization Workflow]{Orientation priors 
    for CM workflow diagram.}
    \label{fig:opcm_workflow}
\end{figure}

Finally, we derive our hybrid objective as
\begin{equation}
    \mathcal{F}(\boldsymbol{\Theta}) = \alpha f_{\text{rel}}(\boldsymbol{\Theta}) + \beta_{\text{lin}} g_{\text{lin}}(\boldsymbol{\Theta}) + \beta_{\text{ang}} g_{\text{ang}}(\boldsymbol{\Theta}),
\end{equation}
where $\alpha$, $\beta_{\text{lin}}$, and $\beta_{\text{ang}}$ are balancing coefficients.
Therefore, our optimization problem is formulated as 
\begin{equation}
    \boldsymbol{\Theta}^{\ast} =
    \argmax_{\boldsymbol{\Theta}} \Big(\mathcal{F}(\boldsymbol{\Theta}) + \lambda\ \text{inv}\bigl(\mathcal{R}(\boldsymbol{\Theta})\bigr)\Big),
\end{equation}
where $\mathcal{R}(\boldsymbol{\Theta})$ is a regularizer and
$\text{inv}(\cdot)$ denotes an additive or multiplicative inverse, \ie, `$-$' or
`$/$', respectively\footnote{The regularization term is included for
completeness. It was set to zero during our method's evaluation. Prior works
\cite{shiba2022secrets,karmokar2025secrets} use the L1 total variation of
$\boldsymbol{\Theta}$ to measure this term.} of the regularizer.
\cref{fig:opcm_workflow} depicts the overall workflow of our proposed framework.

\subsubsection{Multiple Scales} 
\label{subsubsec:mutliple_refs_scales_handover}
Similar to other optimization methodologies, the CM framework exhibits enhanced
performance when provided with effective initial starting points. Previous
state-of-the-art CM methods (\eg,~\cite{shiba2022secrets}) harnessed this
prevalent phenomenon by integrating multi-scale techniques into their
algorithms, thereby significantly improving both the convergence rate and
overall efficacy. The incorporation of multiple scales facilitates a more rapid
convergence process while simultaneously generating superior solutions. Drawing
inspiration from these prior works, we employ a \textit{handover}-based
\cite{karmokar2025secrets} multiple scaling strategy in our implementation.

\section{Experiments}
\label{sec:experiments}
\subsection{Experimental Setup}
\label{subsec:experimental_setup}
All experiments were executed using Python 3.11.11 on a machine with an
Intel(R) Xeon(R) Gold 6330 CPU (2.00 GHz), 512 GB RAM, and two NVIDIA A40 45 GB
GPUs running Ubuntu 22.04. Image processing was performed using OpenCV 4.11.
For the optimization tasks, we employed JAX
0.6.0~\cite{jax2018github,frostig2018highleveltracing} with its support library
\texttt{jaxlib} 0.6.0 and JAXopt 0.8.5~\cite{jaxopt_implicit_diff} using CUDA
toolkit 12.8. We used a quasi-Newton full batch gradient-based optimization
algorithm BFGS~\cite{shanno1970conditioning,shanno1985example}, which is
available in JAXopt through its SciPy wrapper.

\subsection{Datasets}
\label{subsec:datasets}
MVSEC~\cite{zhu2018evflownet} and DSEC~\cite{gehrig2021eraft} are publicly
available real-world datasets primarily for benchmarking event-based optical
flow estimation algorithms. ECD~\cite{mueggler2017event} is a legacy dataset
without optical flow ground truth. However, it has been employed to compare IWE
sharpness scores (\eg,~\cite{shiba2022secrets},~\cite{karmokar2025secrets}).
ECD was created using the DAVIS240C camera, which acquires both images and
events through a single pinhole of a colocated camera system with a hybrid
imaging sensor, offering perfect frame-event spatial alignment. Similarly,
sequences within MVSEC contain stereo frames and events with perfect spatial
alignment due to the use of DAVIS346B cameras. On the other hand, sequences in
DSEC include stereo frames and events captured using separate cameras for each
modality. MVSEC and DSEC provide optical flow ground truth that was generated
from additional sensing modalities such as motion capture and LiDAR. For
example, MVSEC provides odometry data containing 3D linear and angular
velocities of the camera, while DSEC provides LiDAR and IMU data.

The deficiencies in both primary datasets, MVSEC and DSEC, are well known. For
instance, MVSEC sequences contain smaller displacements and lack temporal
synchronization with the ground truth, leading to a deficiency in accuracy.
Nevertheless, MVSEC has perfect frame-event spatial alignment. On the other
hand, DSEC sequences have superior temporal synchronization and larger
displacements, but have imperfect frame-event spatial alignment due to the use
of separate cameras. Nonetheless, both MVSEC and DSEC are established datasets
for benchmarking event-based optical flow evaluations by the research
community. Since our approach makes use of 3D camera velocities, we directly
use the available velocities while evaluating on MVSEC. However, for DSEC we
estimate the 3D camera velocities by registering point clouds using the
iterative closest point algorithm and applying Kalman filtering. For additional
details, please refer to the supplementary material.

\subsection{Metrics, Scaling Strategy, Hyperparameters, and Runtimes}
\label{subsec:metrics_scaling_strategy_and_hyperparameters}
To facilitate reliable comparisons with prior works
(\eg,~\cite{zhu2018evflownet,gehrig2021eraft,brebion2021realtimeflow,shiba2022secrets,karmokar2025secrets}),
we made use of the average endpoint error (AEE), the average $n$-pixel error
percentage (outlier percentage), and the flow warp loss (FWL) as metrics for
quantitative evaluation. For the ECD and MVSEC evaluations, four pyramid levels
were utilized to optimize the motion parameters, where the coarse-to-fine
resolutions ranged from $1\times1$ to $8\times8$. Similarly, five pyramid
levels were used for DSEC such that the native spatial resolution for the
motion parameters at the finest scale was $16\times16$. Bilinear upsampling was
used to upscale the finest scale motion parameters to the sensor size. During
optimizations on MVSEC, we employed 30 K and 40 K events for indoor and outdoor
scenes, respectively. For DSEC and ECD, we used 1.5 M and 30 K events,
respectively. To balance the \textit{contrast} objective
$f_{\text{rel}}(\boldsymbol{\Theta})$, we set $\alpha=20$ for ECD and MVSEC,
while for DSEC $\alpha=5000$. For ECD and MVSEC, $\beta_{\text{lin}}=1$ and
$\beta_{\text{ang}}=0.1$ was used for all experiments to balance
$g_{\text{lin}}(\boldsymbol{\Theta})$ and
$g_{\text{ang}}(\boldsymbol{\Theta})$, respectively, except for
\texttt{outdoor\_day1}, where $\beta_{\text{lin}}=2$ and
$\beta_{\text{ang}}=0.05$. For the DSEC experiments, $\beta_{\text{lin}}=500$
and $\beta_{\text{ang}}=100$. Our JAX-based implementation runs at
$\approx20\,$ms, $\approx38\,$ms, and $\approx140\,$ms per data sample for ECD,
MVSEC, and DSEC sequences, respectively.  Note that throughout our experiments,
we do not use the regularization term $\mathcal{R}(\boldsymbol{\Theta})$ by
explicitly setting $\lambda=0$.

\subsection{MVSEC Evaluations}
Quantitative evaluations against SL, SSL, USL, and model-based methods on MVSEC
are reported in \cref{tab:mvsec_accuracies}. Comparisons for the $\emph{dt}=1$
case, indicating grayscale time intervals corresponding to a single frame
($\approx22\,$ms for \texttt{outdoor\_day1}), are provided in the upper section
of \cref{tab:mvsec_accuracies}, while for the $\emph{dt}=4$ case, indicating
grayscale time intervals corresponding to 4 frames ($\approx89\,$ms for
\texttt{outdoor\_day1}), is provided in the lower section of
\cref{tab:mvsec_accuracies}. Additionally, we report IWE sharpness comparisons
in \cref{tab:fwl_scores}. Our method, OPCM, shows a significant improvement in
accuracy over the previous baseline and establishes new benchmarks with
superior accuracy performance over both learning-based and model-based
approaches. In particular, OPCM evaluates to higher accuracies over the
previous state-of-the-art model-based
approaches~\cite{shiba2022secrets,karmokar2025secrets} while being comparable
to learning-based techniques. 

\begin{figure*}
    \centering
    \begin{adjustbox}{max width=\boxscaler\textwidth}
    \begin{tabular}{@{}c@{\thinspace}c@{\thinspace}c@{\thinspace}c@{\thinspace}|@{\thinspace}c@{\thinspace}c@{\thinspace}c@{}}
        \vspace{-2pt}
        \multirow{2}{*}{\rotatebox[origin=c]{90}{\begin{adjustbox}{max width=\labelscaler\textwidth} \texttt{indoor\_flying1} \end{adjustbox}}} &
        \begin{tabular}{@{}c@{}} \includegraphics[height=0.15\textwidth, cfbox=gray 0pt 0pt]{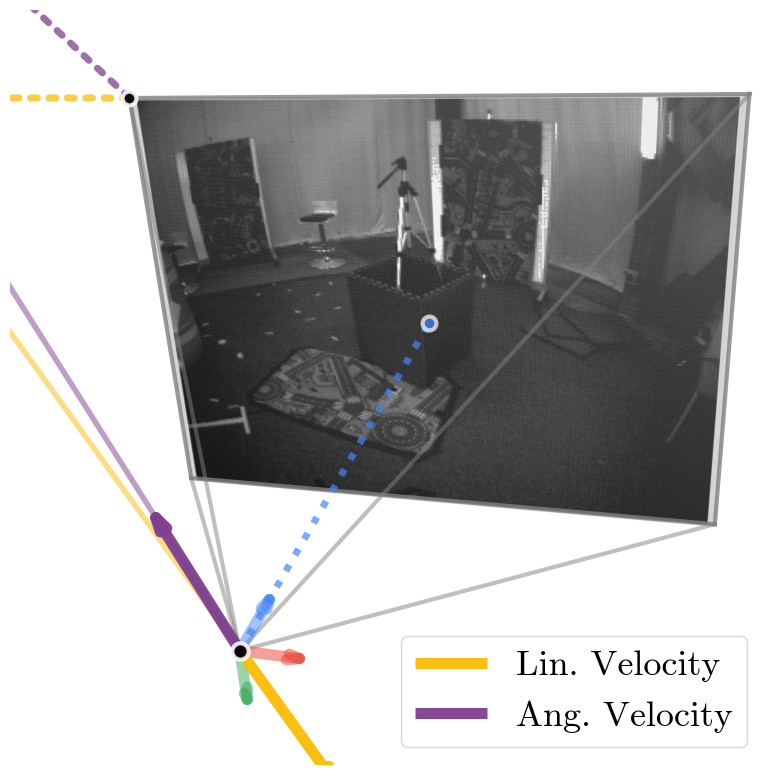} \end{tabular} &      
        \begin{tabular}{@{}c@{}} \includegraphics[width=0.2\textwidth, cfbox=gray 0.1pt 0pt]{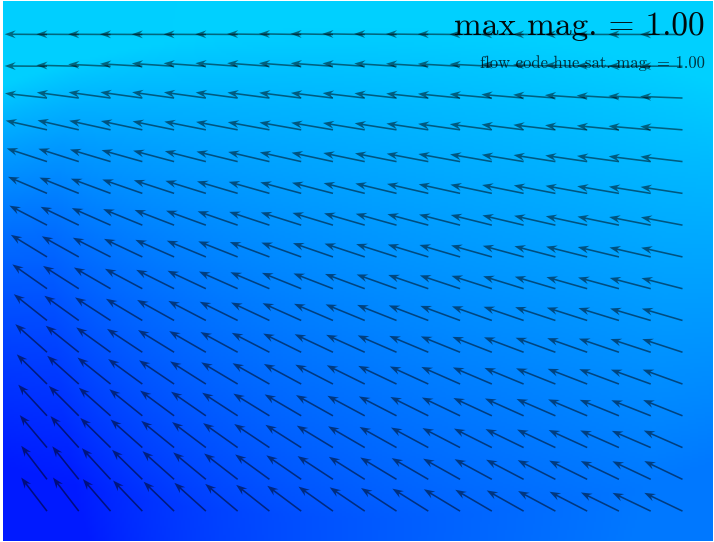} \end{tabular} &      
        \begin{tabular}{@{}c@{}} \includegraphics[width=0.2\textwidth, cfbox=gray 0.1pt 0pt]{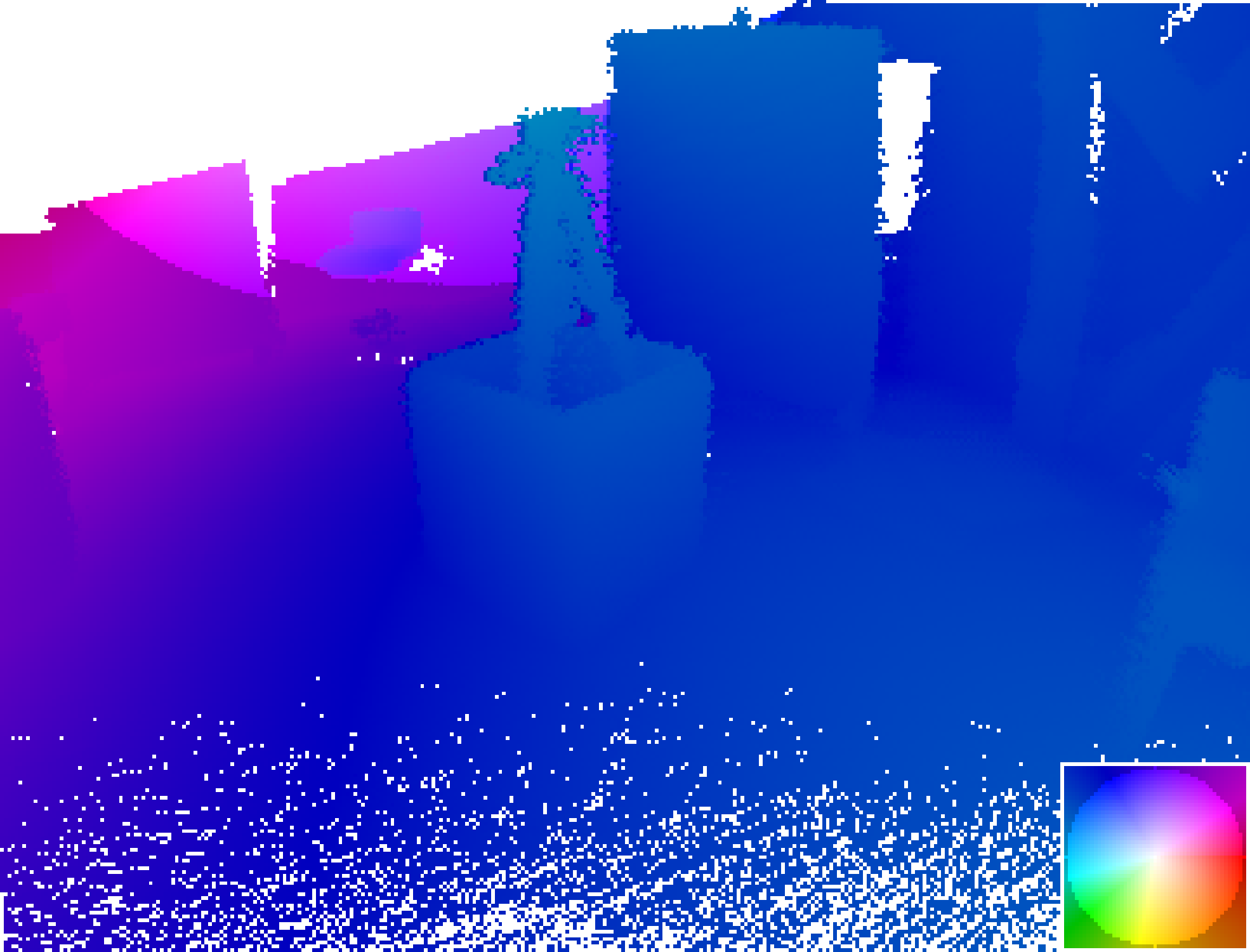} \end{tabular} &        
        \begin{tabular}{@{}c@{}} \includegraphics[width=0.2\textwidth, cfbox=gray 0.1pt 0pt]{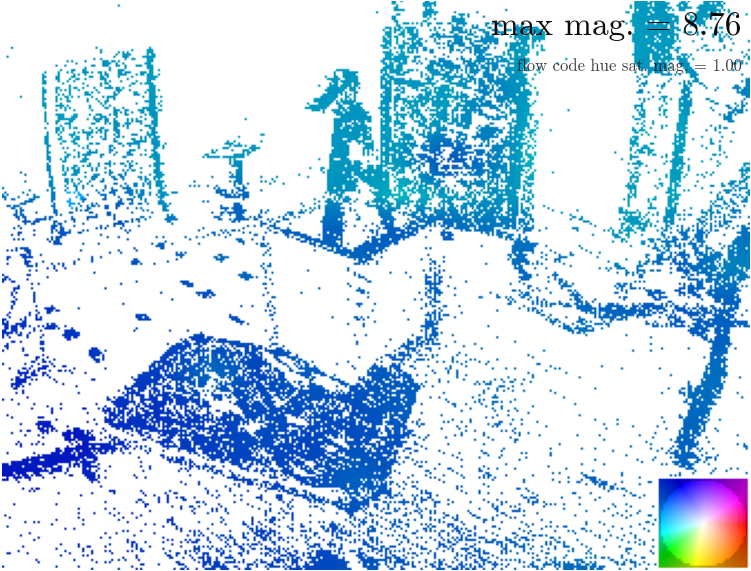} \end{tabular} &  
        \begin{tabular}{@{}c@{}} \includegraphics[width=0.203\textwidth, cfbox=gray 0.1pt 0pt]{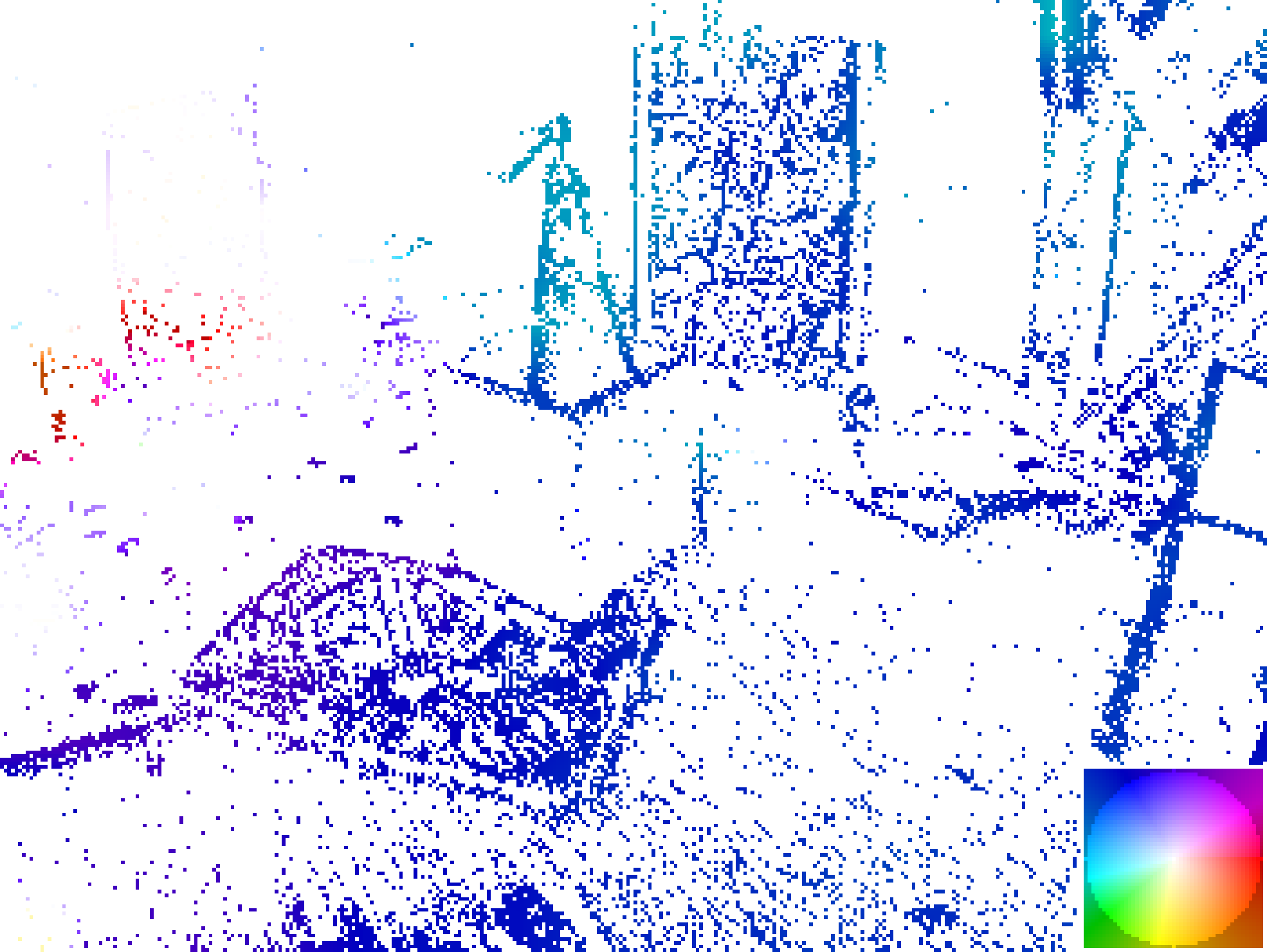} \end{tabular} &                         
        \begin{tabular}{@{}c@{}} \includegraphics[width=0.152\textwidth, cfbox=gray 0.1pt 0pt]{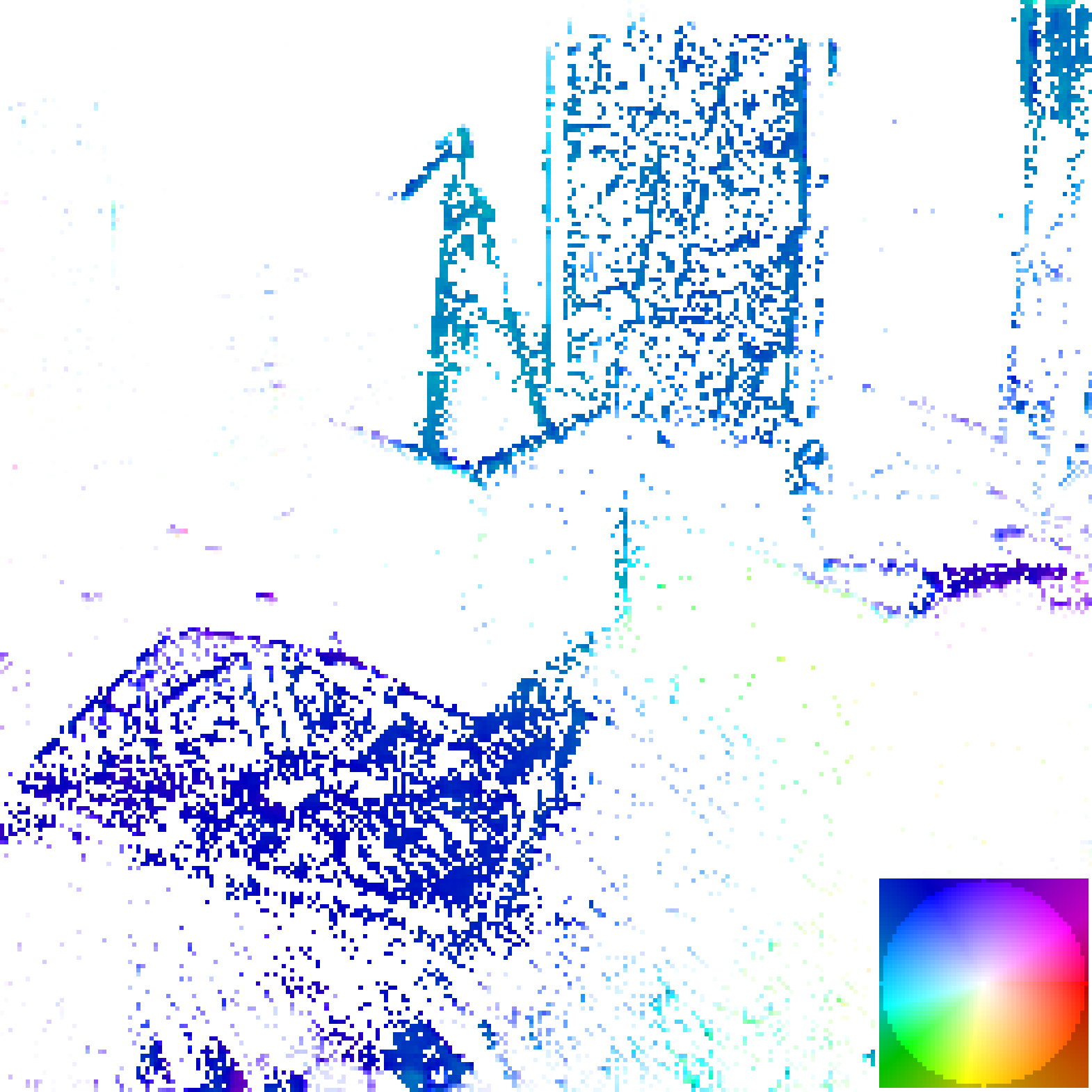} \end{tabular} \\ \vspace{-2pt}          

        &
        \begin{tabular}{@{}c@{}} \includegraphics[width=0.2\textwidth, cfbox=gray 0.1pt 0pt]{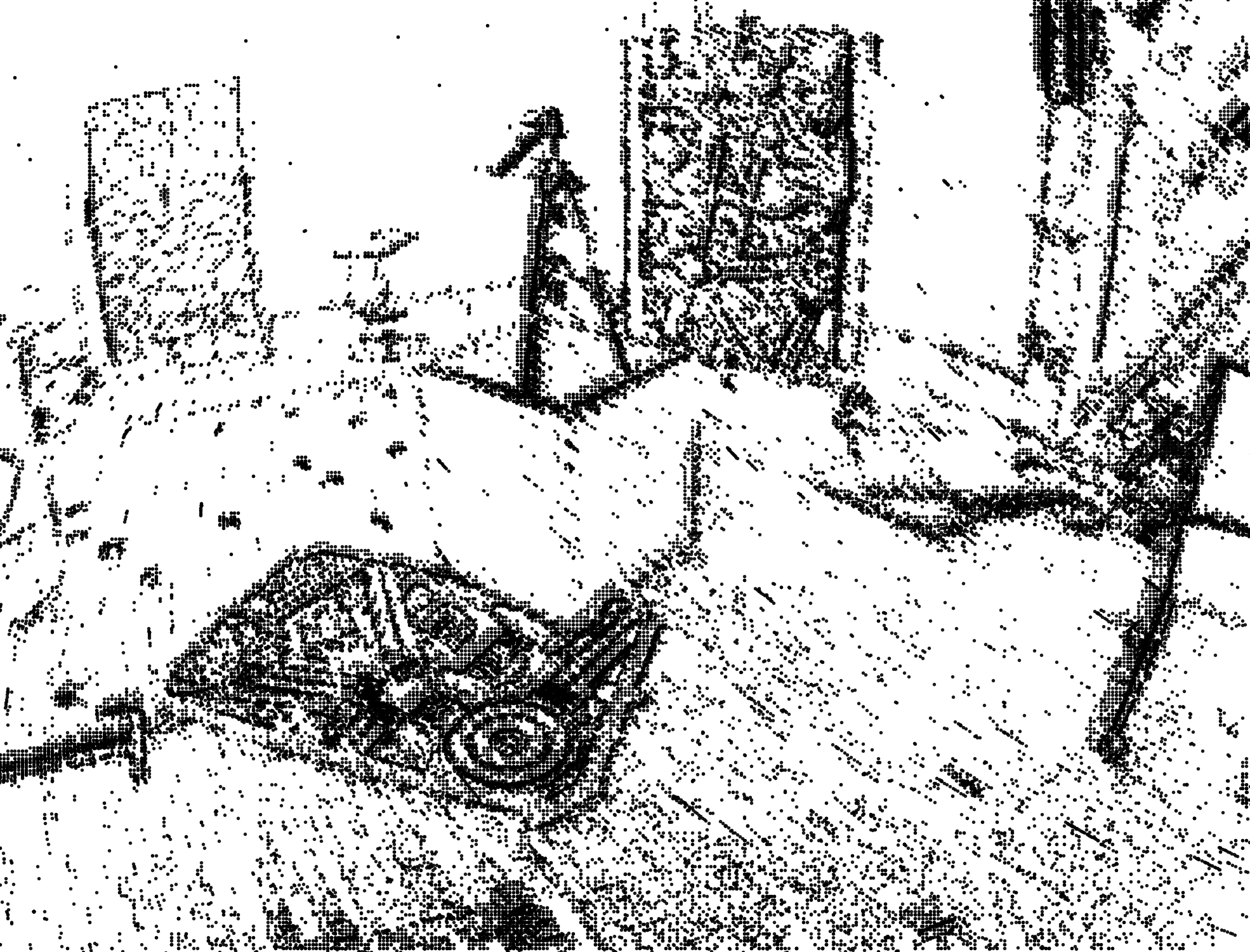} \end{tabular} &       
        \begin{tabular}{@{}c@{}} \includegraphics[width=0.2\textwidth, cfbox=gray 0.1pt 0pt]{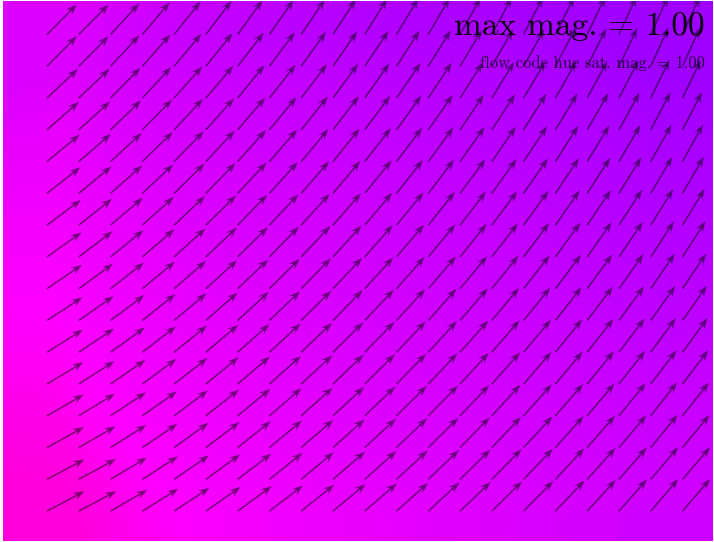} \end{tabular} &       
        \begin{tabular}{@{}c@{}} \includegraphics[width=0.2\textwidth, cfbox=gray 0.1pt 0pt]{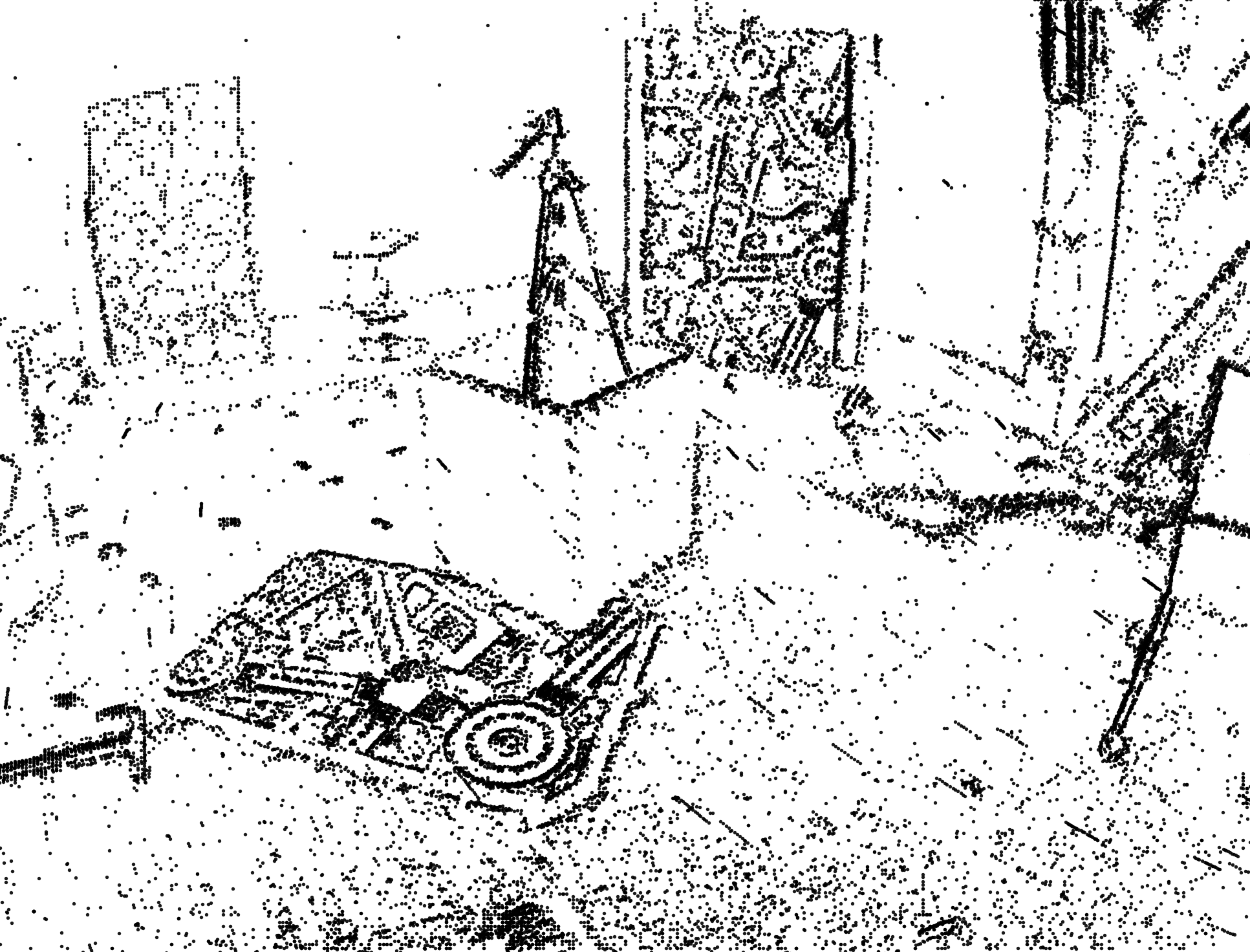} \end{tabular} &    
        \begin{tabular}{@{}c@{}} \includegraphics[width=0.2\textwidth, cfbox=gray 0.1pt 0pt]{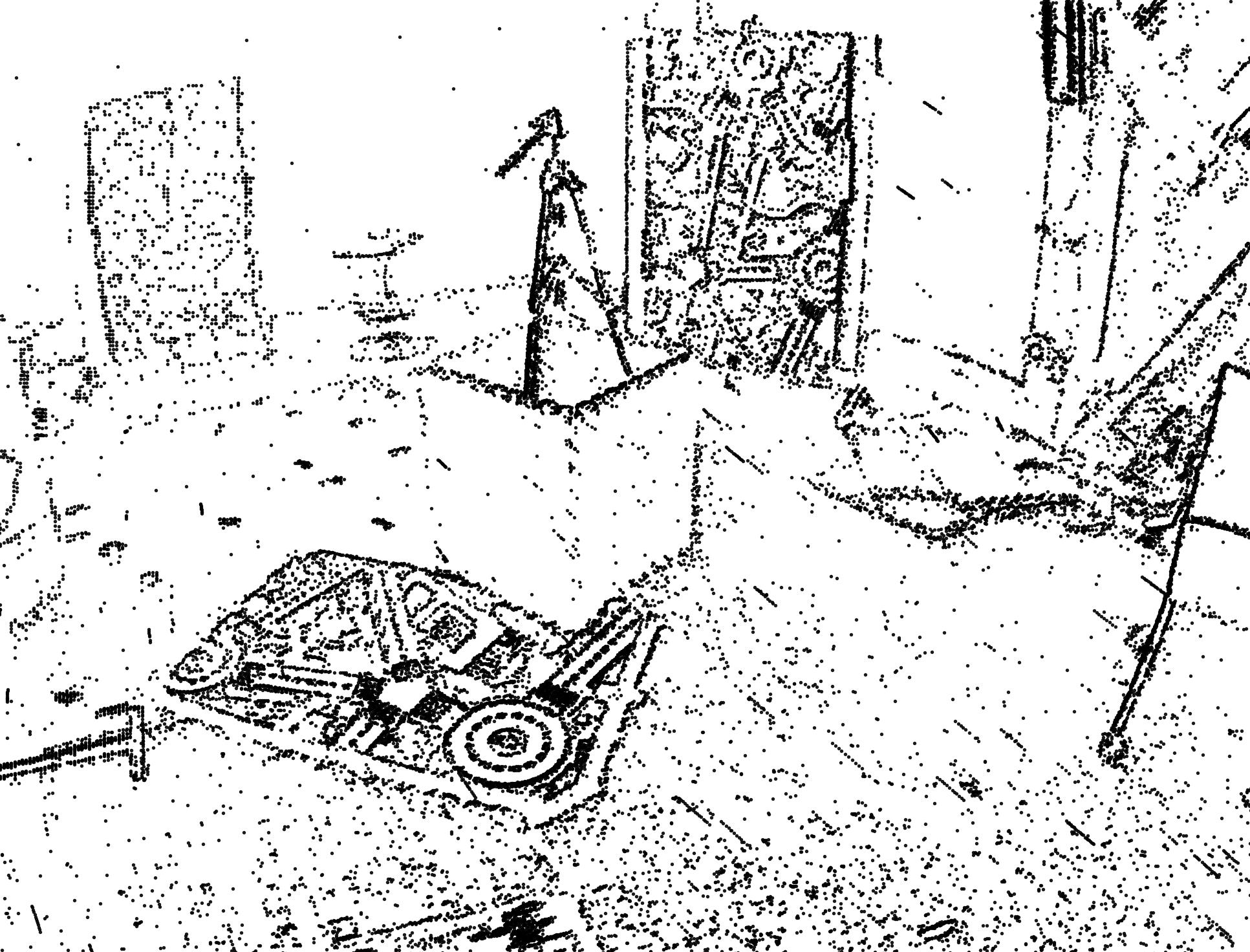} \end{tabular} & 
        \begin{tabular}{@{}c@{}} \includegraphics[width=0.2\textwidth, cfbox=gray 0.1pt 0pt]{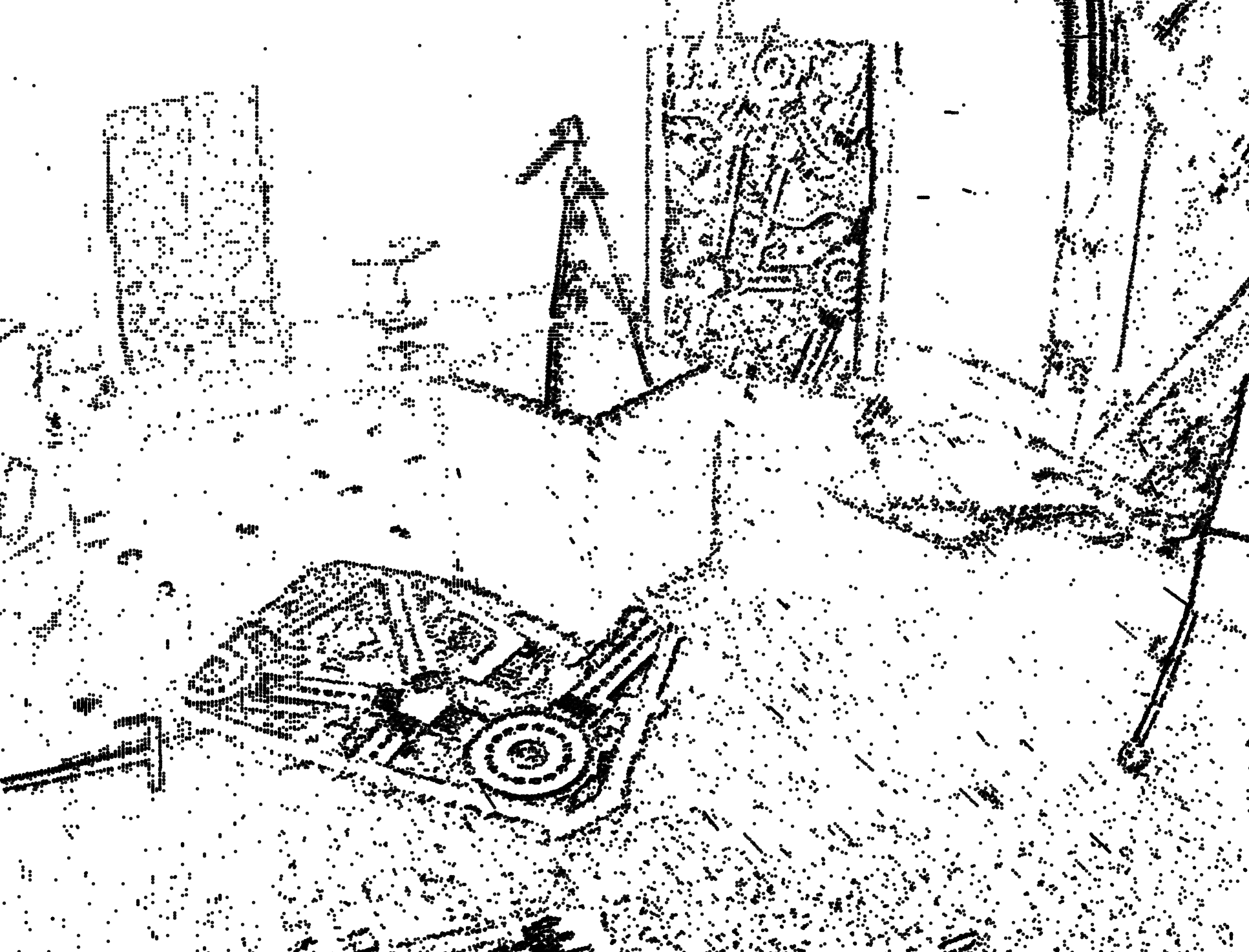} \end{tabular} &                         
        \begin{tabular}{@{}c@{}} \includegraphics[width=0.152\textwidth, cfbox=gray 0.1pt 0pt]{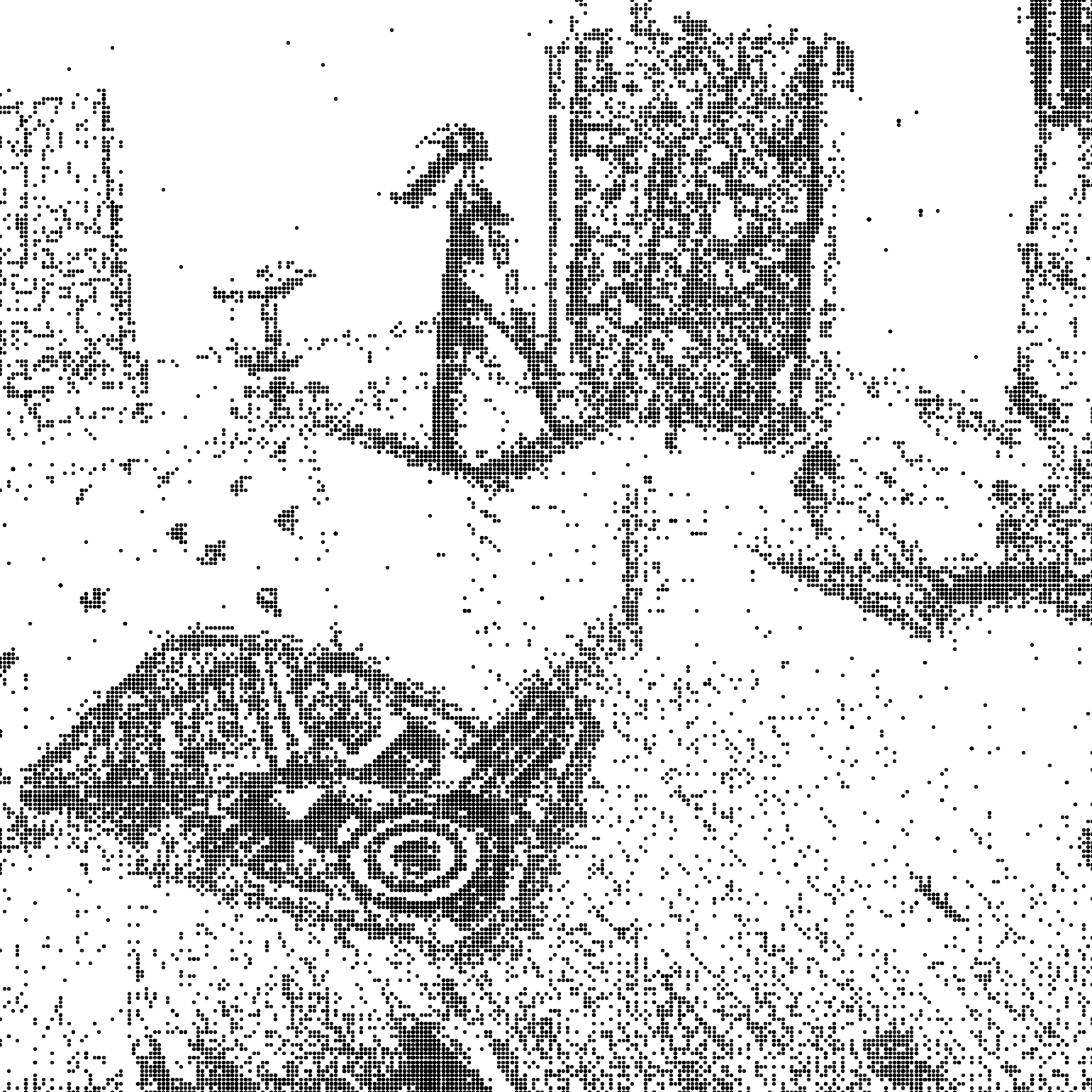} \end{tabular} \\ \vspace{-2pt}          

        \multirow{2}{*}{\rotatebox[origin=c]{90}{\begin{adjustbox}{max width=\labelscaler\textwidth} \texttt{indoor\_flying2} \end{adjustbox}}} &
        \begin{tabular}{@{}c@{}} \includegraphics[height=0.15\textwidth, cfbox=gray 0pt 0pt]{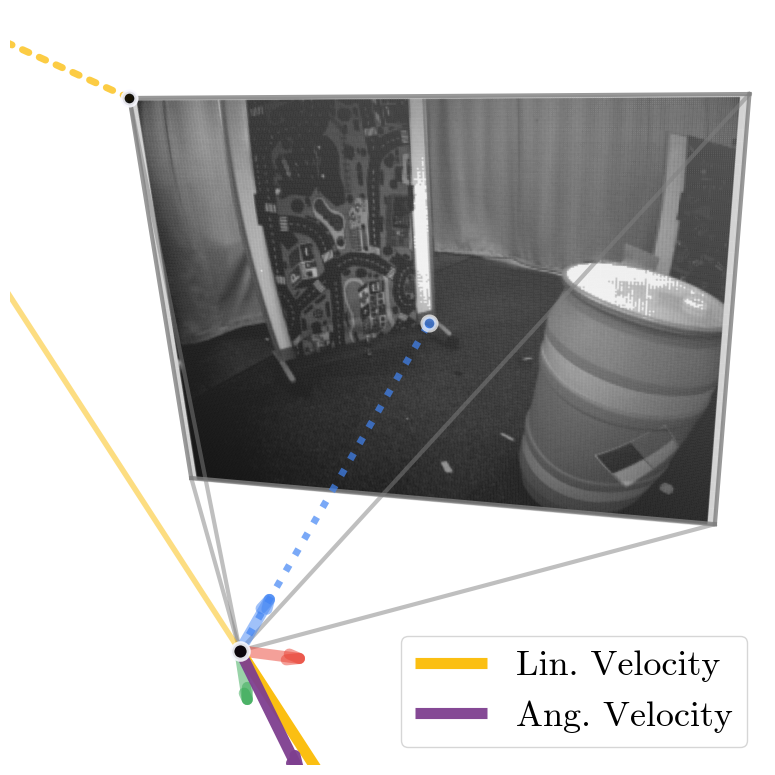} \end{tabular} &      
        \begin{tabular}{@{}c@{}} \includegraphics[width=0.2\textwidth, cfbox=gray 0.1pt 0pt]{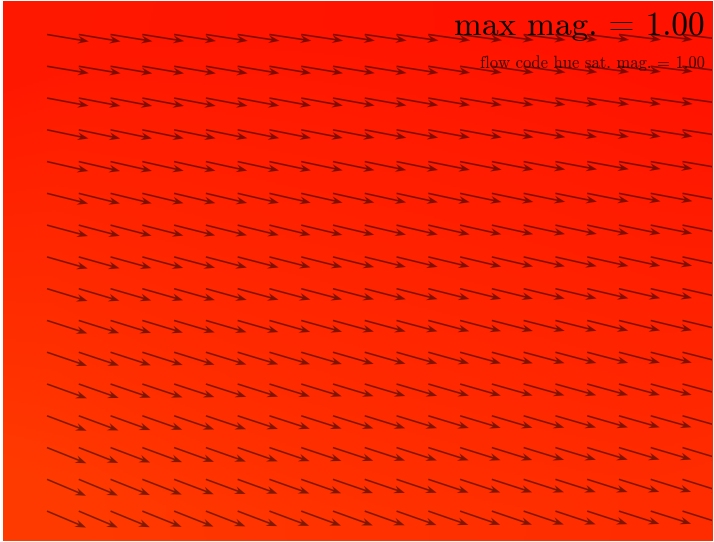} \end{tabular} &      
        \begin{tabular}{@{}c@{}} \includegraphics[width=0.2\textwidth, cfbox=gray 0.1pt 0pt]{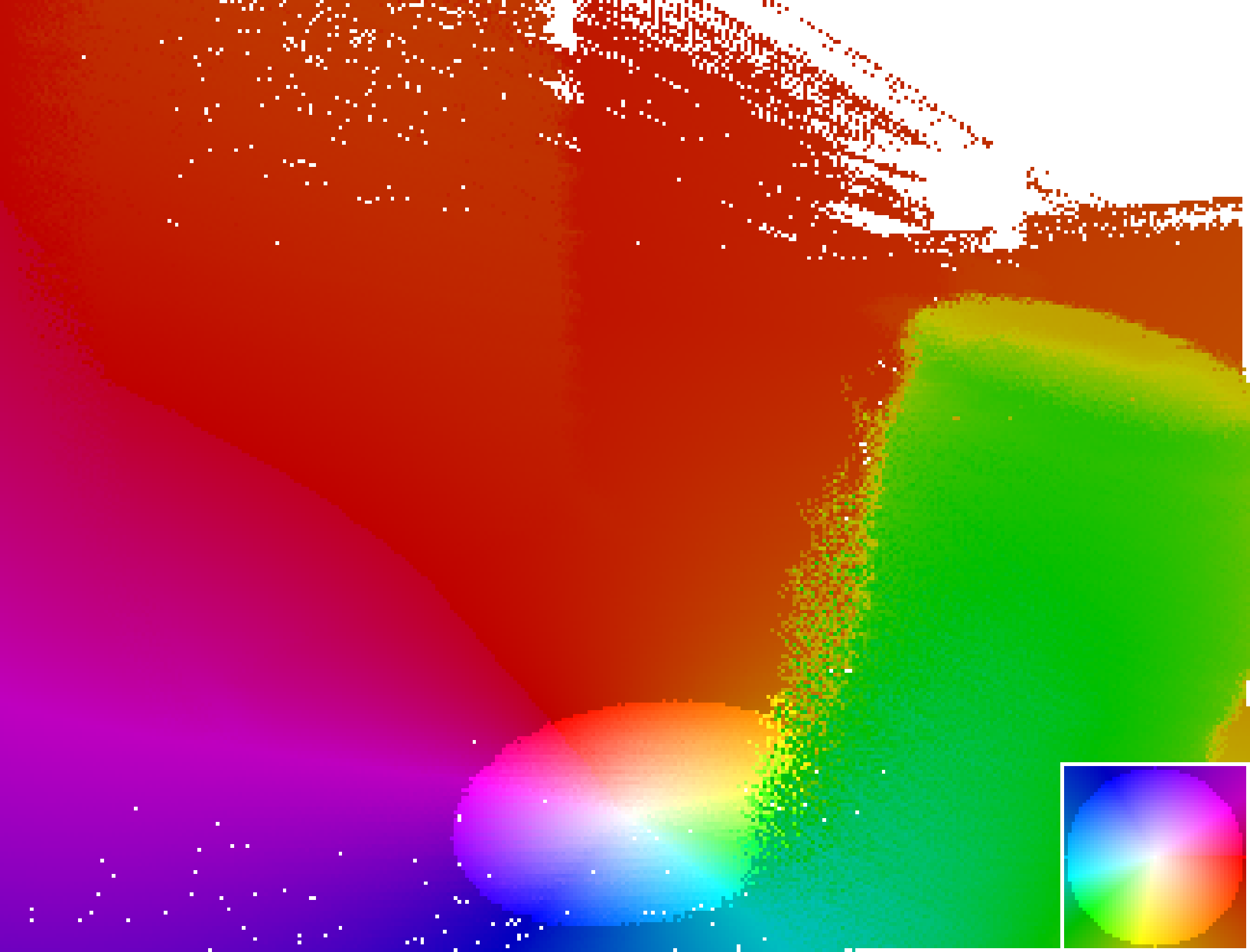} \end{tabular} &        
        \begin{tabular}{@{}c@{}} \includegraphics[width=0.2\textwidth, cfbox=gray 0.1pt 0pt]{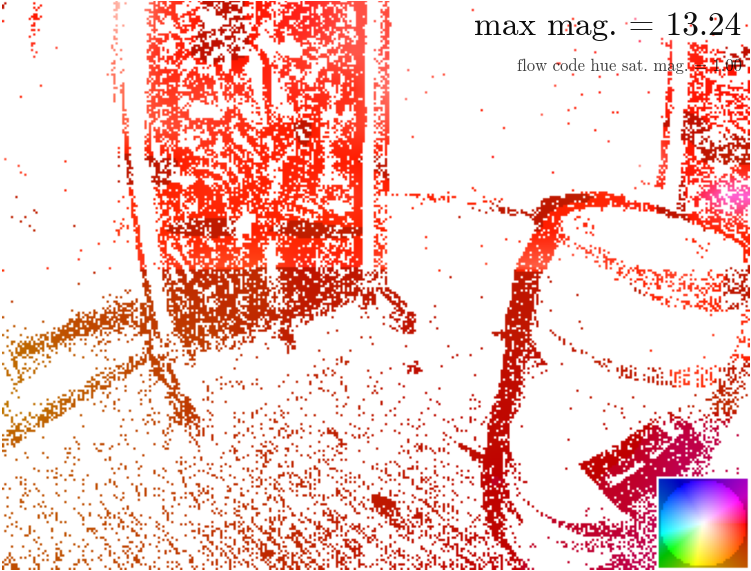} \end{tabular} &  
        \begin{tabular}{@{}c@{}} \includegraphics[width=0.203\textwidth, cfbox=gray 0.1pt 0pt]{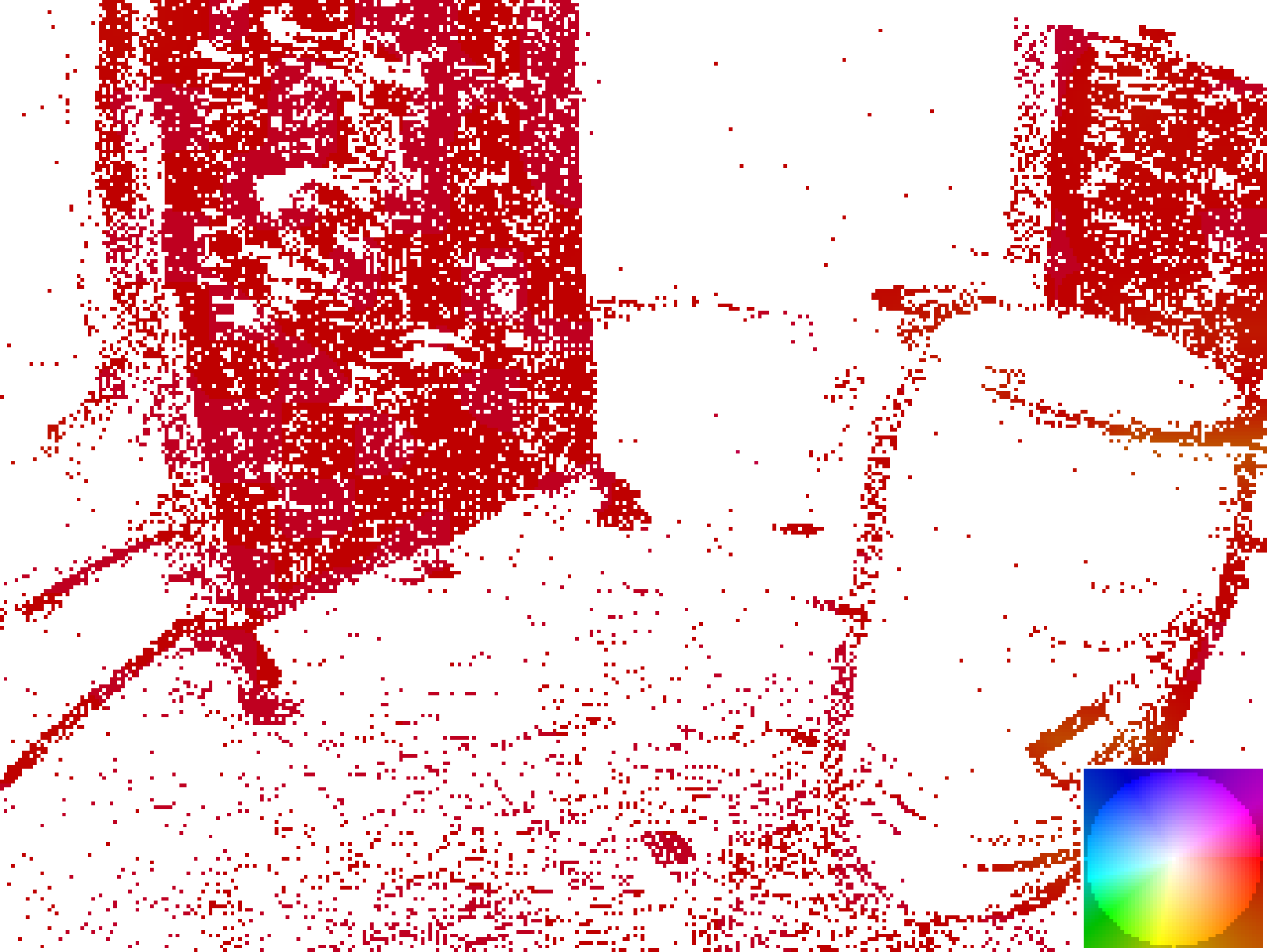} \end{tabular} &                         
        \begin{tabular}{@{}c@{}} \includegraphics[width=0.152\textwidth, cfbox=gray 0.1pt 0pt]{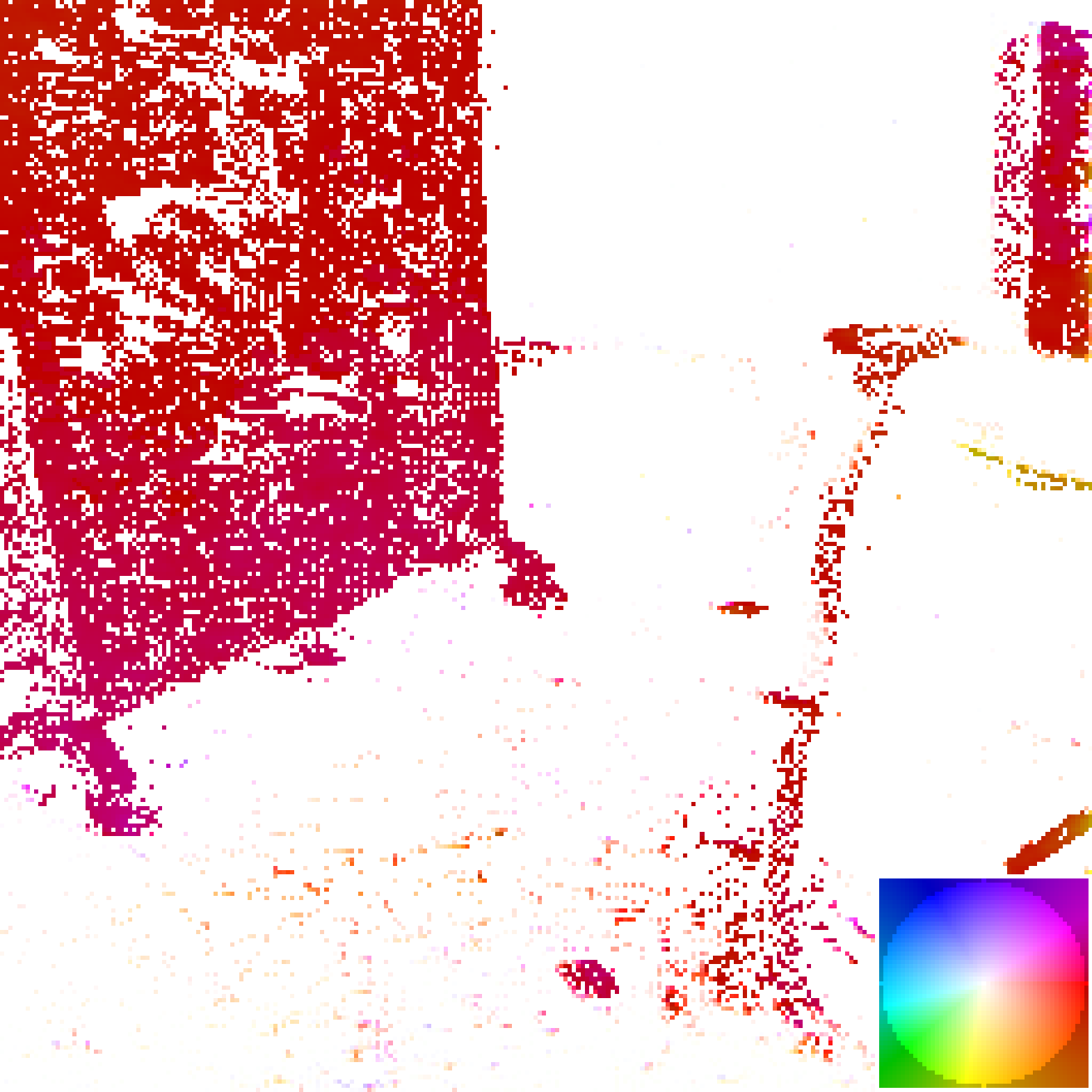} \end{tabular} \\ \vspace{-2pt}          

        &
        \begin{tabular}{@{}c@{}} \includegraphics[width=0.2\textwidth, cfbox=gray 0.1pt 0pt]{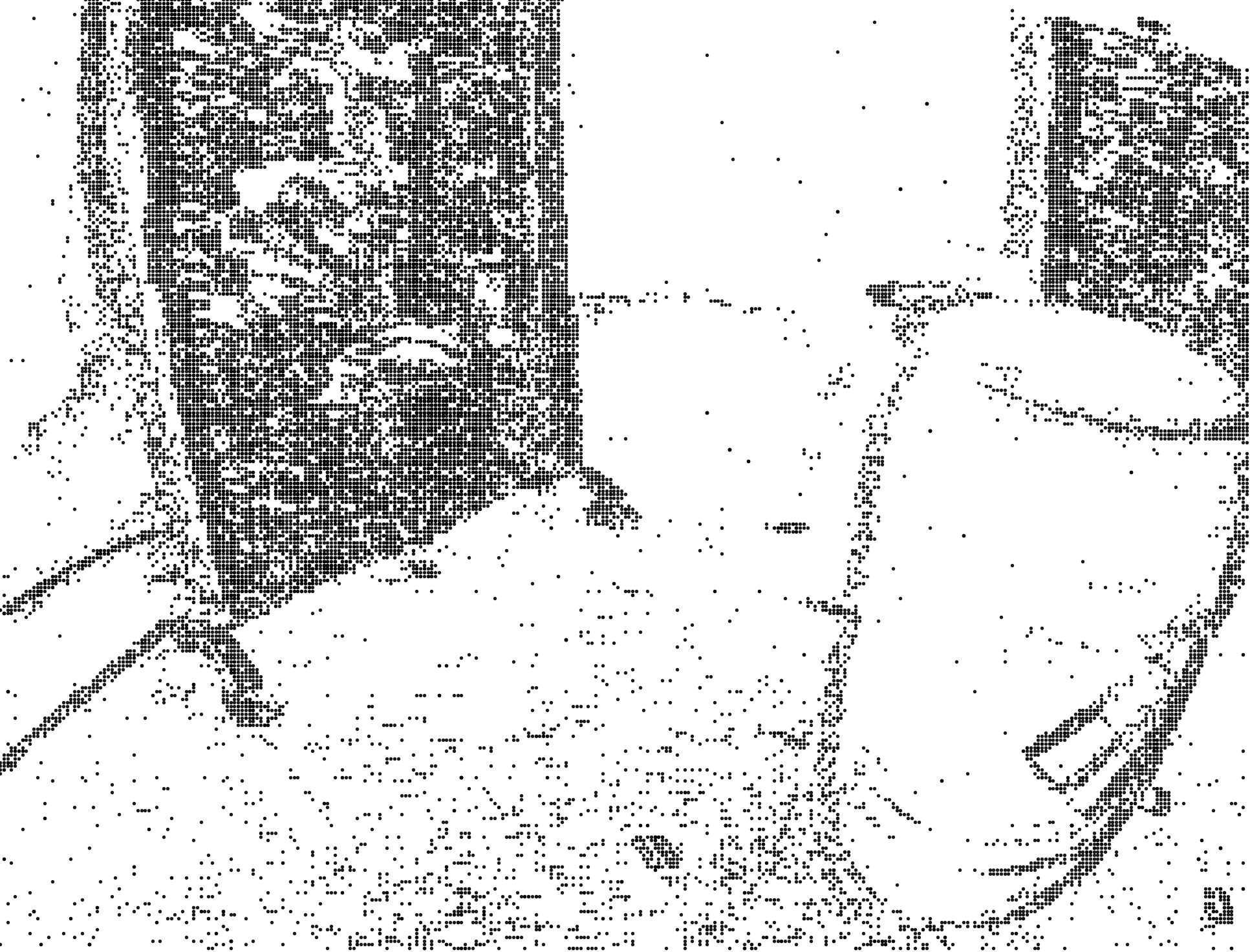} \end{tabular} &       
        \begin{tabular}{@{}c@{}} \includegraphics[width=0.2\textwidth, cfbox=gray 0.1pt 0pt]{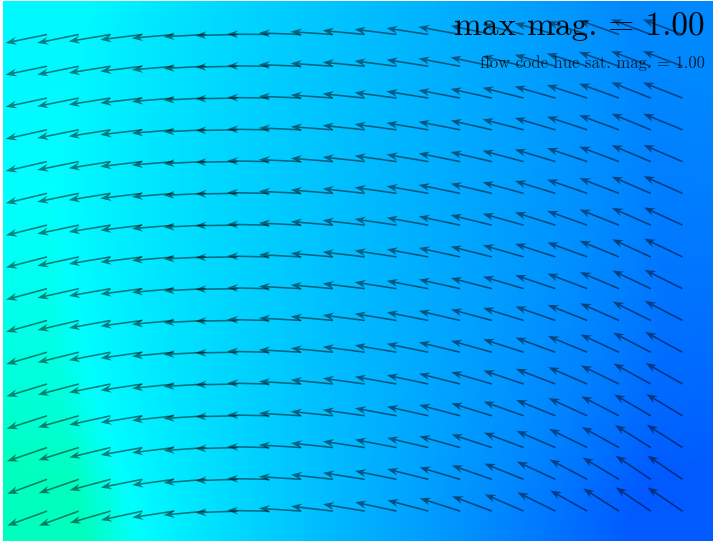} \end{tabular} &       
        \begin{tabular}{@{}c@{}} \includegraphics[width=0.2\textwidth, cfbox=gray 0.1pt 0pt]{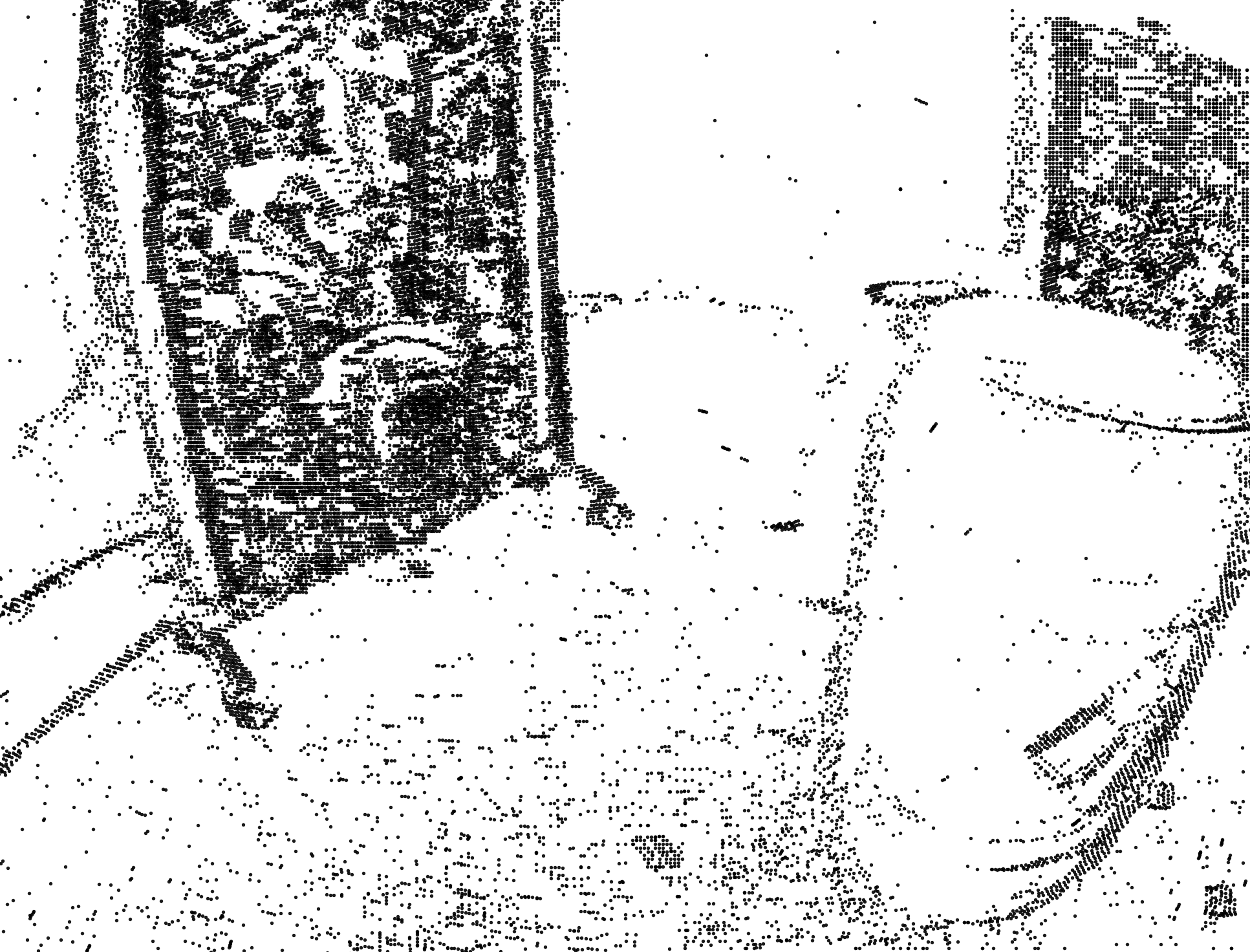} \end{tabular} &    
        \begin{tabular}{@{}c@{}} \includegraphics[width=0.2\textwidth, cfbox=gray 0.1pt 0pt]{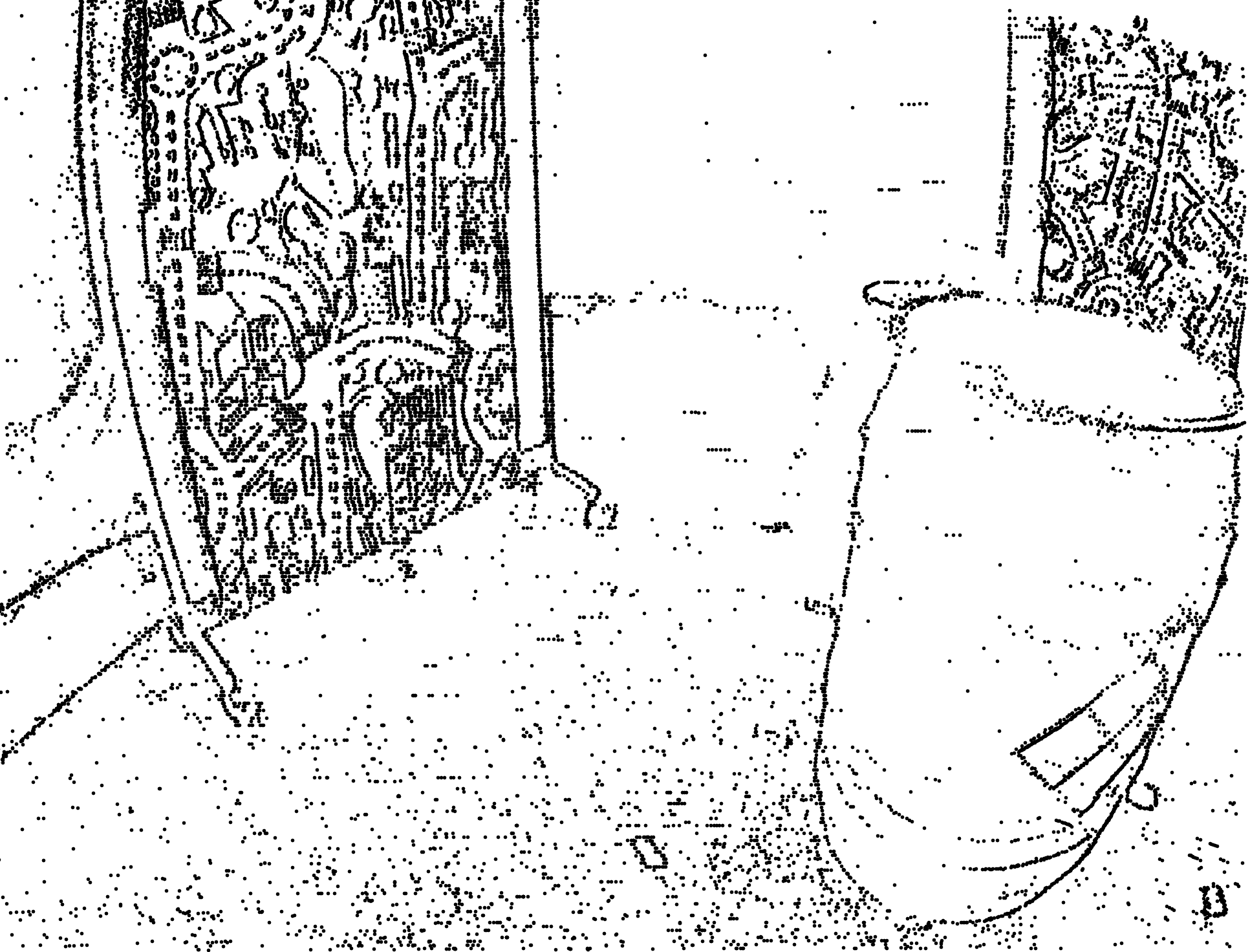} \end{tabular} & 
        \begin{tabular}{@{}c@{}} \includegraphics[width=0.2\textwidth, cfbox=gray 0.1pt 0pt]{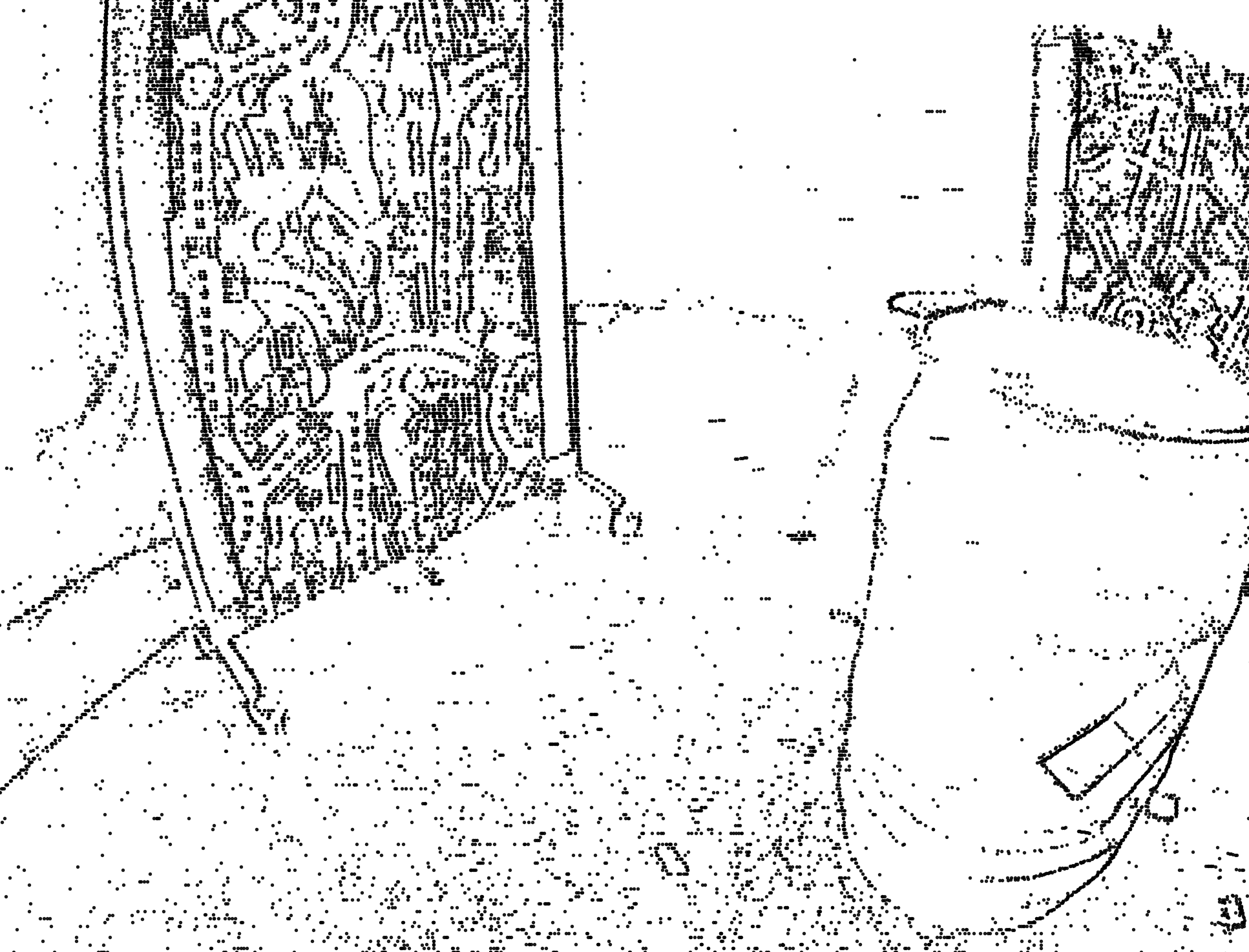} \end{tabular} &                         
        \begin{tabular}{@{}c@{}} \includegraphics[width=0.152\textwidth, cfbox=gray 0.1pt 0pt]{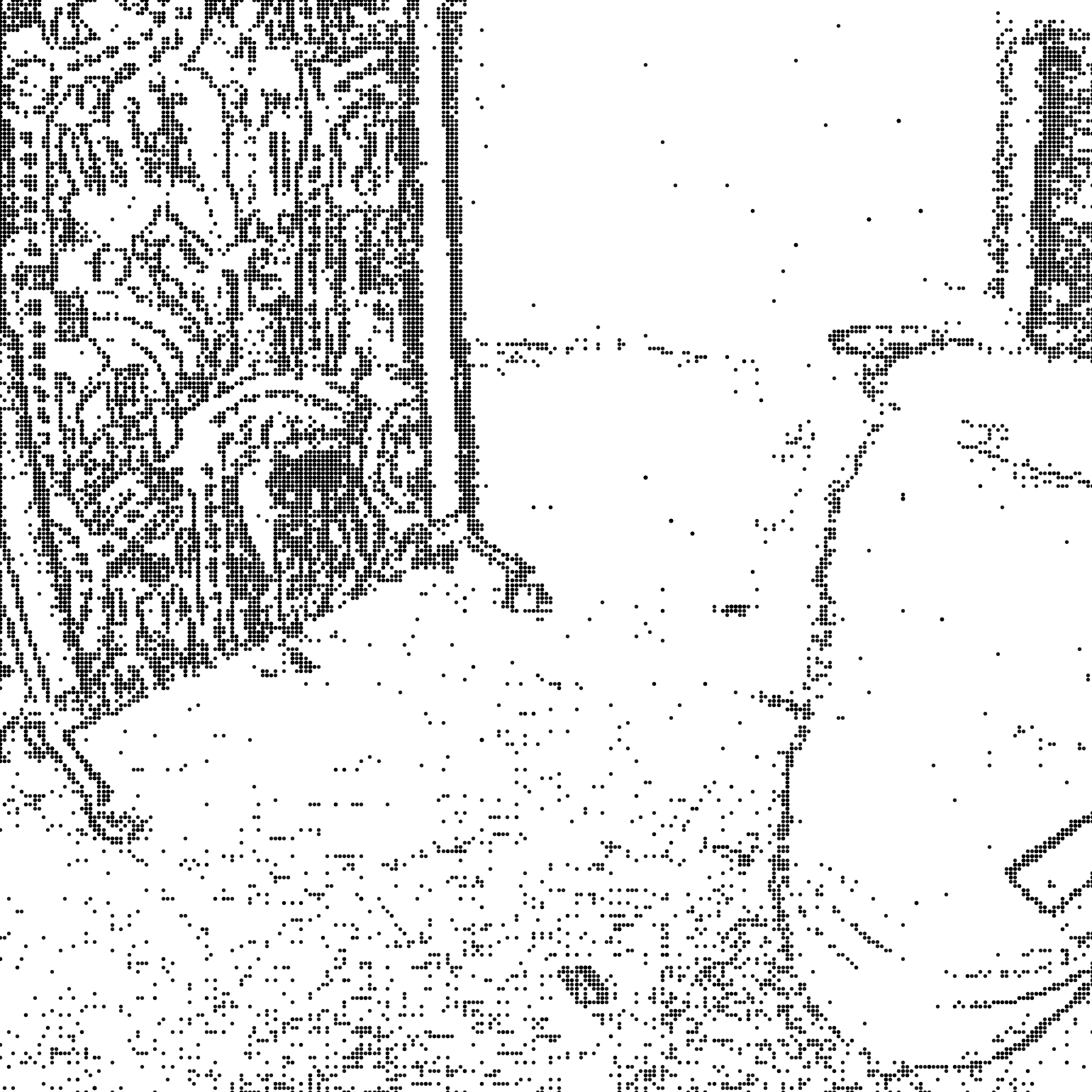} \end{tabular} \\ \vspace{-2pt}          

        \multirow{2}{*}{\rotatebox[origin=c]{90}{\begin{adjustbox}{max width=\labelscaler\textwidth} \texttt{indoor\_flying3} \end{adjustbox}}} &
        \begin{tabular}{@{}c@{}} \includegraphics[height=0.15\textwidth, cfbox=gray 0pt 0pt]{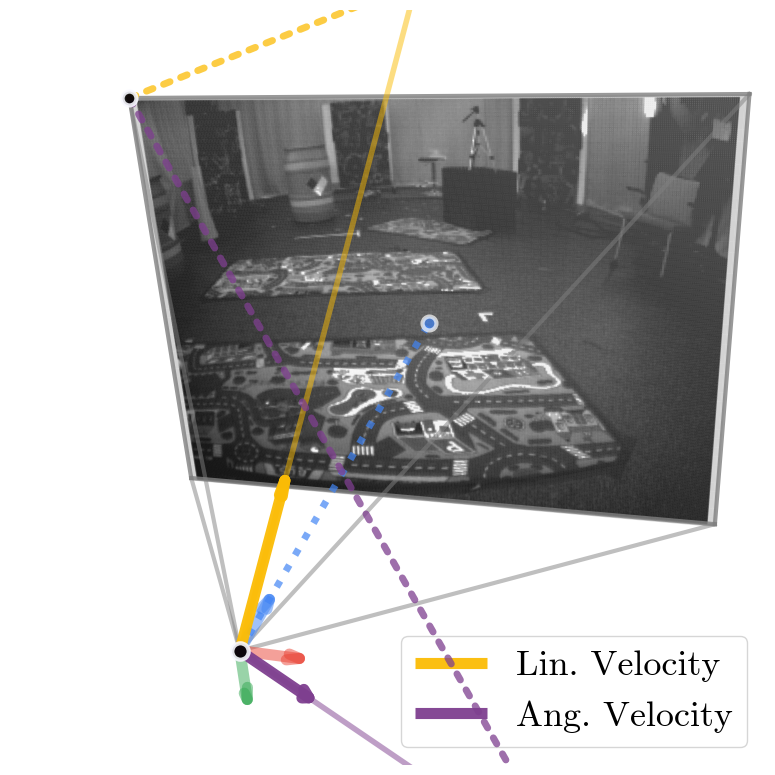} \end{tabular} &      
        \begin{tabular}{@{}c@{}} \includegraphics[width=0.2\textwidth, cfbox=gray 0.1pt 0pt]{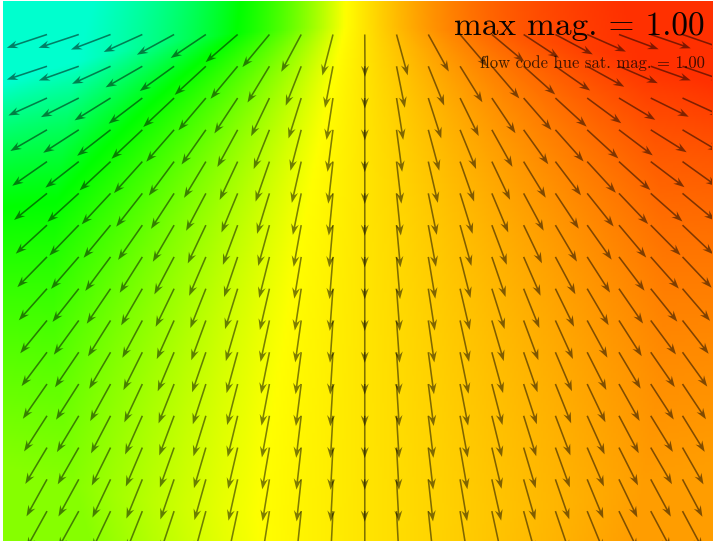} \end{tabular} &      
        \begin{tabular}{@{}c@{}} \includegraphics[width=0.2\textwidth, cfbox=gray 0.1pt 0pt]{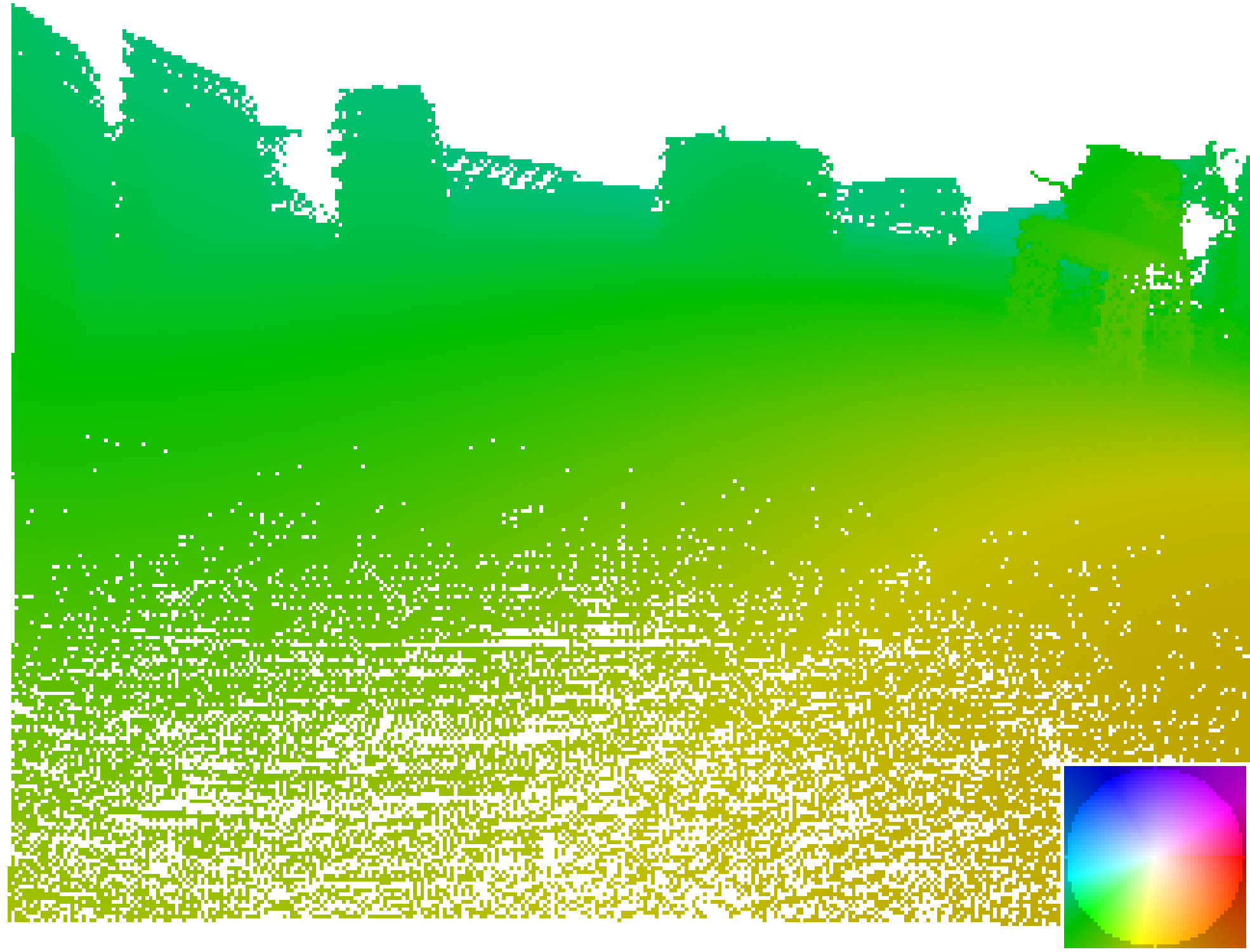} \end{tabular} &        
        \begin{tabular}{@{}c@{}} \includegraphics[width=0.2\textwidth, cfbox=gray 0.1pt 0pt]{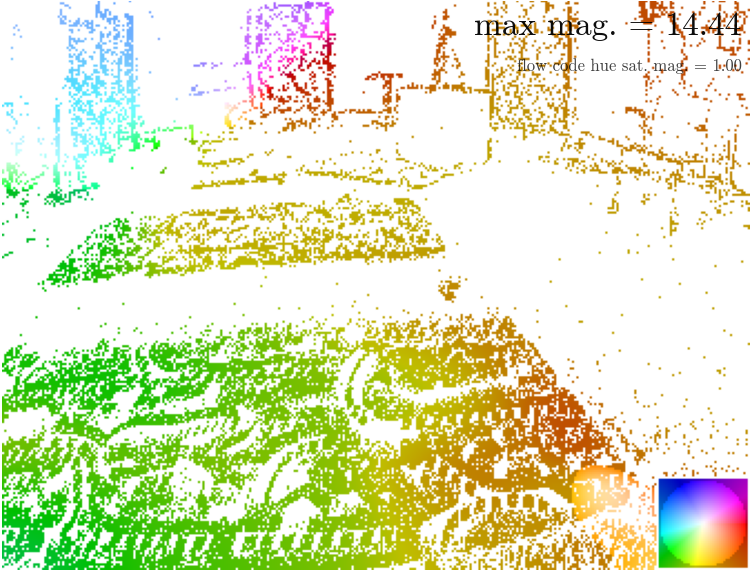} \end{tabular} &  
        \begin{tabular}{@{}c@{}} \includegraphics[width=0.203\textwidth, cfbox=gray 0.1pt 0pt]{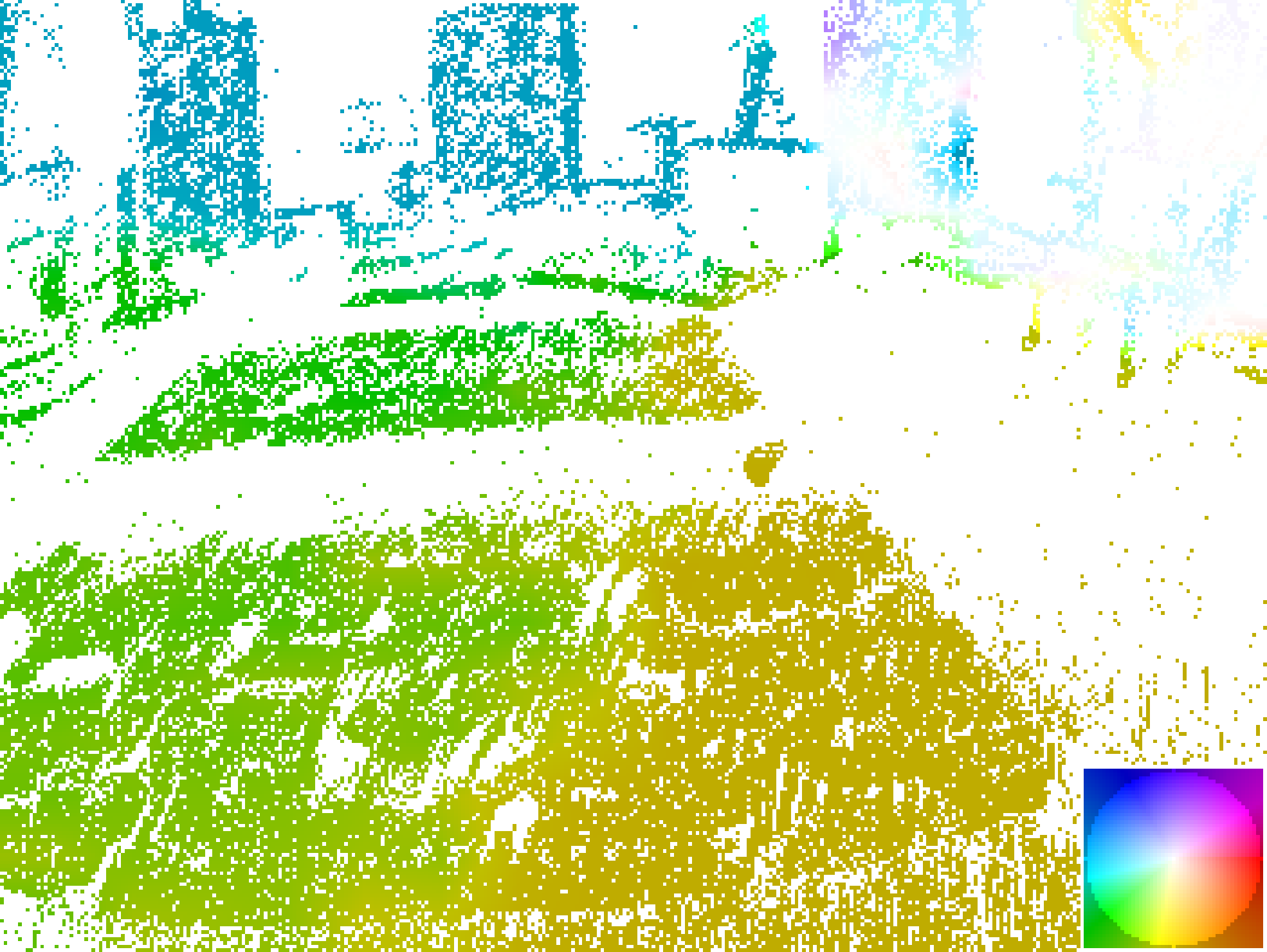} \end{tabular} &                         
        \begin{tabular}{@{}c@{}} \includegraphics[width=0.152\textwidth, cfbox=gray 0.1pt 0pt]{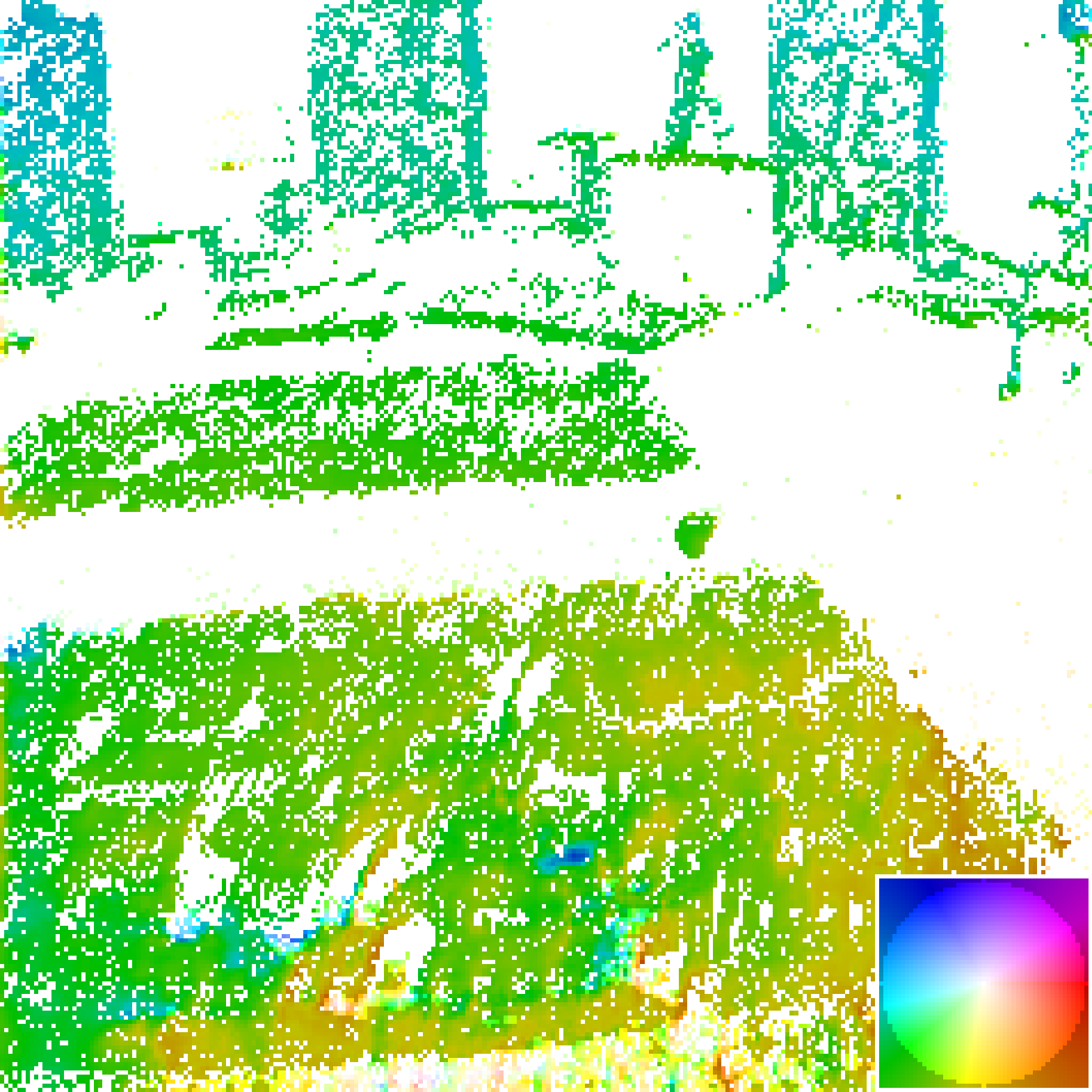} \end{tabular} \\ \vspace{-2pt}          

        &
        \begin{tabular}{@{}c@{}} \includegraphics[width=0.2\textwidth, cfbox=gray 0.1pt 0pt]{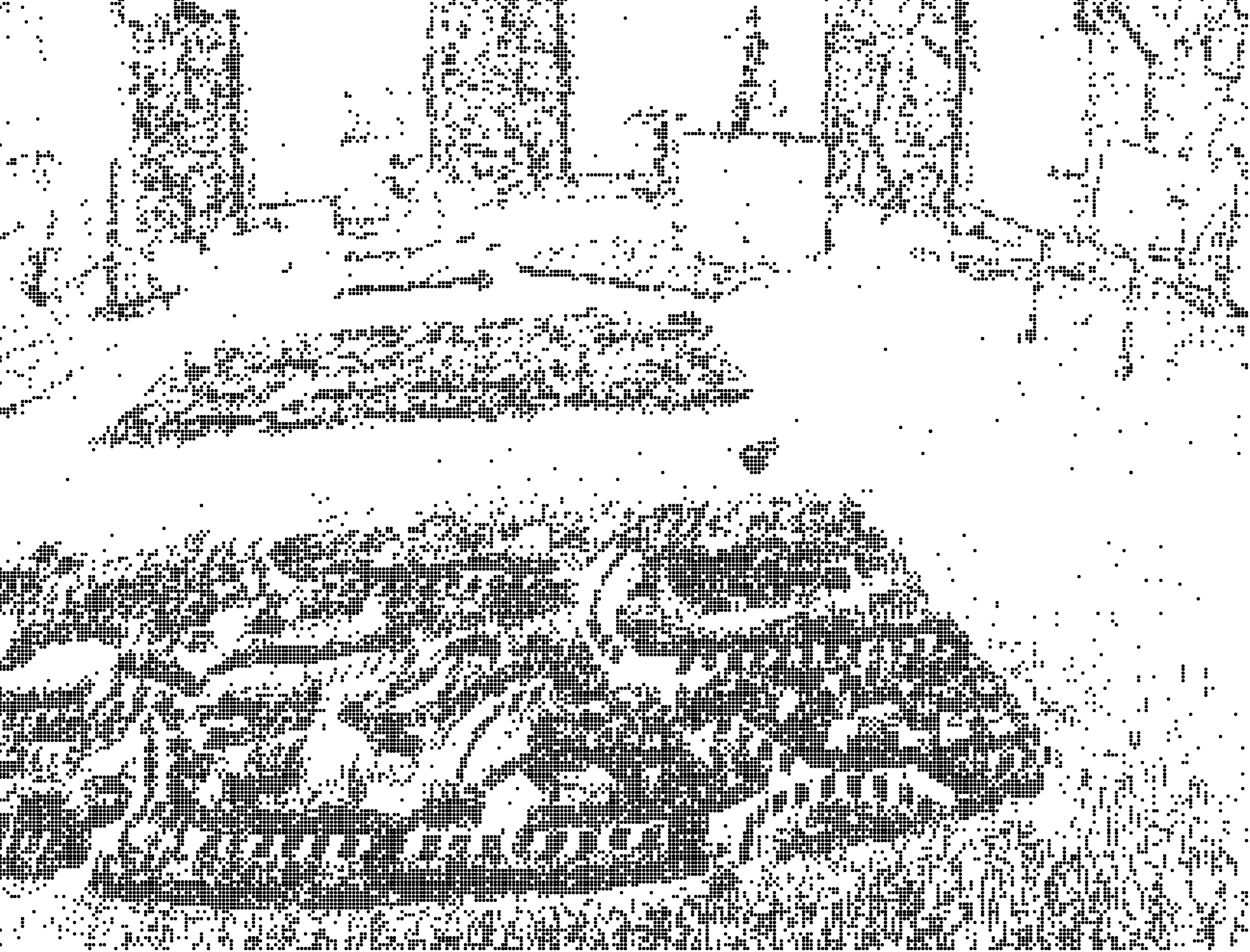} \end{tabular} &       
        \begin{tabular}{@{}c@{}} \includegraphics[width=0.2\textwidth, cfbox=gray 0.1pt 0pt]{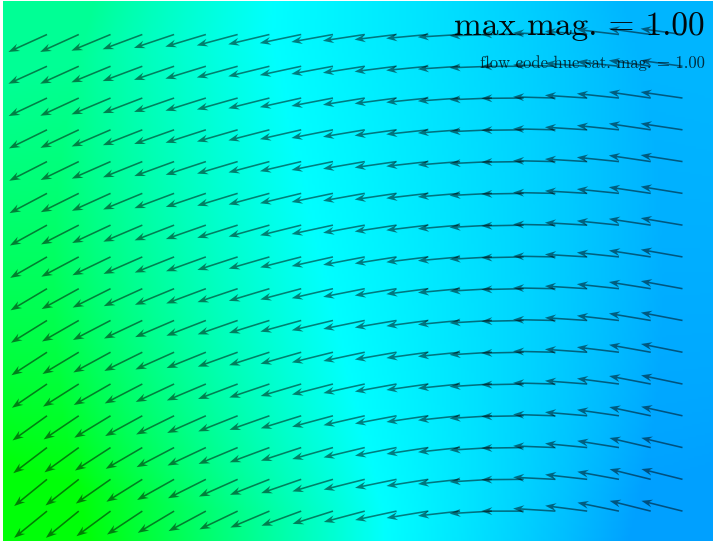} \end{tabular} &       
        \begin{tabular}{@{}c@{}} \includegraphics[width=0.2\textwidth, cfbox=gray 0.1pt 0pt]{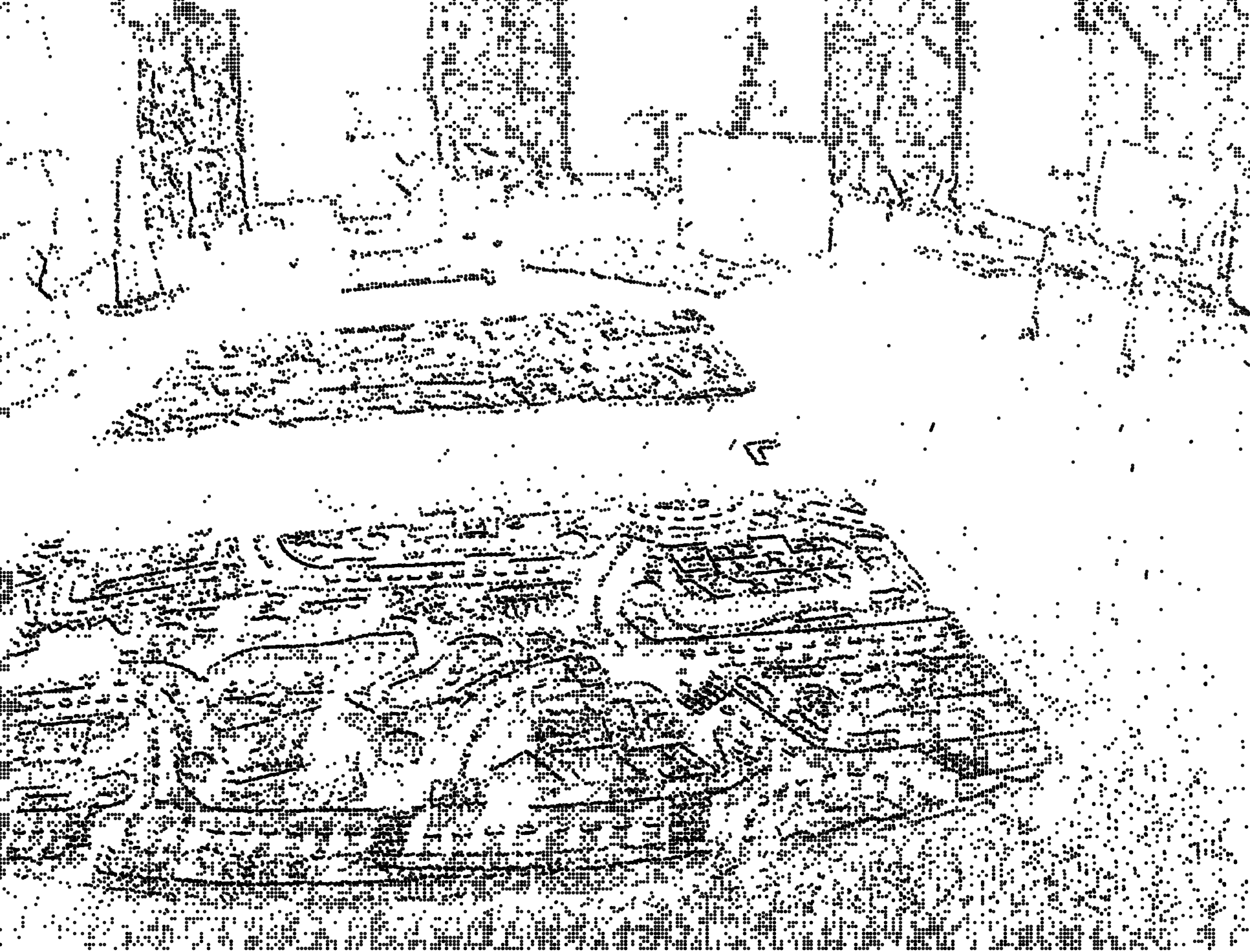} \end{tabular} &    
        \begin{tabular}{@{}c@{}} \includegraphics[width=0.2\textwidth, cfbox=gray 0.1pt 0pt]{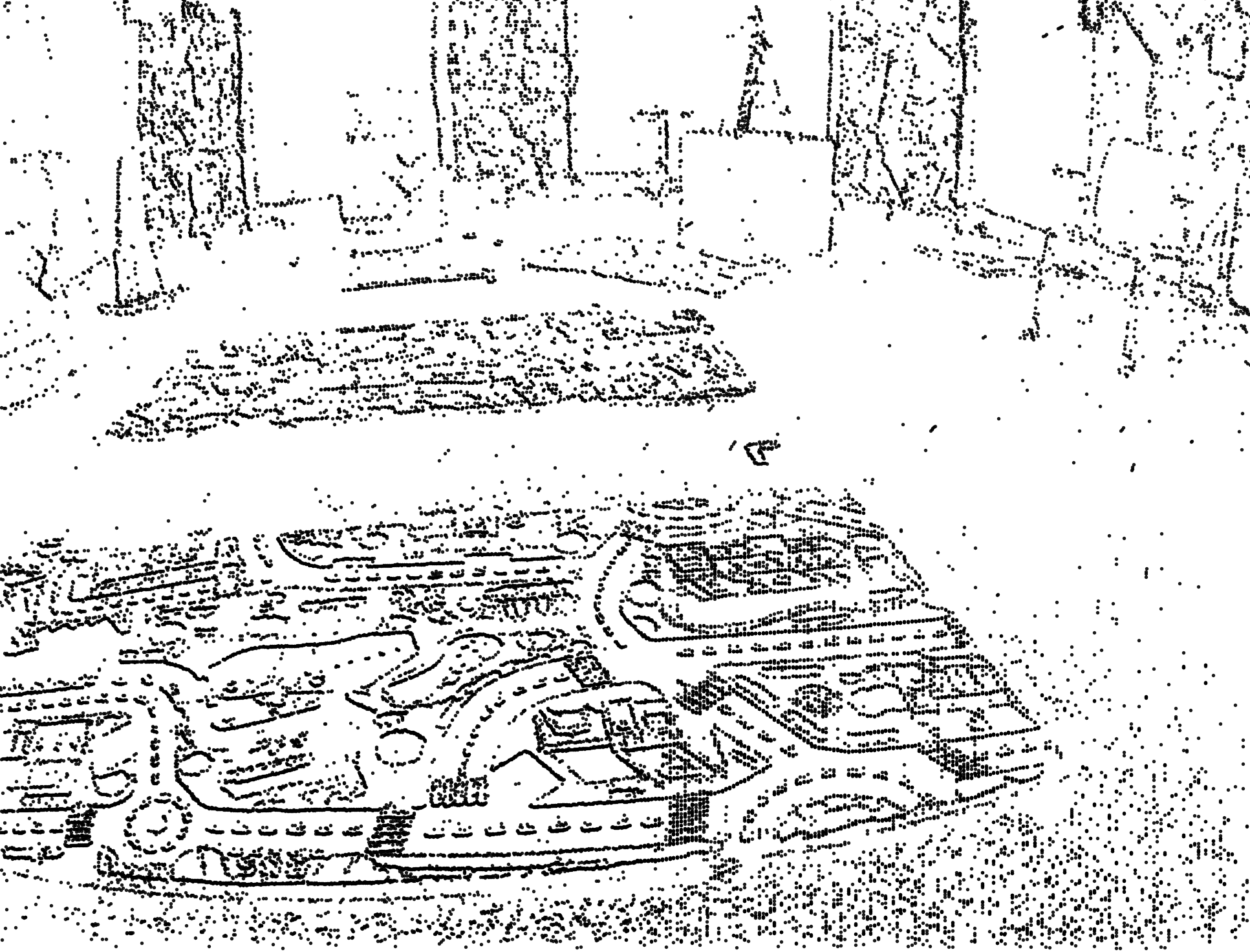} \end{tabular} & 
        \begin{tabular}{@{}c@{}} \includegraphics[width=0.2\textwidth, cfbox=gray 0.1pt 0pt]{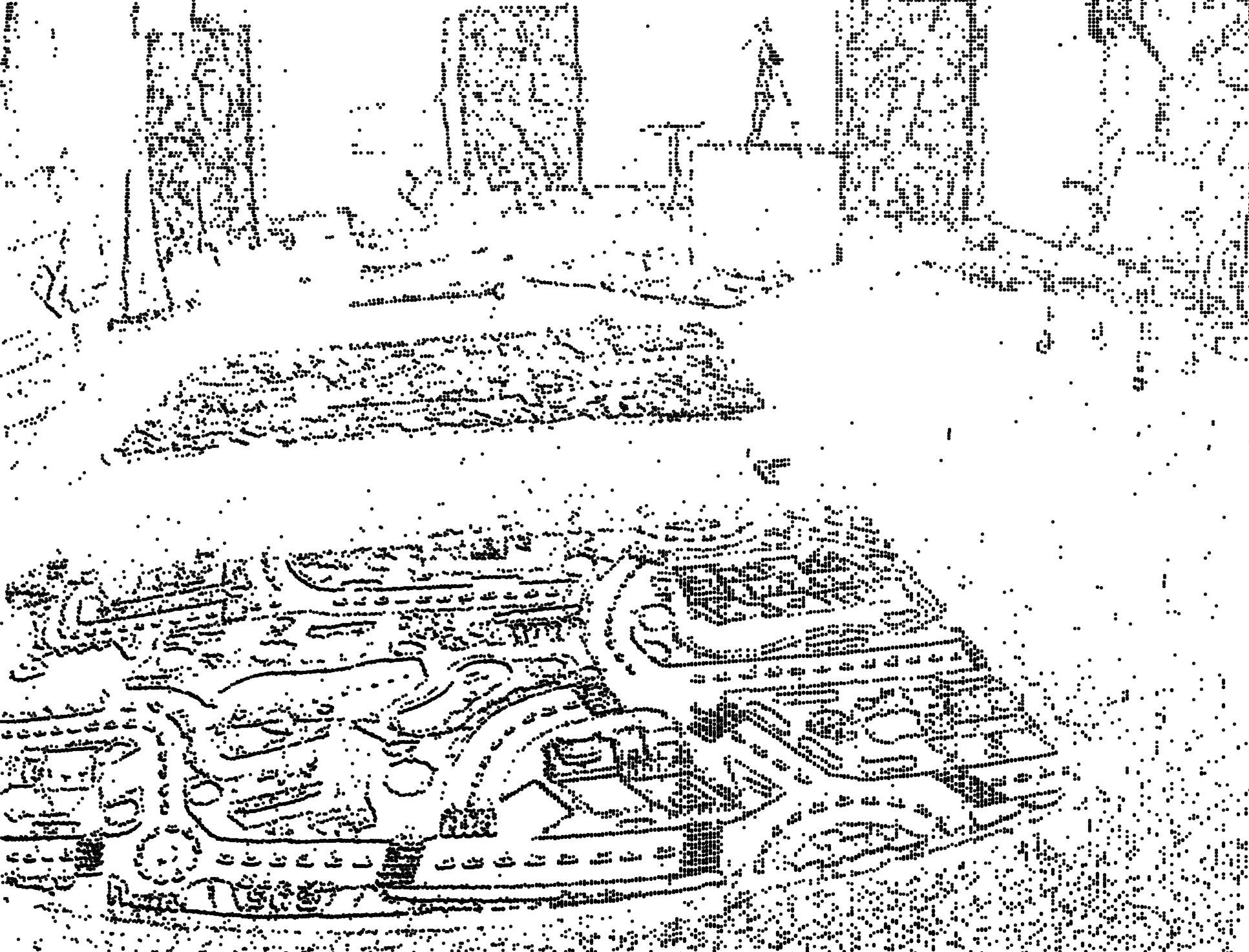} \end{tabular} &                         
        \begin{tabular}{@{}c@{}} \includegraphics[width=0.152\textwidth, cfbox=gray 0.1pt 0pt]{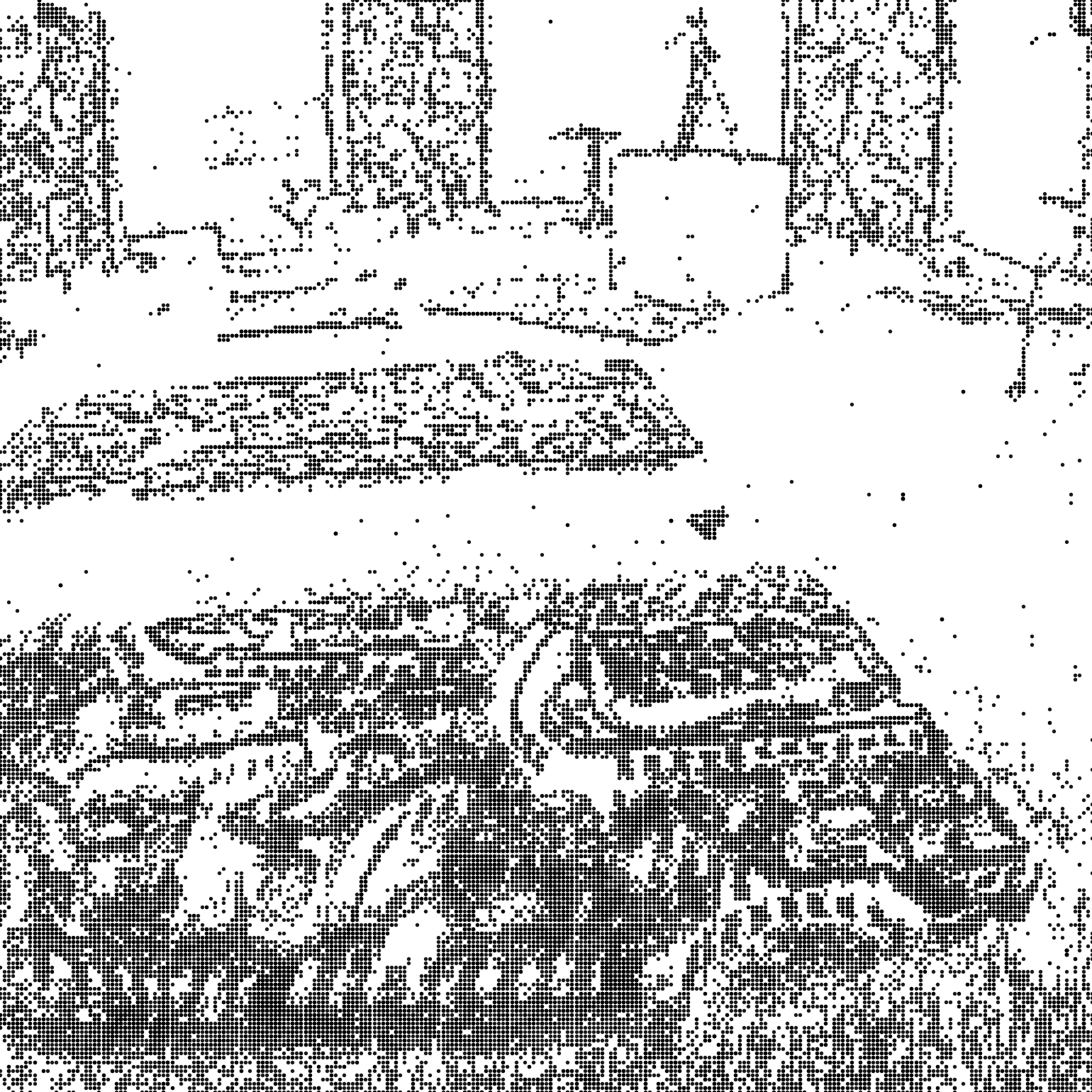} \end{tabular} \\ \vspace{-2pt}          

        \multirow{2}{*}{\rotatebox[origin=c]{90}{\begin{adjustbox}{max width=\labelscaler\textwidth} \texttt{outdoor\_day1} \end{adjustbox}}} &
        \begin{tabular}{@{}c@{}} \includegraphics[height=0.15\textwidth, cfbox=gray 0pt 0pt]{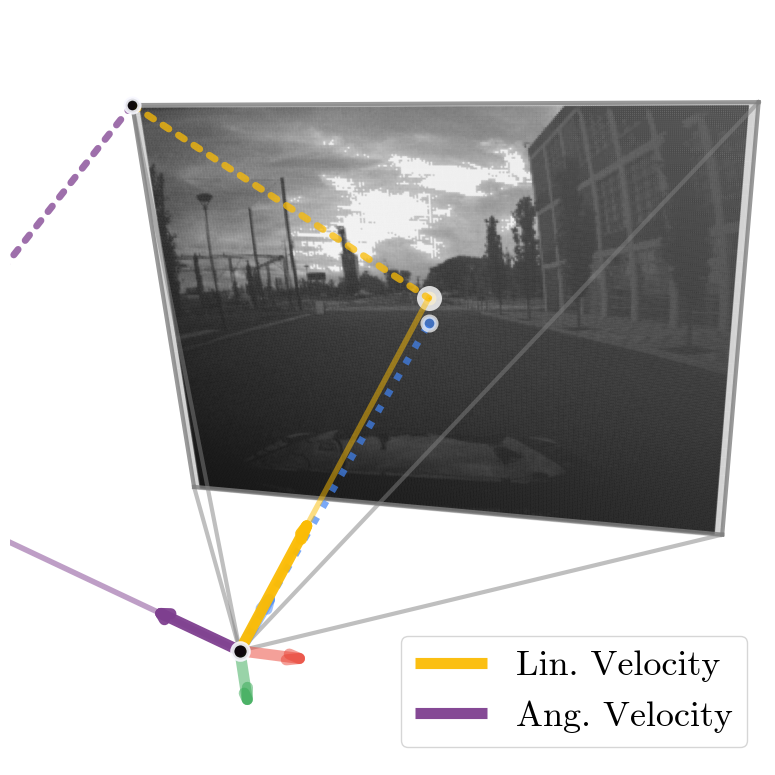} \end{tabular} &      
        \begin{tabular}{@{}c@{}} \includegraphics[width=0.2\textwidth, cfbox=gray 0.1pt 0pt]{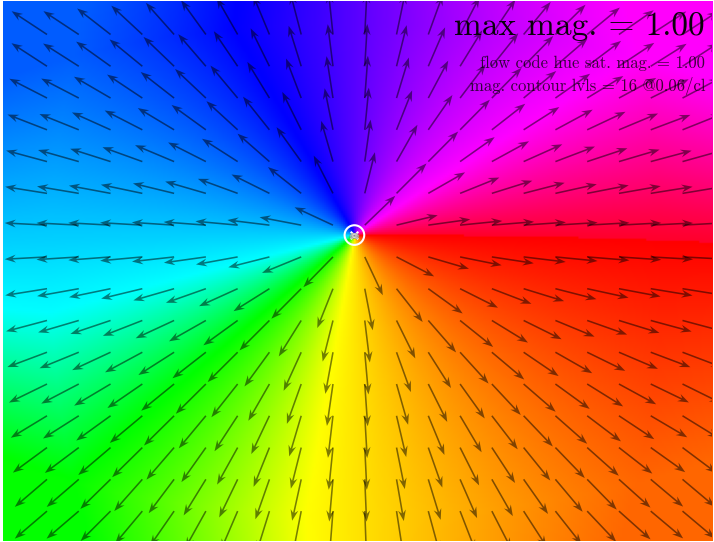} \end{tabular} &      
        \begin{tabular}{@{}c@{}} \includegraphics[width=0.2\textwidth, cfbox=gray 0.1pt 0pt]{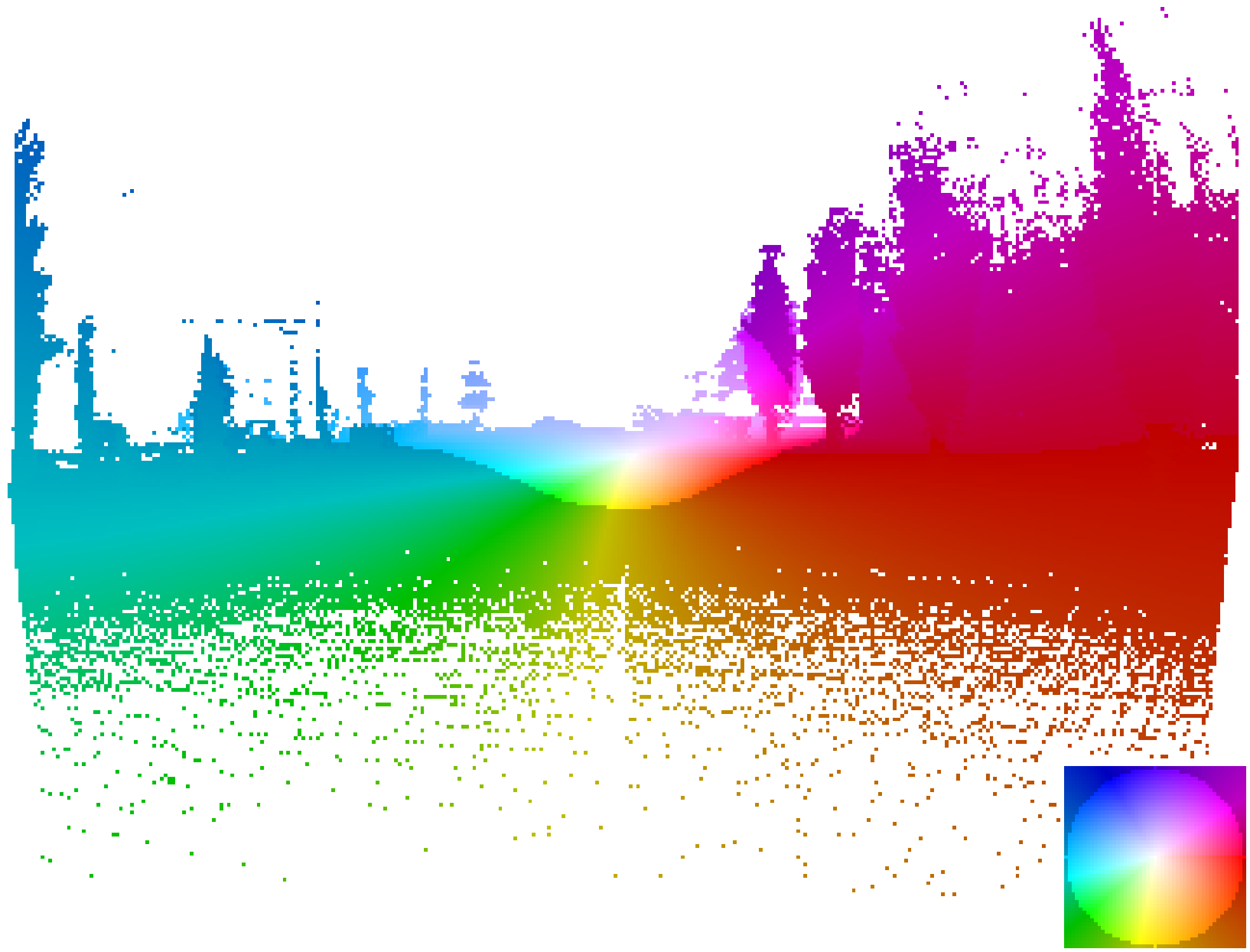} \end{tabular} &        
        \begin{tabular}{@{}c@{}} \includegraphics[width=0.2\textwidth, cfbox=gray 0.1pt 0pt]{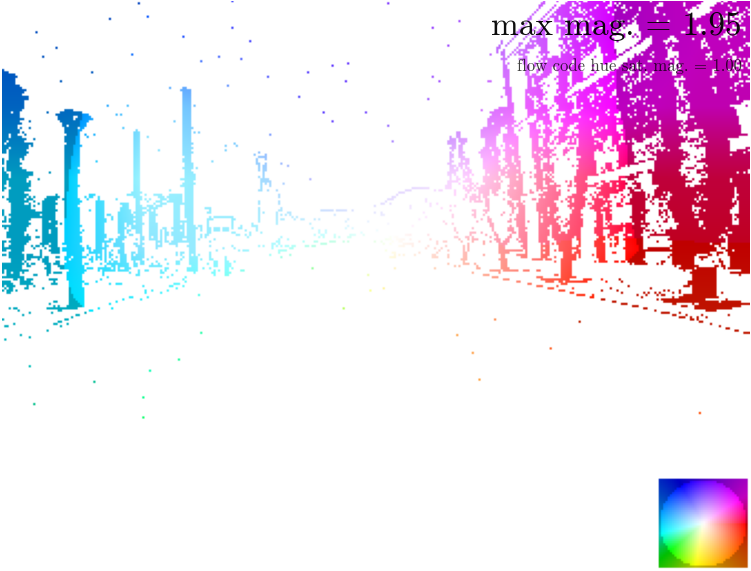} \end{tabular} &  
        \begin{tabular}{@{}c@{}} \includegraphics[width=0.203\textwidth, cfbox=gray 0.1pt 0pt]{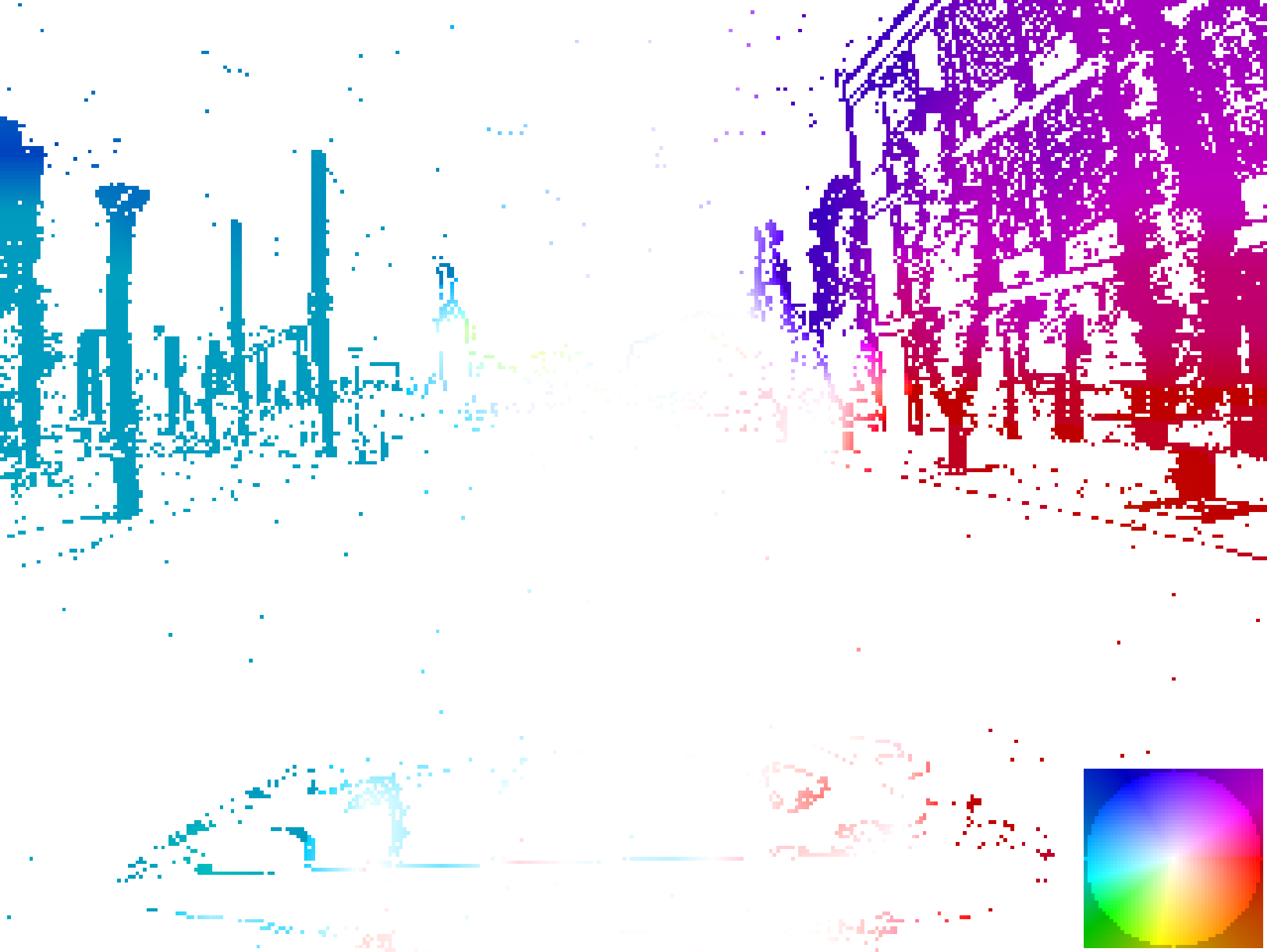} \end{tabular} &                         
        \begin{tabular}{@{}c@{}} \includegraphics[width=0.152\textwidth, cfbox=gray 0.1pt 0pt]{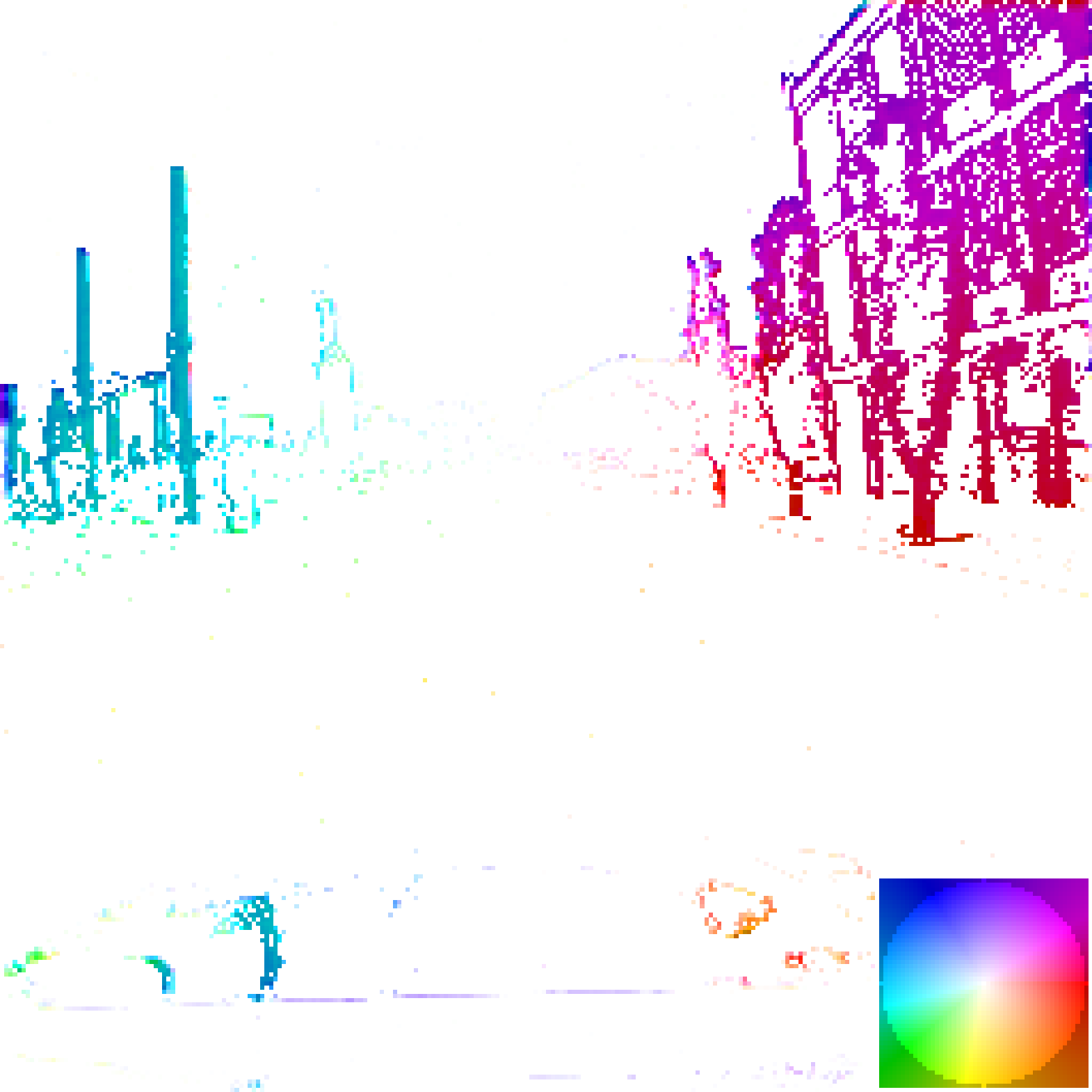} \end{tabular} \\                        

        &
        \begin{tabular}{@{}c@{}} \includegraphics[width=0.2\textwidth, cfbox=gray 0.1pt 0pt]{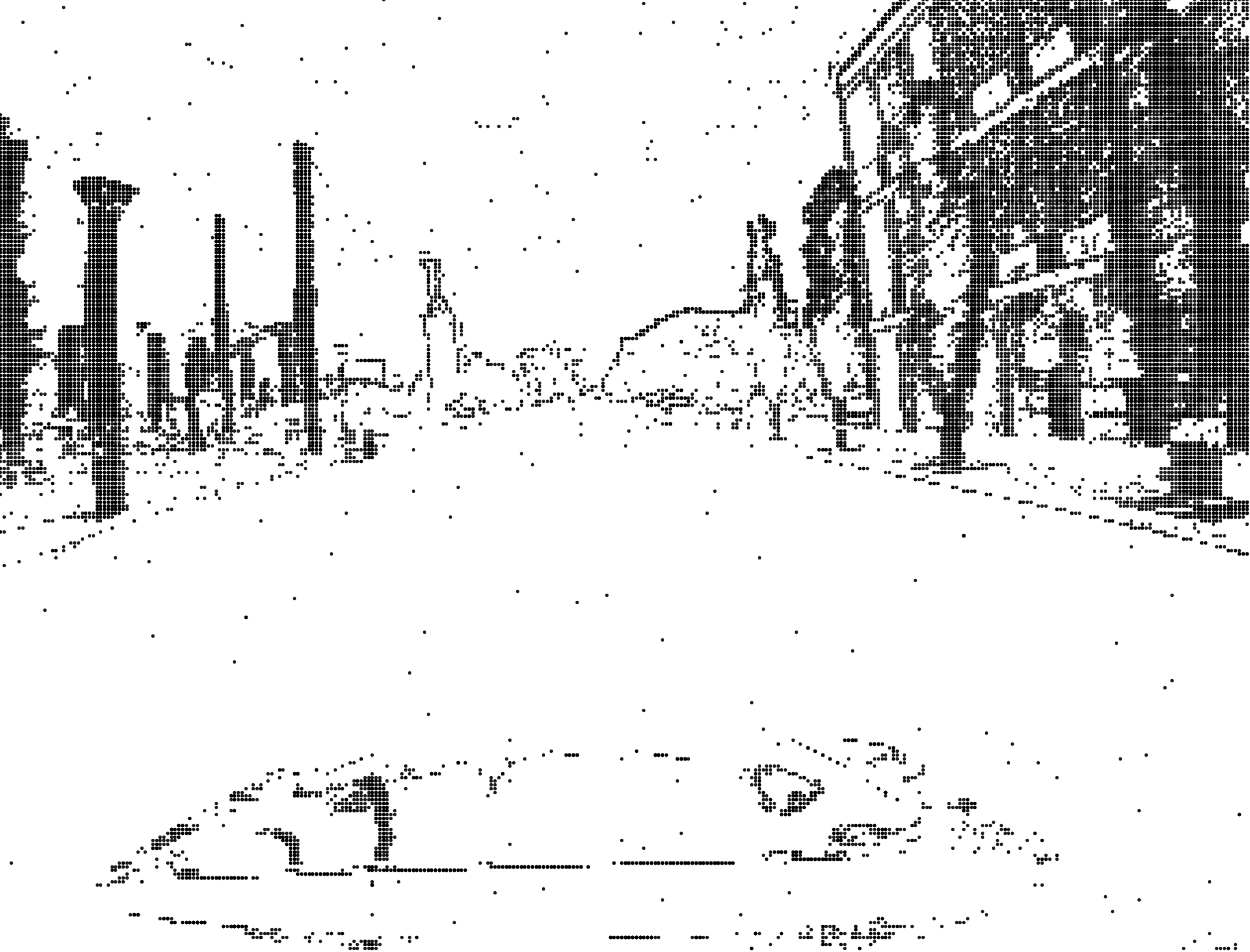} \end{tabular} &       
        \begin{tabular}{@{}c@{}} \includegraphics[width=0.2\textwidth, cfbox=gray 0.1pt 0pt]{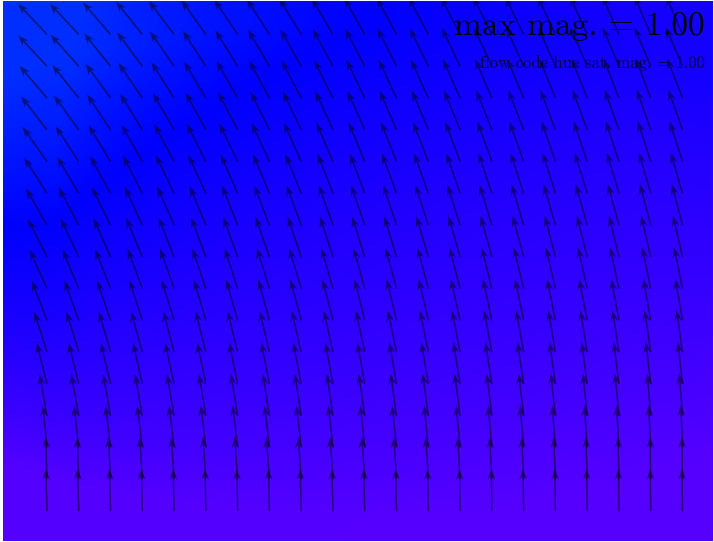} \end{tabular} &       
        \begin{tabular}{@{}c@{}} \includegraphics[width=0.2\textwidth, cfbox=gray 0.1pt 0pt]{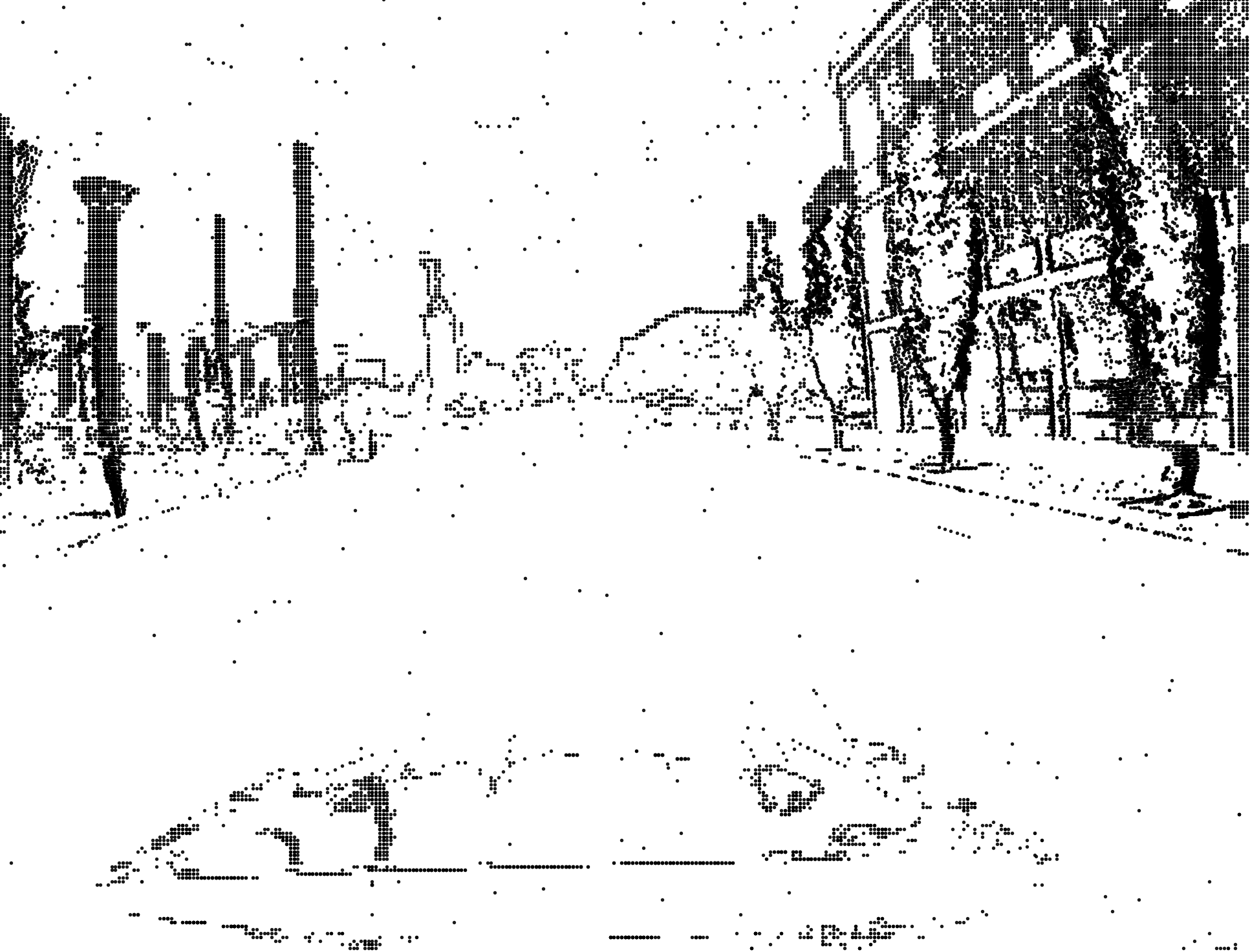} \end{tabular} &    
        \begin{tabular}{@{}c@{}} \includegraphics[width=0.2\textwidth, cfbox=gray 0.1pt 0pt]{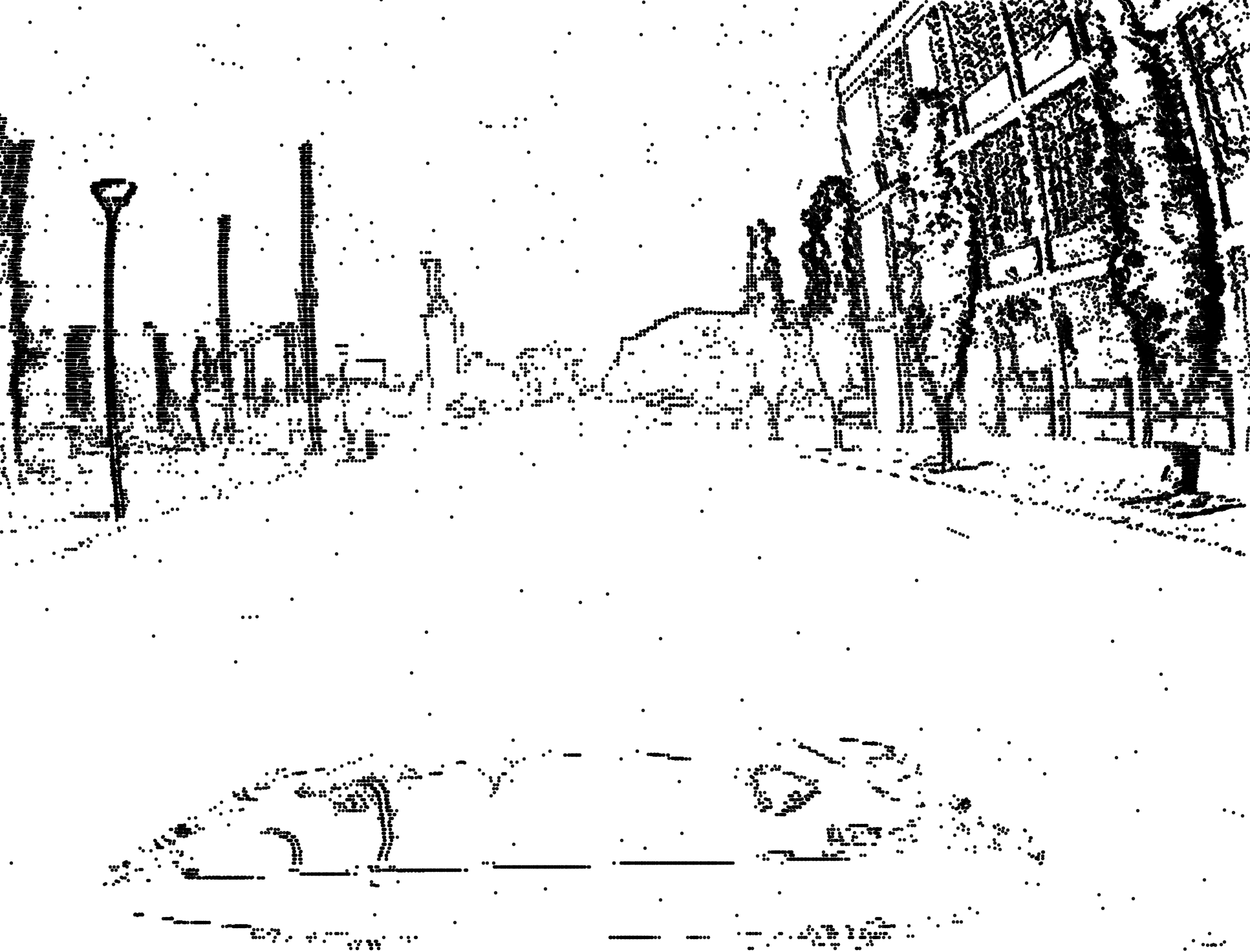} \end{tabular} & 
        \begin{tabular}{@{}c@{}} \includegraphics[width=0.2\textwidth, cfbox=gray 0.1pt 0pt]{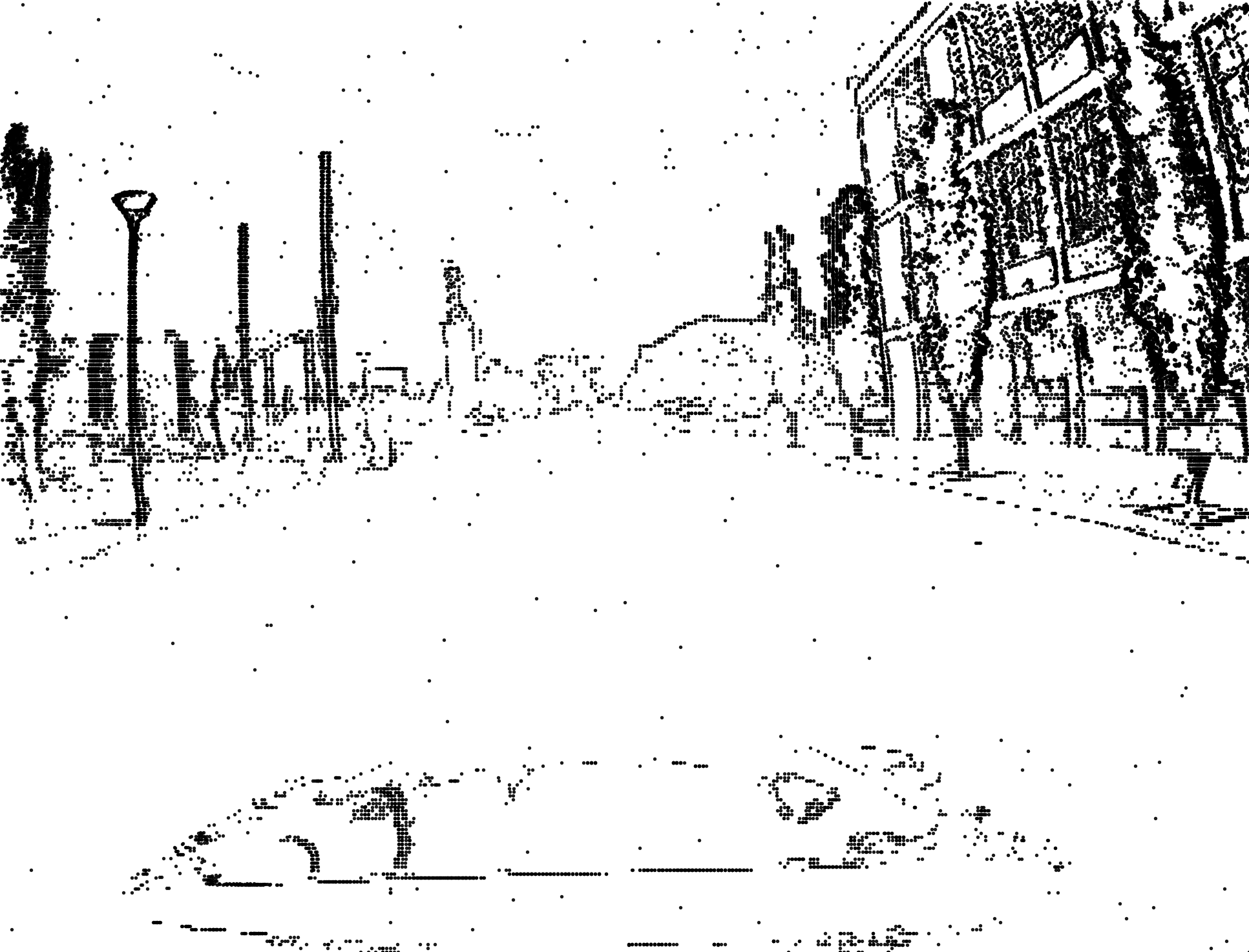} \end{tabular} &                         
        \begin{tabular}{@{}c@{}} \includegraphics[width=0.152\textwidth, cfbox=gray 0.1pt 0pt]{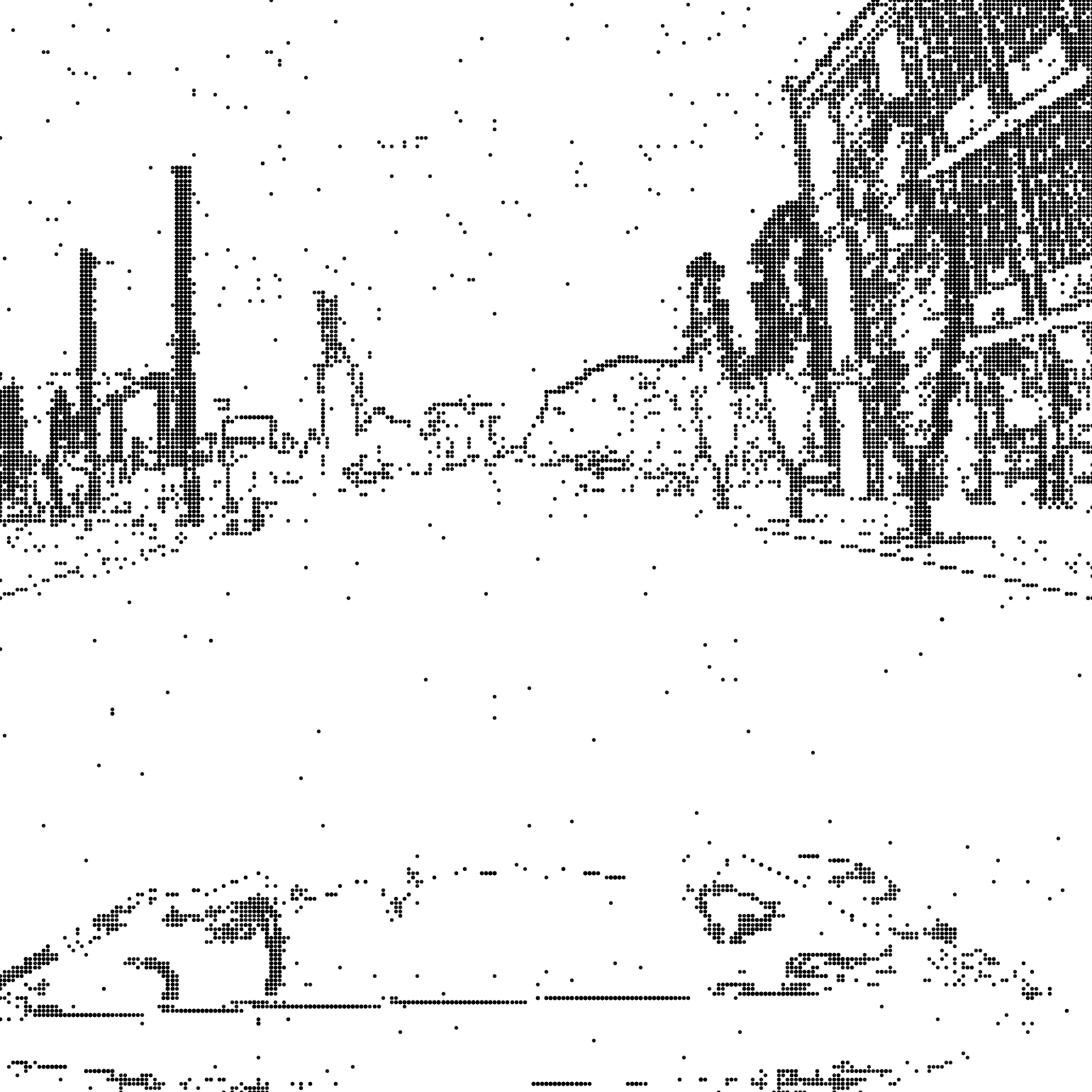} \end{tabular} \\                        

        &
        \begin{tabular}{@{}c@{}} \begin{adjustbox}{max width=\labelscaler\textwidth}(a) $\boldsymbol{v}$, $\boldsymbol{\omega}$ | $I_{\text{events}}$     \end{adjustbox}\end{tabular} & 
        \begin{tabular}{@{}c@{}} \begin{adjustbox}{max width=\labelscaler\textwidth}(b) $\hat{\mathcal{O}}_{\text{lin}}$ | $\hat{\mathcal{O}}_{\text{ang}}$                             \end{adjustbox}\end{tabular} &
        \begin{tabular}{@{}c@{}} \begin{adjustbox}{max width=\labelscaler\textwidth}(c) GT                     \end{adjustbox}\end{tabular} &
        \begin{tabular}{@{}c@{}} \begin{adjustbox}{max width=\labelscaler\textwidth}(d) Ours (OPCM)                      \end{adjustbox}\end{tabular} &
        \begin{tabular}{@{}c@{}} \begin{adjustbox}{max width=\labelscaler\textwidth}(e) \cite{shiba2022secrets} (MB)   \end{adjustbox}\end{tabular} &
        \begin{tabular}{@{}c@{}} \begin{adjustbox}{max width=\labelscaler\textwidth}(f) \cite{zhu2018evflownet} (SSL)  \end{adjustbox}\end{tabular} \\
    \end{tabular}
    \end{adjustbox}
    \caption{Qualitative comparisons ($dt=4$) of our method (OPCM) against two
    prominent approaches~\cite{zhu2018evflownet,shiba2022secrets} on MVSEC. The
    results for each sequence are depicted in two subsequent rows. Column (a)
    displays the vision-motion geometries (upper rows) and the images of the
    original events (lower rows). Column (b) depicts the generated orientation
    maps, $\hat{\mathcal{O}}_{\text{lin}}$ (upper rows) and
    $\hat{\mathcal{O}}_{\text{ang}}$ (lower rows). Column (c) shows the
    available ground-truth (GT) flows (upper rows) and the corresponding IWEs
    (lower rows). Columns (d-f) display the predicted flows masked by the
    original events and the constructed IWEs for each technique.}
    \label{fig:mvsec_comparisons}
\end{figure*}

We report qualitative results for MVSEC in \cref{fig:mvsec_comparisons}.
Additionally, we display visualizations of the vision and motion geometry along
with the generated orientation maps $\boldsymbol{\mathcal{O}}$ in
\cref{fig:mvsec_comparisons}(a) and \cref{fig:mvsec_comparisons}(b),
respectively. Rows 7 and 8 of \cref{fig:mvsec_comparisons} accurately showcase
how the linear velocity intersects with the image plane, inducing a diverging
optical flow orientation pattern. For non-holonomic mobile ground platforms
(\eg, automobiles), such orientation patterns are common due to the
directionally restricted movements. Therefore, as detailed in
\cref{subsec:metrics_scaling_strategy_and_hyperparameters}, we chose to further
skew the balance by setting a higher value of $\alpha_{\text{lin}}=2$ for
\texttt{outdoor\_day1}. For more details, please see the supplementary
material.

\begin{table}[t]
    \centering
  
    \begin{adjustbox}{max width=\columnwidth}
        \begin{tabular}{@{\thinspace}c@{\medspace}l@{\thickspace}
        c@{\thickspace}c@{\thickspace}c 
        c@{\thickspace}c@{\thickspace}c 
        c@{\thickspace}c@{\thickspace}c 
        c@{\thickspace}c} 
    
            \toprule
            & &  
            \multicolumn{2}{c}{\texttt{indoor\_flying1}}        & \thickspace &     
            \multicolumn{2}{c}{\texttt{indoor\_flying2}}        & \thickspace &     
            \multicolumn{2}{c}{\texttt{indoor\_flying3}}        & \thickspace &     
            \multicolumn{2}{c}{\texttt{outdoor\_day1}}          \\                  
            \cmidrule(){3-4} \cmidrule(){6-7} \cmidrule(){9-10} \cmidrule(){12-13} 
            & \multicolumn{1}{c}{\emph{dt} = 1}  & 
            AEE $\downarrow$         & \%Out $\downarrow$       & \thickspace &     
            AEE $\downarrow$         & \%Out $\downarrow$       & \thickspace &     
            AEE $\downarrow$         & \%Out $\downarrow$       & \thickspace &     
            AEE $\downarrow$         & \%Out $\downarrow$       \\                  
            \midrule 

            \multirow{5}{*}{\rotatebox[origin=c]{90}{SL}} 
            & EV-FlowNet-EST \cite{gehrig2019evflownetest} & 
            0.97                     & 0.91                     & \thickspace &     
            1.38                     & 8.20                     & \thickspace &     
            1.43                     & 6.47                     & \thickspace &     
            --                       & --                       \\                  
            & EV-FlowNet$+$ \cite{stoffregen2020evflownetplus} & 
            0.56                     & 1.00                     & \thickspace &     
            0.66                     & 1.00                     & \thickspace &     
            0.59                     & 1.00                     & \thickspace &     
            0.68                     & 0.99                     \\                  
            & E-RAFT \cite{gehrig2021eraft} & 
            --                       & --                       & \thickspace &     
            --                       & --                       & \thickspace &     
            --                       & --                       & \thickspace &     
            0.24                     & 1.70                     \\                  
            & DCEIFlow \cite{wan2022learning} & 
            0.75                     & 1.55                     & \thickspace &     
            0.90                     & 2.10                     & \thickspace &     
            0.80                     & 1.77                     & \thickspace &     
            \textbf{0.22}            & \textbf{0.00}            \\                  
            & TMA \cite{liu2023tma} & 
            1.06                     & 3.63                     & \thickspace &     
            1.81                     & 27.29                    & \thickspace &     
            1.58                     & 23.26                    & \thickspace &     
            0.25                     & 0.07                     \\                  
            \midrule

            \multirow{3}{*}{\rotatebox[origin=c]{90}{SSL}} 
            & EV-FlowNet (original) \cite{zhu2018evflownet} & 
            1.03                     & 2.20                     & \thickspace &     
            1.72                     & 15.1                     & \thickspace &     
            1.53                     & 11.90                    & \thickspace &     
            0.49                     & 0.20                     \\                  
            & Spike-FlowNet \cite{lee2020spikeflownet} & 
            0.84                     & --                       & \thickspace &     
            1.28                     & --                       & \thickspace &     
            1.11                     & --                       & \thickspace &     
            0.49                     & --                       \\                  
            & STE-FlowNet \cite{ding2022steflownet} & 
            0.57                     & 0.10                     & \thickspace &     
            0.79                     & 1.60                     & \thickspace &     
            0.72                     & 1.30                     & \thickspace &     
            0.42                     & \textbf{0.00}            \\                  
            \midrule

            \multirow{3}{*}{\rotatebox[origin=c]{90}{USL}} 
            & EV-FlowNet \cite{zhu2019unsupervised} & 
            0.58                     & \textbf{0.00}            & \thickspace &     
            1.02                     & 4.00                     & \thickspace &     
            0.87                     & 3.00                     & \thickspace &     
            \underline{0.32}         & \textbf{0.00}            \\                  
            & FireFlowNet \cite{paredes2021back} & 
            0.97                     & 2.60                     & \thickspace &     
            1.67                     & 15.30                    & \thickspace &     
            1.43                     & 11.00                    & \thickspace &     
            1.06                     & 6.60                     \\                  
            & ConvGRU-EV-FlowNet \cite{hagenaars2021convgruevflownet} & 
            0.60                     & 0.51                     & \thickspace &     
            1.17                     & 8.06                     & \thickspace &     
            0.93                     & 5.64                     & \thickspace &     
            0.47                     & 0.25                     \\                  
            \midrule

            \multirow{6}{*}{\rotatebox[origin=c]{90}{MB}} 
            & Nagata~\etal \cite{nagata2021matchtimesurf} & 
            0.62                     & --                       & \thickspace &     
            0.93                     & --                       & \thickspace &     
            0.84                     & --                       & \thickspace &     
            0.77                     & --                       \\                  
            & Akolkar~\etal \cite{akolkar2020real} & 
            1.52                     & --                       & \thickspace &     
            1.59                     & --                       & \thickspace &     
            1.89                     & --                       & \thickspace &     
            2.75                     & --                       \\                  
            & RTEF \cite{brebion2021realtimeflow} & 
            0.52                     & 0.10                     & \thickspace &     
            0.98                     & 5.50                     & \thickspace &     
            0.71                     & 2.10                     & \thickspace &     
            0.53                     & {0.20}         \\                  
            & MultiCM \cite{shiba2022secrets} & 
            0.42                     & 0.09                     & \thickspace &     
            0.60                     & 0.59                     & \thickspace &     
            0.50                     & 0.29                     & \thickspace &     
            0.72\tiny{5}             & 5.49\tiny{4}             \\                  
            & EINCM \cite{karmokar2025secrets} & 
            \underline{0.37\tiny{2}}    & {0.04\tiny{6}} & \thickspace &     
            \underline{0.50\tiny{5}}    & \underline{0.15\tiny{5}}    & \thickspace &     
            \underline{0.43\tiny{7}}    & \underline{0.02\tiny{9}}    & \thickspace &     
            {0.61\tiny{8}}           & {3.69\tiny{3}}           \\                  
            & Ours (CM events only) & 
            {0.37\tiny{8}} & 0.05\tiny{5}             & \thickspace &     
            {0.50\tiny{9}} & {0.17\tiny{2}} & \thickspace &     
            {0.44\tiny{3}} & {0.05\tiny{7}} & \thickspace &     
            {0.64\tiny{9}}           & {4.22\tiny{6}}           \\                  
            & Ours (OPCM) & 
            \textbf{0.33\tiny{39}}    & \underline{0.04\tiny{0}} & \thickspace &     
            \textbf{0.47\tiny{7}}    & \textbf{0.11\tiny{8}}    & \thickspace &     
            \textbf{0.40\tiny{6}}    & \textbf{0.01\tiny{9}}    & \thickspace &     
            {0.38\tiny{3}}           & \underline{0.14\tiny{3}}           \\                  
            \bottomrule
            
            &                                   & & & \thickspace & & & \thickspace & & & \thickspace & & \\
            & \multicolumn{1}{c}{\emph{dt} = 4} & & & \thickspace & & & \thickspace & & & \thickspace & & \\
            \midrule

            \multirow{3}{*}{\rotatebox[origin=c]{90}{SL}} 
            & E-RAFT \cite{gehrig2021eraft} + TMA \cite{liu2023tma} & 
            2.81                     & 40.25                    & \thickspace &     
            5.09                     & 64.19                    & \thickspace &     
            4.46                     & 57.11                    & \thickspace &     
            \underline{0.72}         & \underline{1.12}         \\                  
            & DCEIFlow \cite{wan2022learning} & 
            2.08                     & 21.47                    & \thickspace &     
            3.48                     & 42.05                    & \thickspace &     
            2.51                     & 29.73                    & \thickspace &     
            0.89                     & 3.19                     \\                  
            & TMA \cite{liu2023tma} & 
            2.43                     & 3.63                     & \thickspace &     
            4.32                     & 27.29                    & \thickspace &     
            3.60                     & 23.26                    & \thickspace &     
            \textbf{0.70}            & \textbf{1.08}            \\                  
            \midrule

            \multirow{3}{*}{\rotatebox[origin=c]{90}{SSL}}
            & EV-FlowNet (original) \cite{zhu2018evflownet} & 
            2.25                     & 24.70                     & \thickspace &    
            4.05                     & 45.30                     & \thickspace &    
            3.45                     & 39.70                     & \thickspace &    
            1.23                     & 7.30                      \\                 
            & Spike-FlowNet~\cite{lee2020spikeflownet} & 
            2.24                     & --                        & \thickspace &    
            3.83                     & --                        & \thickspace &    
            3.18                     & --                        & \thickspace &    
            1.09                     & --                        \\                 
            & STE-FlowNet~\etal \cite{ding2022steflownet} & 
            1.77                     & 14.70                     & \thickspace &    
            2.52                     & 26.10                     & \thickspace &    
            2.23                     & 22.10                     & \thickspace &    
            0.99                     & 3.90                      \\                 
            \midrule

            \multirow{2}{*}{\rotatebox[origin=c]{90}{USL}}
            & EV-FlowNet~\cite{zhu2019unsupervised} & 
            2.18                     & 24.20                     & \thickspace &    
            3.85                     & 46.80                     & \thickspace &    
            3.18                     & 47.80                     & \thickspace &    
            1.30                     & 9.70                      \\                 
            & ConvGRU-EV-FlowNet~\cite{hagenaars2021convgruevflownet} & 
            2.16                     & 21.50                     & \thickspace &    
            3.90                     & 40.72                     & \thickspace &    
            3.00                     & 29.60                     & \thickspace &    
            1.69                     & 12.50                     \\                 
            \midrule

            \multirow{3}{*}{\rotatebox[origin=c]{90}{MB}}
            & MultiCM~\etal \cite{shiba2022secrets} & 
            1.68                     & 12.79                     & \thickspace &    
            2.49                     & 26.31                     & \thickspace &    
            2.06                     & 18.93                     & \thickspace &    
            {2.07}                   & {19.99}                   \\                 
            & EINCM~\cite{karmokar2025secrets} & 
            \underline{1.43\tiny{9}}    & \underline{7.78\tiny{9}}     & \thickspace &    
            \underline{1.97\tiny{3}}    & \underline{17.18\tiny{3}}    & \thickspace &    
            \underline{1.70\tiny{6}}    & \underline{12.33\tiny{9}}    & \thickspace &    
            {1.70\tiny{4}}           & {16.01\tiny{3}}           \\                 
            & Ours (CM events only) & 
            {1.53\tiny{9}} & {9.48\tiny{2}}  & \thickspace &    
            {2.38\tiny{2}} & {21.84\tiny{9}} & \thickspace &    
            {1.92\tiny{3}} & {15.37\tiny{8}} & \thickspace &    
            {2.83\tiny{1}}           & {25.01\tiny{1}}           \\                 
            & Ours (OPCM) & 
            \textbf{1.33\tiny{3}}    & \textbf{6.34\tiny{1}}     & \thickspace &    
            \textbf{1.95\tiny{2}}    & \textbf{17.07\tiny{3}}    & \thickspace &    
            \textbf{1.60\tiny{3}}    & \textbf{11.17\tiny{7}}    & \thickspace &    
            {1.60\tiny{4}}           & {15.07\tiny{5}}           \\                 
            \bottomrule
        
        \end{tabular}
    \end{adjustbox}
    \caption{Quantitative results on MVSEC. The approaches are categorized as
    model-based (MB) and the following learning-based categories: (i) supervised
    learning (SL) through supervision via ground-truth flow, (ii)
    semi-supervised learning (SSL) using supervision by frames, and (iii)
    unsupervised learning (USL). \emph{Bold} and \emph{underline} typefaces are
    used to indicate the \textbf{best} and the \underline{second best},
    respectively. \%Out indicates a 3-pixel error percentage.}
    \label{tab:mvsec_accuracies}
\end{table}


\begin{table}[!htb]
    \centering
  
    \begin{adjustbox}{max width=\columnwidth}
        \begin{tabular}{@{\thinspace}l@{\thickspace}c@{\thickspace}
        c@{\thickspace}c@{\thickspace}c@{\thickspace}c@{\thickspace}c@{\thickspace}c@{\thickspace}c@{\thickspace}c@{\thickspace}
        c@{\thickspace}c@{\thickspace}
        c@{\thickspace}c@{\thickspace}c}
    
            \toprule
            & \thickspace &
            \multicolumn{7}{c}{MVSEC ($dt=4$)}              & \thickspace &     
            \multicolumn{1}{c}{ECD}                       & \thickspace &     
            \multicolumn{3}{c}{DSEC (train)}              \\                  
            \cmidrule(){3-9} \cmidrule(){11-11} \cmidrule(){13-15} 

            & \thickspace & 
            \texttt{ind\_fly1}                      & \thickspace &     
            \texttt{ind\_fly2}                      & \thickspace &     
            \texttt{ind\_fly3}                      & \thickspace &     
            \texttt{out\_day1}                      & \thickspace &     
            \texttt{slider\_depth}                  & \thickspace &     
            \texttt{thu\_00\_a}                     & \thickspace &     
            \texttt{zur\_07\_a}                     \\                  
            \midrule 
            
            Ground truth & \thickspace &
            1.09                                          & \thickspace &     
            1.20                                          & \thickspace &     
            1.12                                          & \thickspace &     
            1.07                                          & \thickspace &     
            --                                            & \thickspace &     
            1.01                                          & \thickspace &     
            1.04                                          \\                  
            Shiba \etal \cite{shiba2022secrets} & \thickspace &
            1.17                                          & \thickspace &     
            \underline{1.30}                              & \thickspace &     
            1.23                                          & \thickspace &     
            1.11                                          & \thickspace &     
            \underline{1.93}                              & \thickspace &     
            1.42                                          & \thickspace &     
            \underline{1.63}                              \\                  
            EINCM~\cite{karmokar2025secrets} & \thickspace &
            \textbf{1.33\tiny{3}}                         & \thickspace &     
            \textbf{1.45\tiny{5}}                         & \thickspace &     
            \textbf{1.39\tiny{2}}                         & \thickspace &     
            \textbf{1.23\tiny{0}}                         & \thickspace &     
            \textbf{1.97}                                 & \thickspace &     
            \textbf{1.53\tiny{2}}                         & \thickspace &     
            \textbf{1.63\tiny{6}}                         \\                  
            Ours (OPCM) & \thickspace &
            \underline{1.24\tiny{2}}                      & \thickspace &     
            {1.22\tiny{8}}                                & \thickspace &     
            \underline{1.30\tiny{9}}                      & \thickspace &     
            \underline{1.19\tiny{4}}                      & \thickspace &     
            \underline{1.93}                              & \thickspace &     
            \underline{1.48\tiny{7}}                      & \thickspace &     
            {1.61\tiny{6}}                               \\                  
            \bottomrule
        
        \end{tabular}
    \end{adjustbox}
    \caption{Flow warp loss (FWL) on MVSEC, ECD, and DSEC (train). \emph{Bold} and \emph{underline} indicates
    \textbf{best} and \underline{second best}, respectively.}
    \label{tab:fwl_scores}
\end{table}

\subsection{DSEC Evaluations}
Shown in \cref{tab:dsec_testset_results} are the quantitative evaluations on
the DSEC test set sequences. We observe that OPCM demonstrates competitive or
superior performance relative to other MB approaches. Despite this, OPCM,
similar to other MB methods, is markedly outperformed by learning-based
techniques in terms of accuracy. This discrepancy in performance between MB and
learning-based methods is well known and attributed to the following. DSEC
reports at least 20\% of the pixel displacements greater than 22 pixels with
maximum displacements up to 210 pixels, which is much larger than MVSEC
(between 4-20 pixels of maximum displacements in a data sample for
$\emph{dt}=4$ indoor scenes). 

This makes DSEC a more challenging dataset for all methods. Furthermore, CM is
a model-based optimization framework, and CM-based approaches require good
initializations. In the case of contiguous evaluation points, current
optimizations can leverage previous solutions. The DSEC benchmark evaluates
non-contiguously, \ie, time windows of $100\,$ms at every $500\,$ms.
Additionally, the evaluation points also contain jumps within the sequence,
introducing larger intra-sequence discontinuities. Also, the DSEC ground-truth
flow is evaluated for dense flow. However, all model-based methods can only
reliably estimate optical flow where events exist.

\begin{table}[!htb]
    \centering
    \begin{adjustbox}{max width=\columnwidth}
        \begin{tabular}{@{\thinspace}c@{\medspace}l@{\thickspace}c@{\thickspace}
        c@{\thickspace}c@{\thickspace}c@{\thickspace}c@{\thickspace}c@{\thickspace}c
        c@{\thickspace}c@{\thickspace}c@{\thickspace}c@{\thickspace}c@{\thickspace}c
        c@{\thickspace}c@{\thickspace}c@{\thickspace}c@{\thickspace}c} 
    
            \toprule
            & & &
            \multicolumn{5}{c}{\texttt{thun\_01\_a}}              & \thickspace &     
            \multicolumn{5}{c}{\texttt{thun\_01\_b}}              & \thickspace &     
            \multicolumn{5}{c}{\texttt{zurich\_city\_15\_a}}      \\                  
            \cmidrule(){4-8} \cmidrule(){10-14} \cmidrule(){16-20}

            & & &
            AEE $\downarrow$         & \thickspace & \%Out $\downarrow$         & \thickspace & FWL $\uparrow$        & \thickspace &     
            AEE $\downarrow$         & \thickspace & \%Out $\downarrow$         & \thickspace & FWL $\uparrow$        & \thickspace &     
            AEE $\downarrow$         & \thickspace & \%Out $\downarrow$         & \thickspace & FWL $\uparrow$        \\                  
            \midrule 
            
            \multirow{1}{*}{\rotatebox[origin=c]{90}{SL}}
            & E-RAFT \cite{gehrig2021eraft} & &
            \textbf{0.65}             & \thickspace & \textbf{1.87}            & \thickspace & 1.20                       & \thickspace &     
            \textbf{0.58}             & \thickspace & \textbf{1.52}            & \thickspace & 1.18                       & \thickspace &     
            \textbf{0.59}             & \thickspace & \textbf{1.30}            & \thickspace & 1.34                       \\                  
            \midrule

            \multirow{4}{*}{\rotatebox[origin=c]{90}{MB}} 
            & Brebion \etal \cite{brebion2021realtimeflow} & &
            3.01                     & \thickspace & 29.69\tiny{7}             & \thickspace & --                         & \thickspace &     
            3.91\tiny{3}             & \thickspace & 34.69                     & \thickspace & --                         & \thickspace &     
            3.78\tiny{1}             & \thickspace & 37.98\tiny{7}             & \thickspace & --                         \\                  
            
            & MultiCM \cite{shiba2022secrets} & &
            2.12                     & \thickspace & 17.68                     & \thickspace & 1.24                       & \thickspace &     
            {2.48}                   & \thickspace & {23.56}                   & \thickspace & 1.124                      & \thickspace &     
            \underline{2.35}         & \thickspace & \underline{20.99}         & \thickspace & 1.41                       \\                  
            
            & EINCM \cite{karmokar2025secrets} & &
            {2.01\tiny{5}}           & \thickspace & \underline{16.17\tiny{4}} & \thickspace & \textbf{1.40}              & \thickspace &     
            2.77\tiny{8}             & \thickspace & 26.56                     & \thickspace & \textbf{1.39\tiny{6}}      & \thickspace &     
            3.00\tiny{5}             & \thickspace & 26.63\tiny{3}             & \thickspace & \textbf{1.60\tiny{3}}      \\                  
            
            & Ours (OPCM) & &
            \underline{2.01\tiny{4}} & \thickspace & {16.47\tiny{4}}           & \thickspace & \underline{1.36\tiny{9}}   & \thickspace &     
            \underline{2.43\tiny{6}} & \thickspace & \underline{21.04\tiny{8}} & \thickspace & \underline{1.34\tiny{9}}   & \thickspace &     
            {2.82\tiny{7}} & \thickspace & {25.02\tiny{5}}           & \thickspace & \underline{1.51\tiny{8}}              \\                  
            \bottomrule
        \end{tabular}
    \end{adjustbox}
    \caption{Accuracy and loss scores on the DSEC test sequences. \emph{Bold}
    and \emph{underline} typefaces indicate the \textbf{best} and the
    \underline{second best}, respectively.}
    \label{tab:dsec_testset_results}
\end{table}


\section{Discussion}
\label{sec:discussion}

Orientation priors have a strict causal relationship with inertia. This poses a
natural limitation for guidance due to its strong reliance on self-motion.
Since these priors are derived from 3D camera velocities, they are effective
primarily in scenarios where the motion originates from the camera itself.  In
cases involving independently moving objects or static scenes with external
motion, such priors may not provide meaningful cues for the flow field. While
the core CM framework still operates in the absence of egomotion, the benefits
of orientation guidance are naturally diminished in these cases.

Due to the high non-convexity of the traditional CM framework, all previous CM
approaches heavily rely on an effective regularization mechanism. Typically,
the process of fine-tuning the balancing coefficient ($\lambda$) is somewhat
arbitrary, even when the relative dynamics in the scene are well-understood. In
this work, we have illustrated that this reliance on regularization is not
indispensable. The use of orientation priors allows for a more informed and
methodical approach to coefficient tuning ($\beta_{\text{lin}}$ and
$\beta_{\text{ang}}$). Unlike the traditional regularization term, which merely
imposes a global smoothness constraint on the motion field, orientation priors
can be leveraged to impose fine-grained directional alignment constraints.

Given linear and angular velocities, their corresponding flow orientation
fields are fully determined. However, combining directions is not trivial and
additionally requires accurate scene depth information.  To maintain simplicity
and general applicability, we constrained our priors to directional orientation
information only. While full-flow vectors can be theoretically inferred from
known depth and velocities under rigid scene assumptions, our aim in this work
is not to perform global motion estimation. Instead, our focus is on using
egomotion cues to assist the CM process.  Extending this idea to incorporate
richer priors, such as coarse depth, scene semantics, or confidence-weighted
guidance, offers promising directions for future work. 


\section{Conclusion}
\label{sec:conclusion}
In this work we proposed OPCM, a biologically inspired extension to the CM
framework that enhances event-based optical flow estimation through guidance
from orientation priors derived from 3D camera velocities. Motivated by the
visual-vestibular coupling observed in nature, our approach leverages
first-order motion cues to regularize the CM optimization process in a modular
and interpretable way. Employing multiple reference times and multiple scales
while incorporating these priors with \textit{contrast}, our methodology
achieves stable convergence and improved accuracy in event-based optical flow
estimation even in low-texture regions. Although the benefits of our method are
naturally contingent upon the presence of camera motion, OPCM still establishes
competitive or superior performance on public dataset benchmarks.

\section*{Acknowledgments} 
This material is based upon work supported by the Air Force Research Laboratory
under award number FA8571-23-C-0041. Pritam P. Karmokar was supported by a
University of Texas at Arlington Dissertation Fellowship.

{
\small
\bibliographystyle{ieeenat_fullname}
\bibliography{references}

@String(AAAI = {AAAI})

@article{akolkar2020real,
  title={Real-time high speed motion prediction using fast aperture-robust event-driven visual flow},
  author={Akolkar, Himanshu and Ieng, Sio-Hoi and Benosman, Ryad},
  journal={IEEE Transactions on Pattern Analysis and Machine Intelligence},
  volume={44},
  number={1},
  pages={361--372},
  year={2020},
  doi={10.1109/TPAMI.2020.3010468}
}

@article{bertolini2016moving,
  title={Moving in a moving world: A review on vestibular motion sickness},
  author={Bertolini, Giovanni and Straumann, Dominik},
  journal={Frontiers in Neurology},
  volume={7},
  pages={14},
  year={2016},
  publisher={Frontiers Media SA},
  doi={10.3389/fneur.2016.00014}
}

@article{brebion2021realtimeflow,
  title={Real-time optical flow for vehicular perception with low-and high-resolution event cameras},
  author={Brebion, Vincent and Moreau, Julien and Davoine, Franck},
  journal={IEEE Transactions on Intelligent Transportation Systems},
  volume={23},
  number={9},
  pages={15066--15078},
  year={2021},
  doi={10.1109/TITS.2021.3136358}
}

@article{buchanan2022deep,
  title={Deep imu bias inference for robust visual-inertial odometry with factor graphs},
  author={Buchanan, Russell and Agrawal, Varun and Camurri, Marco and Dellaert, Frank and Fallon, Maurice},
  journal={IEEE Robotics and Automation Letters},
  volume={8},
  number={1},
  pages={41--48},
  year={2022},
  doi={10.1109/LRA.2022.3222956}
}

@article{butler2010bayesian,
  title={Bayesian integration of visual and vestibular signals for heading},
  author={Butler, John S and Smith, Stuart T and Campos, Jennifer L and B{\"u}lthoff, Heinrich H},
  journal={Journal of Vision},
  volume={10},
  number={11},
  pages={23--23},
  year={2010},
  publisher={The Association for Research in Vision and Ophthalmology},
  doi={10.1167/10.11.23}
}

@inbook{campos2012multimodal,
  title={Multimodal integration during self-motion in virtual reality},
  author={Campos, Jennifer L and B{\"u}lthoff, Heinrich H},
  booktitle={The Neural Bases of Multisensory Processes},
  pages={603-628},
  year={2012},
  publisher={CRC Press},
  doi={10.1201/B11092-38}
}

@article{campos2014contributions,
  title={Contributions of visual and proprioceptive information to travelled distance estimation during changing sensory congruencies},
  author={Campos, Jennifer L and Butler, John S and B{\"u}lthoff, Heinrich H},
  journal={Experimental Brain Research},
  volume={232},
  number={10},
  pages={3277--3289},
  year={2014},
  publisher={Springer},
  doi={10.1007/s00221-014-4011-0}
}

@article{cohen2003critical,
  title={The critical role of velocity storage in production of motion sickness},
  author={Cohen, Bernard and Dai, Mingjia and Raphan, Theodore},
  journal={Annals of the New York Academy of Sciences},
  volume={1004},
  number={1},
  pages={359--376},
  year={2003},
  publisher={Wiley Online Library},
  doi={10.1196/annals.1303.034}
}

@article{deng2018visual,
  title={Visual-inertial estimation of velocity for multicopters based on vision motion constraint},
  author={Deng, Heng and Arif, Usman and Fu, Qiang and Xi, Zhiyu and Quan, Quan and Cai, Kai-Yuan},
  journal={Robotics and Autonomous Systems},
  volume={107},
  pages={262--279},
  year={2018},
  publisher={Elsevier},
  doi={10.1016/j.robot.2018.06.010}
}

@inproceedings{ding2022steflownet,
  title={Spatio-temporal recurrent networks for event-based optical flow estimation},
  author={Ding, Ziluo and Zhao, Rui and Zhang, Jiyuan and Gao, Tianxiao and Xiong, Ruiqin and Yu, Zhaofei and Huang, Tiejun},
  booktitle={Proceedings of the AAAI Conference on Artificial Intelligence},
  volume={36},
  pages={525--533},
  year={2022},
  doi={10.1609/aaai.v36i1.19931}
}

@article{duffy2000optic,
  title={Optic flow analysis for self-movement perception},
  author={Duffy, Charles J},
  journal={International Review of Neurobiology},
  volume={44},
  pages={199--218},
  year={2000},
  publisher={Elsevier},
  doi={10.1016/S0074-7742(08)60743-6}
}

@article{feigl2020rnn,
  title={RNN-aided human velocity estimation from a single IMU},
  author={Feigl, Tobias and Kram, Sebastian and Woller, Philipp and Siddiqui, Ramiz H and Philippsen, Michael and Mutschler, Christopher},
  journal={Sensors},
  volume={20},
  number={13},
  pages={3656},
  year={2020},
  publisher={MDPI},
  doi={10.3390/s20133656}
}

@article{fetsch2009dynamic,
  title={Dynamic reweighting of visual and vestibular cues during self-motion perception},
  author={Fetsch, Christopher R and Turner, Amanda H and DeAngelis, Gregory C and Angelaki, Dora E},
  journal={Journal of Neuroscience},
  volume={29},
  number={49},
  pages={15601--15612},
  year={2009},
  publisher={Society for Neuroscience},
  doi={10.1523/JNEUROSCI.2574-09.2009}
}

@article{fetsch2010visual,
  title={Visual-vestibular cue integration for heading perception: applications of optimal cue integration theory},
  author={Fetsch, Christopher R and DeAngelis, Gregory C and Angelaki, Dora E},
  journal={European Journal of Neuroscience},
  volume={31},
  number={10},
  pages={1721--1729},
  year={2010},
  publisher={Wiley Online Library},
  doi={10.1111/j.1460-9568.2010.07207.x}
}

@article{frostig2018highleveltracing,
  title={Compiling machine learning programs via high-level tracing},
  author={Frostig, Roy and Johnson, Matthew James and Leary, Chris},
  journal={Systems for Machine Learning},
  volume={4},
  number={9},
  year={2018},
  publisher={SysML},
  url={https://api.semanticscholar.org/CorpusID:4625928}
}

@inproceedings{gallego2018unifying,
  title={A unifying contrast maximization framework for event cameras, with applications to motion, depth, and optical flow estimation},
  author={Gallego, Guillermo and Rebecq, Henri and Scaramuzza, Davide},
  booktitle={Proceedings of the IEEE/CVF Conference on Computer Vision and Pattern Recognition},
  pages={3867--3876},
  year={2018},
  doi={10.1109/CVPR.2018.00407}
}

@inproceedings{gallego2019focus,
  title={Focus is all you need: Loss functions for event-based vision},
  author={Gallego, Guillermo and Gehrig, Mathias and Scaramuzza, Davide},
  booktitle={Proceedings of the IEEE/CVF Conference on Computer Vision and Pattern Recognition},
  pages={12280--12289},
  year={2019},
  doi={10.1109/CVPR.2019.01256}
}

@inproceedings{gehrig2019evflownetest,
  title={End-to-end learning of representations for asynchronous event-based data},
  author={Gehrig, Daniel and Loquercio, Antonio and Derpanis, Konstantinos G and Scaramuzza, Davide},
  booktitle={Proceedings of the IEEE/CVF International Conference on Computer Vision},
  pages={5633--5643},
  year={2019},
  doi={10.1109/ICCV.2019.00573}
}

@article{gehrig2020eklt,
  title={EKLT: Asynchronous photometric feature tracking using events and frames},
  author={Gehrig, Daniel and Rebecq, Henri and Gallego, Guillermo and Scaramuzza, Davide},
  journal={International Journal of Computer Vision},
  volume={128},
  number={3},
  pages={601--618},
  year={2020},
  publisher={Springer}
}

@inproceedings{gehrig2021eraft,
  title={E-raft: Dense optical flow from event cameras},
  author={Gehrig, Mathias and Millh{\"a}usler, Mario and Gehrig, Daniel and Scaramuzza, Davide},
  booktitle={Proceedings of the International Conference on 3D Vision},
  pages={197--206},
  year={2021},
  organization={IEEE},
  doi={10.1109/3DV53792.2021.00030}
}

@article{gehrig2024dense,
  title={Dense Continuous-Time Optical Flow from Event Cameras},
  author={Gehrig, Mathias and Muglikar, Manasi and Scaramuzza, Davide},
  journal={IEEE Transactions on Pattern Analysis and Machine Intelligence},
  year={2024},
  doi={10.1109/TPAMI.2024.3361671}
}

@article{gibson1950perception,
  title={The perception of the visual world},
  author={Gibson, James J},
  year={1950},
  publisher={Houghton Mifflin}
}

@article{gibson1966senses,
  title={The senses considered as perceptual systems},
  author={Gibson, James J},
  year={1966},
  publisher={Houghton Mifflin}
}

@inproceedings{gluckman1998ego,
  title={Ego-motion and omnidirectional cameras},
  author={Gluckman, Joshua and Nayar, Shree K},
  booktitle={Proceedings of the IEEE International Conference on Computer Vision},
  pages={999--1005},
  year={1998},
  doi={10.1109/ICCV.1998.710838}
}

@article{gu2008neural,
  title={Neural correlates of multisensory cue integration in macaque MSTd},
  author={Gu, Yong and Angelaki, Dora E and DeAngelis, Gregory C},
  journal={Nature Neuroscience},
  volume={11},
  number={10},
  pages={1201--1210},
  year={2008},
  publisher={Nature Publishing Group US New York},
  doi={10.1038/nn.2191}
}

@inproceedings{hagenaars2021convgruevflownet,
  title={Self-supervised learning of event-based optical flow with spiking neural networks},
  author={Hagenaars, Jesse and Paredes-Vall{\'e}s, Federico and De Croon, Guido},
  booktitle={Proceedings of the Advances in Neural Information Processing Systems},
  volume={34},
  pages={7167--7179},
  year={2021},
  doi={10.48550/arXiv.2106.01862}
}

@inproceedings{hamann2024motion,
  title={Motion-prior contrast maximization for dense continuous-time motion estimation},
  author={Hamann, Friedhelm and Wang, Ziyun and Asmanis, Ioannis and Chaney, Kenneth and Gallego, Guillermo and Daniilidis, Kostas},
  booktitle={Proceedings of the European Conference on Computer Vision},
  pages={18--37},
  year={2024},
  organization={Springer},
  doi={10.1007/978-3-031-72646-0_2}
}

@article{henning2022populations,
  title={Populations of local direction-selective cells encode global motion patterns generated by self-motion},
  author={Henning, Miriam and Ramos-Traslosheros, Giordano and G{\"u}r, Burak and Silies, Marion},
  journal={Science Advances},
  volume={8},
  number={31},
  pages={eabi7112},
  year={2022},
  publisher={American Association for the Advancement of Science},
  doi={10.1126/sciadv.abi7112}
}

@article{horst2015reliability,
  title={Reliability-based weighting of visual and vestibular cues in displacement estimation},
  author={ter Horst, Arjan C and Koppen, Mathieu and Selen, Luc PJ and Medendorp, W Pieter},
  journal={PLOS One},
  volume={10},
  number={12},
  pages={e0145015},
  year={2015},
  publisher={Public Library of Science San Francisco, CA USA},
  doi={10.1371/journal.pone.0145015}
}

@misc{jax2018github,
  author={Bradbury, James and Frostig, Roy and Hawkins, Peter and Johnson, Matthew James and Leary, Chris and Maclaurin, Dougal and Necula, George and Paszke, Adam and VanderPlas, Jake and Wanderman-Milne, Skye and Zhang, Qiao},
  title={{JAX}: Composable transformations of {P}ython+{N}um{P}y programs},
  url={http://github.com/google/jax},
  version={0.3.13},
  year={2018}
}

@article{jaxopt_implicit_diff,
  title={Efficient and modular implicit differentiation},
  author={Blondel, Mathieu and Berthet, Quentin and Cuturi, Marco and Frostig, Roy and Hoyer, Stephan and Llinares-L{\'o}pez, Felipe and Pedregosa, Fabian and Vert, Jean-Philippe},
  journal={arXiv preprint arXiv:2105.15183},
  year={2021},
  doi={10.48550/arXiv.2105.15183}
}

@inproceedings{karmokar2025secrets,
  title={Secrets of edge-informed contrast maximization for event-based vision},
  author={Karmokar, Pritam P and Nguyen, Quan H and Beksi, William J},
  booktitle={Proceedings of the IEEE/CVF Winter Conference on Applications of Computer Vision},
  pages={630--639},
  year={2025}
}

@article{lappe1999perception,
  title={Perception of self-motion from visual flow},
  author={Lappe, Markus and Bremmer, Frank and van den Berg, Albert V},
  journal={Trends in Cognitive Sciences},
  volume={3},
  number={9},
  pages={329--336},
  year={1999},
  publisher={Elsevier},
  doi={10.1016/S1364-6613(99)01364-9}
}

@inproceedings{liu2023tma,
  title={Tma: Temporal motion aggregation for event-based optical flow},
  author={Liu, Haotian and Chen, Guang and Qu, Sanqing and Zhang, Yanping and Li, Zhijun and Knoll, Alois and Jiang, Changjun},
  booktitle={Proceedings of the IEEE/CVF International Conference on Computer Vision},
  pages={9685--9694},
  year={2023},
  doi={10.1109/ICCV51070.2023.00888}
}

@inproceedings{lee2020spikeflownet,
  title={Spike-flownet: Event-based optical flow estimation with energy-efficient hybrid neural networks},
  author={Lee, Chankyu and Kosta, Adarsh Kumar and Zhu, Alex Zihao and Chaney, Kenneth and Daniilidis, Kostas and Roy, Kaushik},
  booktitle={Proceedings of the European Conference on Computer Vision},
  pages={366--382},
  year={2020},
  organization={Springer},
  doi={10.1007/978-3-030-58526-6_22}
}

@article{lu2023event,
  title={Event-based visual inertial velometer},
  author={Lu, Xiuyuan and Zhou, Yi and Niu, Junkai and Zhong, Sheng and Shen, Shaojie},
  journal={arXiv preprint arXiv:2311.18189},
  year={2023},
  doi={10.48550/arXiv.2311.18189}
}

@article{mueggler2017event,
  title={The event-camera dataset and simulator: Event-based data for pose estimation, visual odometry, and SLAM},
  author={Mueggler, Elias and Rebecq, Henri and Gallego, Guillermo and Delbruck, Tobi and Scaramuzza, Davide},
  journal={The International Journal of Robotics Research},
  volume={36},
  number={2},
  pages={142--149},
  year={2017},
  publisher={SAGE Publications Sage UK: London, England},
  doi={10.1177/0278364917691115}
}

@article{nagata2021matchtimesurf,
  title={Optical flow estimation by matching time surface with event-based cameras},
  author={Nagata, Jun and Sekikawa, Yusuke and Aoki, Yoshimitsu},
  journal={Sensors},
  volume={21},
  number={4},
  pages={1150},
  year={2021},
  doi={10.3390/s21041150},
  publisher={MDPI}
}

@misc{opcm,
  key={OPCM},
  note={\url{https://github.com/robotic-vision-lab/Inertia-Informed-Contrast-Maximization}}
}

@inproceedings{paredes2021back,
  title={Back to event basics: Self-supervised learning of image reconstruction for event cameras via photometric constancy},
  author={Paredes-Vall{\'e}s, Federico and de Croon, Guido},
  booktitle={Proceedings of the IEEE/CVF Conference on Computer Vision and Pattern Recognition},
  pages={3446--3455},
  year={2021},
  doi={10.1109/CVPR46437.2021.00345}
}

@article{raudies2012review,
  title={A review and evaluation of methods estimating ego-motion},
  author={Raudies, Florian and Neumann, Heiko},
  journal={Computer Vision and Image Understanding},
  volume={116},
  number={5},
  pages={606--633},
  year={2012},
  publisher={Elsevier}
}

@inproceedings{seok2020robust,
  title={Robust feature tracking in dvs event stream using b{\'e}zier mapping},
  author={Seok, Hochang and Lim, Jongwoo},
  booktitle={Proceedings of the IEEE/CVF Winter Conference on Applications of Computer Vision},
  pages={1658--1667},
  year={2020}
}

@article{shanno1970conditioning,
  title={Conditioning of quasi-Newton methods for function minimization},
  author={Shanno, David F},
  journal={Mathematics of Computation},
  volume={24},
  number={111},
  pages={647--656},
  year={1970},
  doi={10.2307/2004840}
}

@article{shanno1985example,
  title={An example of numerical nonconvergence of a variable-metric method},
  author={Shanno, David F},
  journal={Journal of Optimization Theory and Applications},
  volume={46},
  pages={87--94},
  year={1985},
  publisher={Springer},
  doi={10.1007/BF00938762}
}

@article{shiba2022eventcollapse,
  title={Event collapse in contrast maximization frameworks},
  author={Shiba, Shintaro and Aoki, Yoshimitsu and Gallego, Guillermo},
  journal={Sensors},
  volume={22},
  number={14},
  pages={5190},
  year={2022},
  doi={10.3390/s22145190},
  publisher={MDPI}
}

@inproceedings{shiba2022secrets,
  title={Secrets of event-based optical flow},
  author={Shiba, Shintaro and Aoki, Yoshimitsu and Gallego, Guillermo},
  booktitle={Proceedings of the European Conference on Computer Vision},
  pages={628--645},
  year={2022},
  doi={10.1007/978-3-031-19797-0_36},
  organization={Springer}
}

@article{shiba2024secrets,
  title={Secrets of event-based optical flow, depth and ego-motion estimation by contrast maximization},
  author={Shiba, Shintaro and Klose, Yannick and Aoki, Yoshimitsu and Gallego, Guillermo},
  journal={IEEE Transactions on Pattern Analysis and Machine Intelligence},
  year={2024}
}

@article{shiba2025simultaneous,
  title={Simultaneous motion and noise estimation with event cameras},
  author={Shiba, Shintaro and Aoki, Yoshimitsu and Gallego, Guillermo},
  journal={arXiv preprint arXiv:2504.04029},
  year={2025},
  doi={10.48550/arXiv.2504.04029}
}

@inproceedings{srinivasan2013high,
  title={High frame rate egomotion estimation},
  author={Srinivasan, Natesh and Roberts, Richard and Dellaert, Frank},
  booktitle={Proceedings of the International Conference on Computer Vision Systems},
  pages={183--192},
  year={2013},
  organization={Springer}
}

@inproceedings{stoffregen2019analysis,
  title={Event cameras, contrast maximization and reward functions: An analysis},
  author={Stoffregen, Timo and Kleeman, Lindsay},
  booktitle={Proceedings of the IEEE/CVF Conference on Computer Vision and Pattern Recognition},
  pages={12300--12308},
  year={2019},
  doi={10.1109/CVPR.2019.01258}
}

@inproceedings{stoffregen2020evflownetplus,
  title={Reducing the sim-to-real gap for event cameras},
  author={Stoffregen, Timo and Scheerlinck, Cedric and Scaramuzza, Davide and Drummond, Tom and Barnes, Nick and Kleeman, Lindsay and Mahony, Robert},
  booktitle={Proceedings of the European Conference on Computer Vision},
  pages={534--549},
  year={2020},
  organization={Springer},
  doi={10.1007/978-3-030-58583-9_32}
}

@article{sunkara2016joint,
  title={Joint representation of translational and rotational components of optic flow in parietal cortex},
  author={Sunkara, Adhira and DeAngelis, Gregory C and Angelaki, Dora E},
  journal={Proceedings of the National Academy of Sciences},
  volume={113},
  number={18},
  pages={5077--5082},
  year={2016},
  publisher={National Academy of Sciences},
  doi={10.1073/pnas.1604818113}
}

@inproceedings{teed2020raft,
  title={Raft: Recurrent all-pairs field transforms for optical flow},
  author={Teed, Zachary and Deng, Jia},
  booktitle={Proceedings of the European Conference on Computer Vision},
  pages={402--419},
  year={2020},
  organization={Springer},
  doi={10.1007/978-3-030-58536-5_24}
}

@book{trucco1998introductory,
  title={Introductory techniques for 3-D computer vision},
  author={Trucco, Emanuele and Verri, Alessandro},
  volume={201},
  year={1998},
  publisher={Prentice Hall Englewood Cliffs},
  doi={10.5555/551277}
}

@article{verri1989mathematical,
  title={Mathematical properties of the two-dimensional motion field: From singular points to motion parameters},
  author={Verri, Alessandro and Girosi, F and Torre, Vincente},
  journal={Journal of the Optical Society of America A},
  volume={6},
  number={5},
  pages={698--712},
  year={1989},
  publisher={Optical Society of America},
  doi={10.1364/JOSAA.6.000698}
}

@inproceedings{wang2020jointfiltering,
  title={Joint filtering of intensity images and neuromorphic events for high-resolution noise-robust imaging},
  author={Wang, Zihao W and Duan, Peiqi and Cossairt, Oliver and Katsaggelos, Aggelos and Huang, Tiejun and Shi, Boxin},
  booktitle={Proceedings of the IEEE/CVF Conference on Computer Vision and Pattern Recognition},
  pages={1609--1619},
  year={2020},
  doi={10.1109/CVPR42600.2020.00168}
}

@article{wan2022learning,
  title={Learning dense and continuous optical flow from an event camera},
  author={Wan, Zhexiong and Dai, Yuchao and Mao, Yuxin},
  journal={IEEE Transactions on Image Processing},
  volume={31},
  pages={7237--7251},
  year={2022},
  doi={10.1109/TIP.2022.3220938}
}

@article{warren1988direction,
  title={Direction of self-motion is perceived from optical flow},
  author={Warren Jr, William H and Hannon, Daniel J},
  journal={Nature},
  volume={336},
  number={6195},
  pages={162--163},
  year={1988},
  publisher={Nature Publishing Group UK London},
  doi={10.1038/336162a0}
}

@article{winkel2015forced,
  title={Forced fusion in multisensory heading estimation},
  author={De Winkel, Ksander N and Katliar, Mikhail and B{\"u}lthoff, Heinrich H},
  journal={PLOS One},
  volume={10},
  number={5},
  pages={e0127104},
  year={2015},
  publisher={Public Library of Science San Francisco, CA USA},
  doi={10.1371/journal.pone.0127104}
}

@article{xiao2025inertial,
  title={An inertial sequence learning framework for vehicle speed estimation via smartphone IMU},
  author={Xiao, Xuan and Ren, Xiaotong and Li, Haitao},
  journal={arXiv preprint arXiv:2505.18490},
  year={2025},
  doi={10.48550/arXiv.2505.18490}
}

@article{xu2023tight,
  title={Tight fusion of events and inertial measurements for direct velocity estimation},
  author={Xu, Wanting and Peng, Xin and Kneip, Laurent},
  journal={IEEE Transactions on Robotics},
  volume={40},
  pages={240--256},
  year={2023}
}

@inproceedings{ye2020unsupervised,
  title={Unsupervised learning of dense optical flow, depth and egomotion with event-based sensors},
  author={Ye, Chengxi and Mitrokhin, Anton and Ferm{\"u}ller, Cornelia and Yorke, James A and Aloimonos, Yiannis},
  booktitle={Proceedings of the IEEE/RSJ International Conference on Intelligent Robots and Systems},
  pages={5831--5838},
  year={2020},
  doi={10.1109/IROS45743.2020.9341224}
}

@article{yuan2014localization,
  title={Localization and velocity tracking of human via 3 IMU sensors},
  author={Yuan, Qilong and Chen, I-Ming},
  journal={Sensors and Actuators A: Physical},
  volume={212},
  pages={25--33},
  year={2014},
  publisher={Elsevier},
  doi={10.1016/j.sna.2014.03.004}
}

@inproceedings{zhu2017event,
  title={Event-based feature tracking with probabilistic data association},
  author={Zhu, Alex Zihao and Atanasov, Nikolay and Daniilidis, Kostas},
  booktitle={Proceedings of the IEEE International Conference on Robotics and Automation},
  pages={4465--4470},
  year={2017}
}

@inproceedings{zhu2018evflownet, 
  title={EV-FlowNet: Self-supervised optical flow estimation for event-based cameras}, 
  author={Zhu, Alex Zihao and Yuan, Liangzhe and Chaney, Kenneth and Daniilidis, Kostas},
  booktitle={Proceedings of Robotics: Science and Systems}, 
  year={2018}, 
  doi={10.15607/RSS.2018.XIV.062} 
}

@inproceedings{zhu2019unsupervised,
  title={Unsupervised event-based learning of optical flow, depth, and egomotion},
  author={Zhu, Alex Zihao and Yuan, Liangzhe and Chaney, Kenneth and Daniilidis, Kostas},
  booktitle={Proceedings of the IEEE/CVF Conference on Computer Vision and Pattern Recognition},
  pages={989--997},
  year={2019},
  doi={10.1109/CVPR.2019.00108}
}
}

\section*{Supplementary Material}

\appendix

\section{MVSEC Outdoor Evaluations}
\label{sec:mvsec_outdoor_evaluations}
\begin{figure}
    \centering
    \adjincludegraphics[width=0.8\linewidth, trim={{0.07\width} {0.07\height} {0.09\width} {0.045\height}}, clip, cfbox=gray 0pt 0pt]{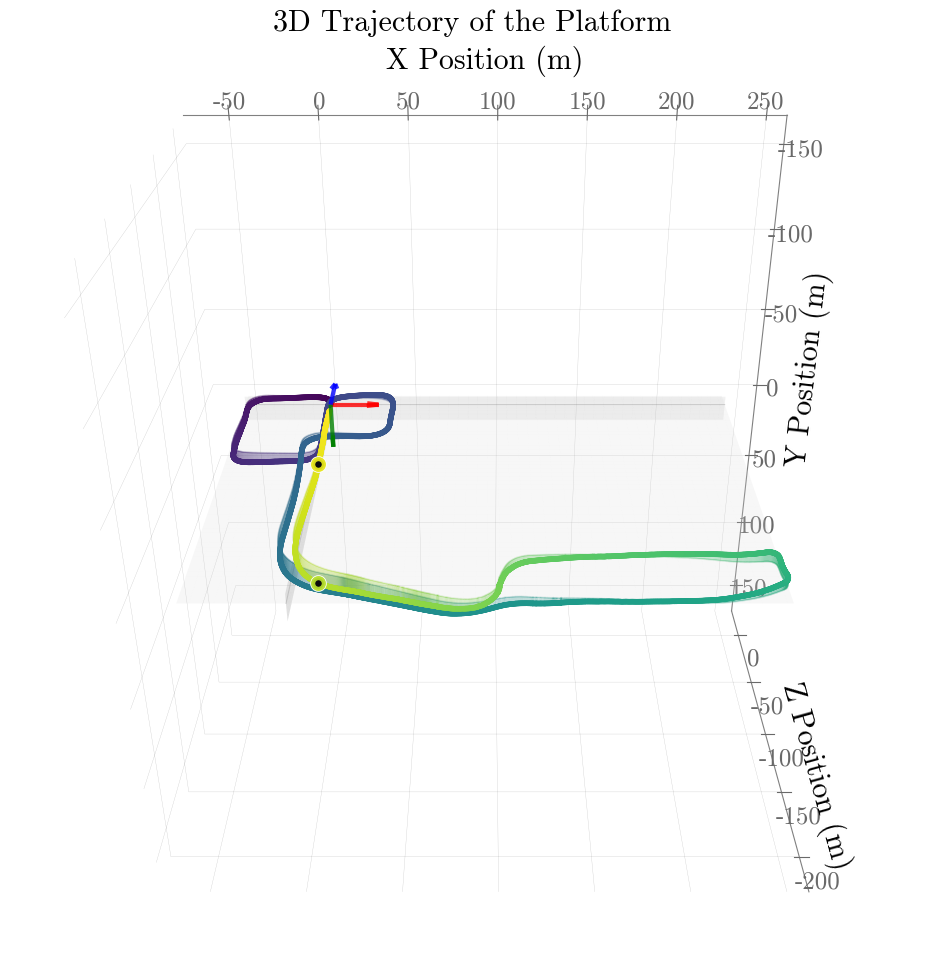}
    \caption[MVSEC \texttt{outdoor\_day1} 3D trajectory]{%
    The 3D trajectory of the moving camera for the \texttt{outdoor\_day1}
    sequence from MVSEC. The colormap \texttt{viridis} from matplotlib is used
    to code the progression of the trajectory from start to end, where purple
    (\protect\tikz[baseline]
    \protect\filldraw[draw=VIRIDIS_START_PURPLE!85!black!65,
    fill=VIRIDIS_START_PURPLE!85, line width=0.5pt] (0ex,-0.25ex) rectangle
    (2.25ex,1.25ex);) indicates the beginning and yellow
    (\protect\tikz[baseline]
    \protect\filldraw[draw=VIRIDIS_END_YELLOW!85!black!65,
    fill=VIRIDIS_END_YELLOW!85, line width=0.5pt] (0ex,-0.25ex) rectangle
    (2.25ex,1.25ex);) indicates the end of the entire sequence. The camera frame
    is drawn at the start position. Additionally, the starting and ending
    evaluation points are indicated by [\protect\tikz[baseline]
    \protect\filldraw[draw=VIRIDIS_START_OD1, fill=RAISIN_BLACK!85, line
    width=2.75pt] (0ex,0.5ex) circle (0.75ex);] and [\protect\tikz[baseline]
    \protect\filldraw[draw=VIRIDIS_END_OD1,fill=RAISIN_BLACK!85,line
    width=2.75pt] (0ex,0.5ex) circle (0.75ex);], respectively.}
    \label{fig:outdoor_day1_3d_traj}
\end{figure}

The MVSEC outdoor sequence \texttt{outdoor\_day1} consists of 11,440 image
frames.  Following related prior works
\cite{zhu2018evflownet,shiba2022secrets,karmokar2025secrets}, our evaluations
on MVSEC \texttt{outdoor\_day1} were performed on the 800 frames corresponding
to the image indices $[10148, 10948]$ (starting at 0), as depicted in
\cref{fig:outdoor_day1_3d_traj}, spanning a time window from $222.4\,$s to
$240.4\,$s.  These start and end times, interpreted as image timestamps,
correspond to $1,506,118,124.7330644\,$s and $1,506,118,142.7177844\,$s,
respectively.

\section{Orientation Map Extraction}
\label{sec:orientation_map_extraction}
\begin{figure*}
    \centering
    \begin{adjustbox}{max width=0.9\textwidth}
    \begin{tabular}{@{}c@{\medspace}c@{\thinspace}|@{\thinspace}c@{\thinspace}|@{\thinspace}c@{\thinspace}c@{\thinspace}c@{\medspace}c@{}}
        
        & 
        \multirow{2}{*}{\begin{tabular}{@{}c@{}} \begin{adjustbox}{max width=0.19\textwidth} Undistorted \end{adjustbox}\end{tabular}} & 
        \multicolumn{4}{c}{\begin{tabular}{@{}c@{}} \begin{adjustbox}{max width=0.19\textwidth} Distorted  \end{adjustbox}\end{tabular}} & \\
        \cmidrule(){3-6}
        & & 
        \begin{tabular}{@{}c@{}} \begin{adjustbox}{max width=0.19\textwidth} No Fill  \end{adjustbox}\end{tabular} &
        \begin{tabular}{@{}c@{}} \begin{adjustbox}{max width=0.19\textwidth} Navier-Stokes Fill     \end{adjustbox}\end{tabular} &
        \begin{tabular}{@{}c@{}} \begin{adjustbox}{max width=0.19\textwidth} Border-Reflect Fill  \end{adjustbox}\end{tabular} &
        \begin{tabular}{@{}c@{}} \begin{adjustbox}{max width=0.19\textwidth} Border-Replicate Fill   \end{adjustbox}\end{tabular} & \\
        
        \multirow{4}{*}{\rotatebox[origin=c]{90}{\begin{adjustbox}{max width=0.6\textwidth}Linear Distorted Orientation Map Extraction \hspace{5ex} \end{adjustbox}}} & 
        \begin{tabular}{@{}c@{}} \includegraphics[width=0.19\textwidth, cfbox=gray 0.1pt 0pt]{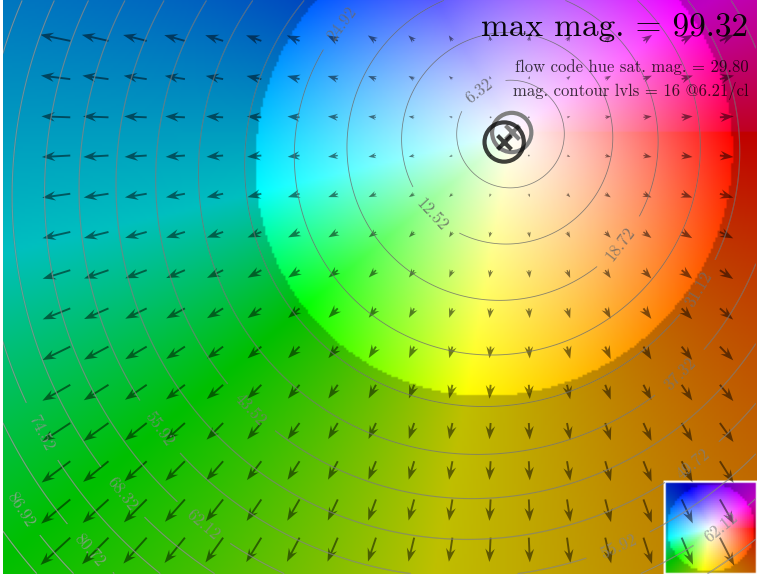} \end{tabular} &  
        \begin{tabular}{@{}c@{}} \includegraphics[width=0.19\textwidth, cfbox=gray 0.1pt 0pt]{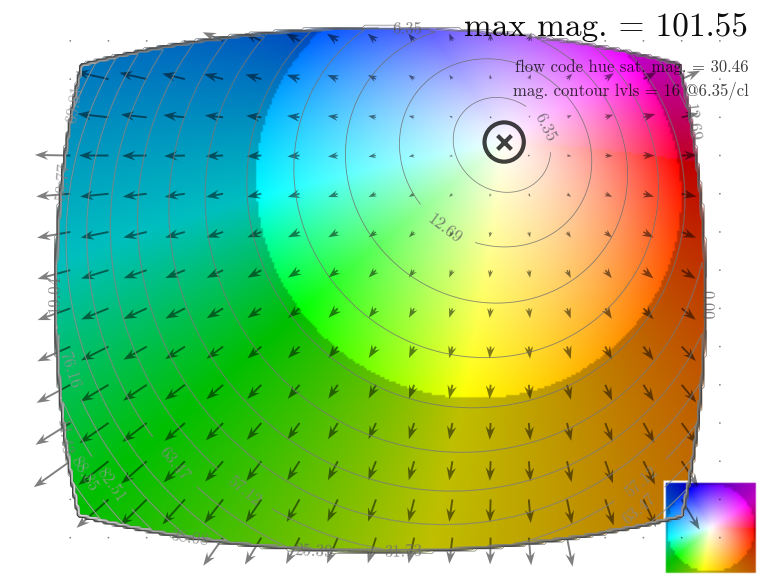} \end{tabular} &  
        \begin{tabular}{@{}c@{}} \includegraphics[width=0.19\textwidth, cfbox=gray 0.1pt 0pt]{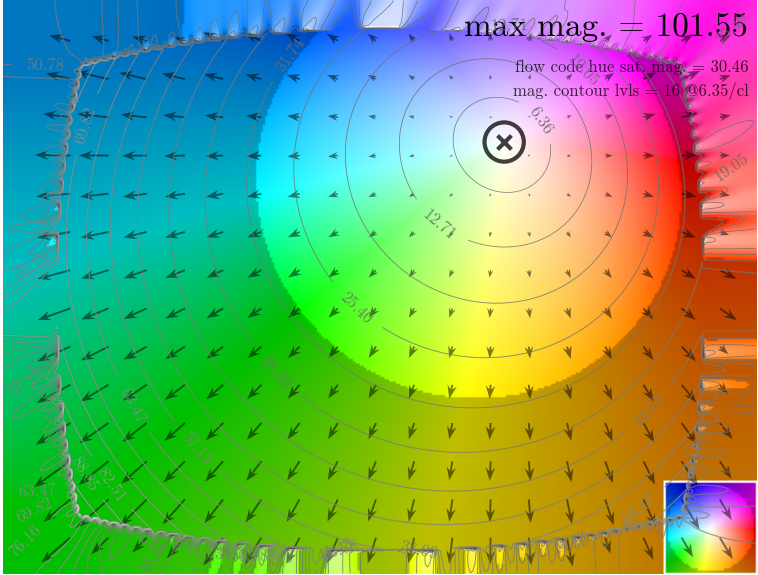} \end{tabular} &
        \begin{tabular}{@{}c@{}} \includegraphics[width=0.19\textwidth, cfbox=gray 0.1pt 0pt]{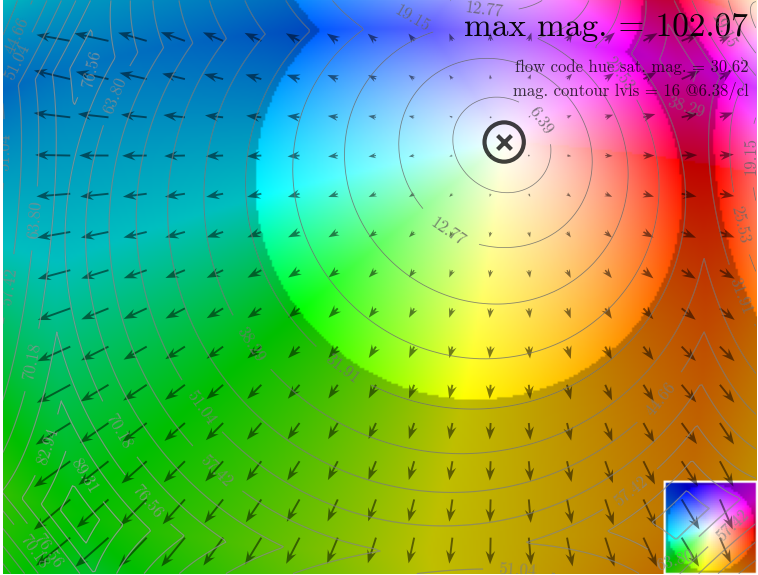} \end{tabular} &
        \begin{tabular}{@{}c@{}} \includegraphics[width=0.19\textwidth, cfbox=gray 0.1pt 0pt]{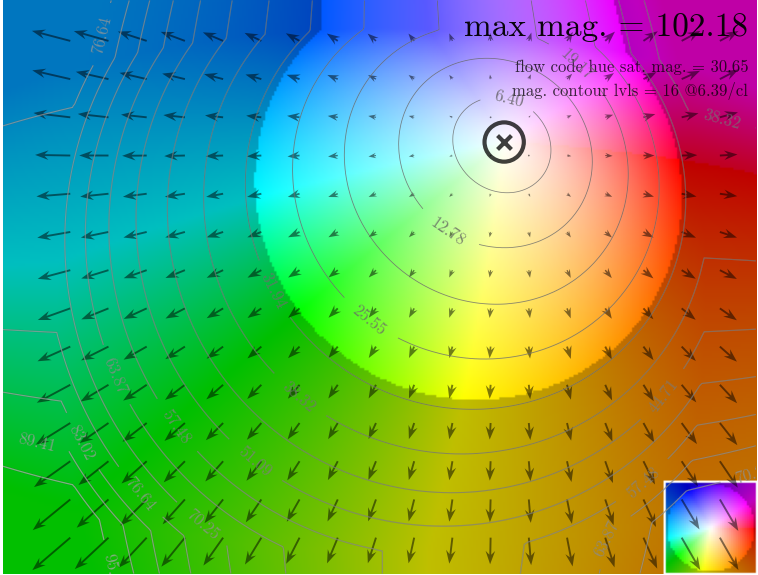} \end{tabular} &
        \rotatebox[origin=c]{-90}{\begin{adjustbox}{max width=0.19\textwidth} Not Normalized \end{adjustbox}}\\

        &
        \begin{tabular}{@{}c@{}} \includegraphics[width=0.19\textwidth, cfbox=gray 0.1pt 0pt]{images/omap_extraction/normed/no_fill/lin_omap_undist.png} \end{tabular} &  
        \begin{tabular}{@{}c@{}} \includegraphics[width=0.19\textwidth, cfbox=gray 0.1pt 0pt]{images/omap_extraction/normed/no_fill/lin_omap_dist.png} \end{tabular} &  
        \begin{tabular}{@{}c@{}} \includegraphics[width=0.19\textwidth, cfbox=gray 0.1pt 0pt]{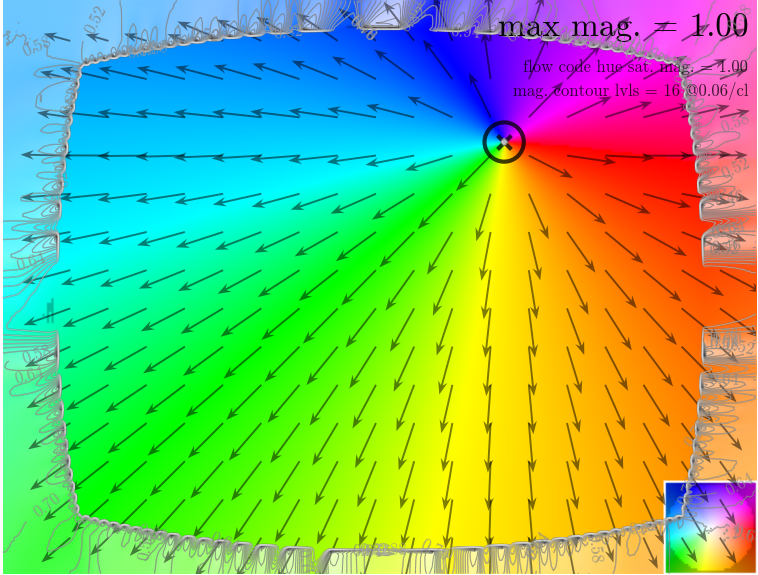} \end{tabular} &
        \begin{tabular}{@{}c@{}} \includegraphics[width=0.19\textwidth, cfbox=gray 0.1pt 0pt]{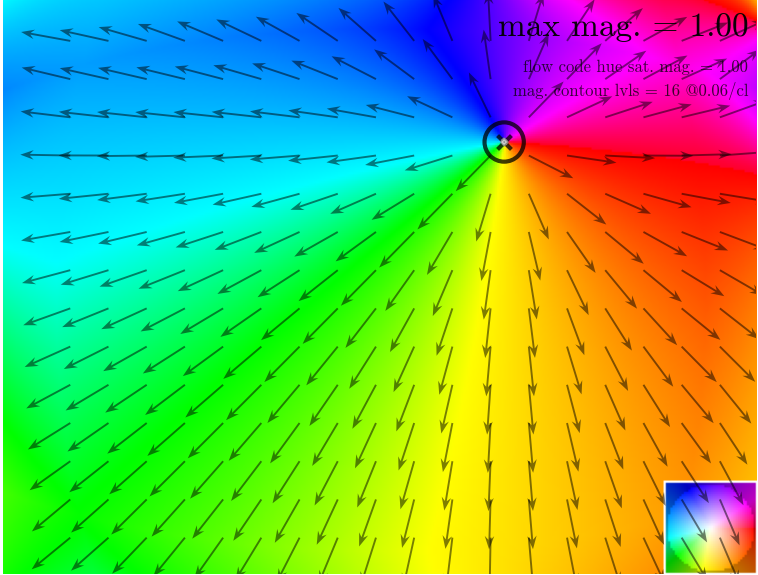} \end{tabular} &
        \begin{tabular}{@{}c@{}} \includegraphics[width=0.19\textwidth, cfbox=gray 0.1pt 0pt]{images/omap_extraction/normed/cv_rep_fill/lin_omap_dist.png} \end{tabular} & 
        \rotatebox[origin=c]{-90}{\begin{adjustbox}{max width=0.19\textwidth} Normalized \end{adjustbox}}\\
        
        &
        \multirow{2}{*}{\begin{tabular}{@{}c@{}} \includegraphics[width=0.18\textwidth, trim={0 2.5cm 0 0}, clip, cfbox=gray 0pt 0pt]{images/vels/vision_geom_lin.pdf} \end{tabular}} &  
        \begin{tabular}{@{}c@{}} \includegraphics[width=0.19\textwidth, cfbox=gray 0.1pt 0pt]{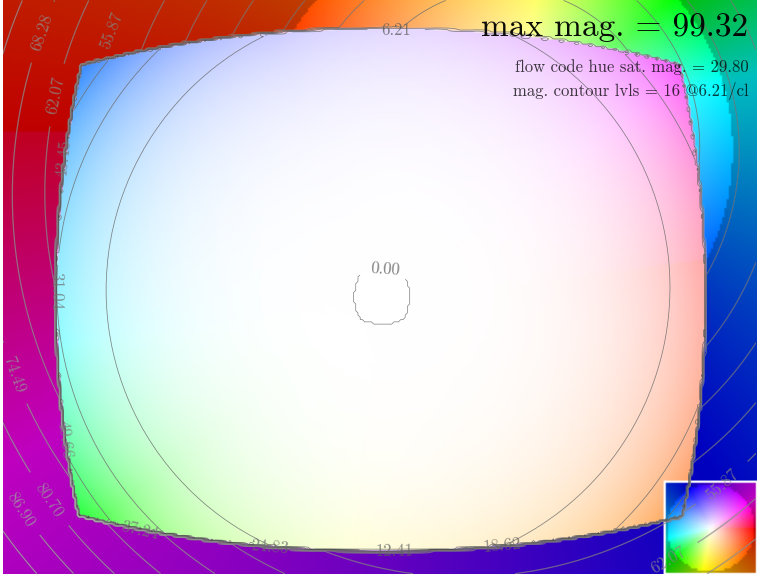} \end{tabular} &  
        \begin{tabular}{@{}c@{}} \includegraphics[width=0.19\textwidth, cfbox=gray 0.1pt 0pt]{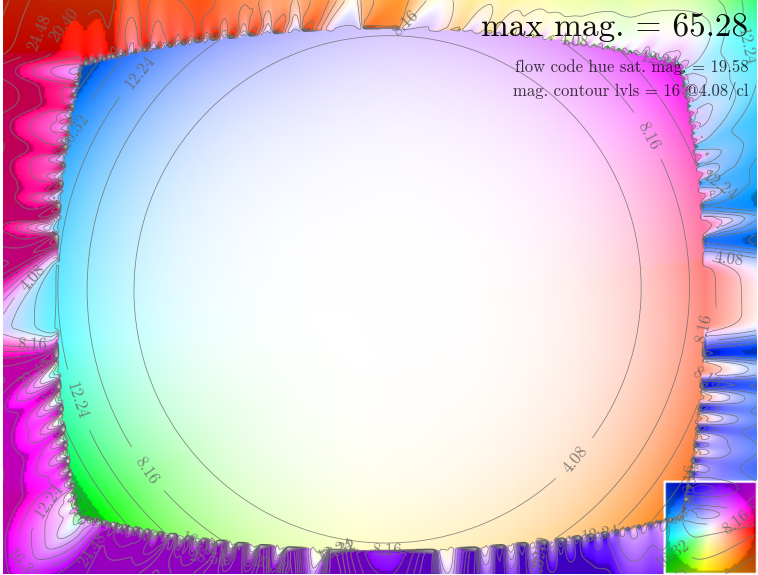} \end{tabular} &
        \begin{tabular}{@{}c@{}} \includegraphics[width=0.19\textwidth, cfbox=gray 0.1pt 0pt]{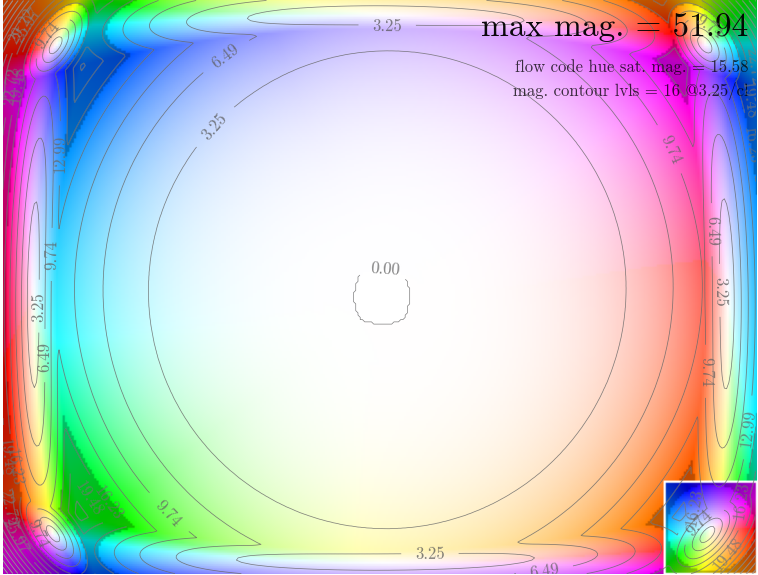} \end{tabular} &
        \begin{tabular}{@{}c@{}} \includegraphics[width=0.19\textwidth, cfbox=gray 0.1pt 0pt]{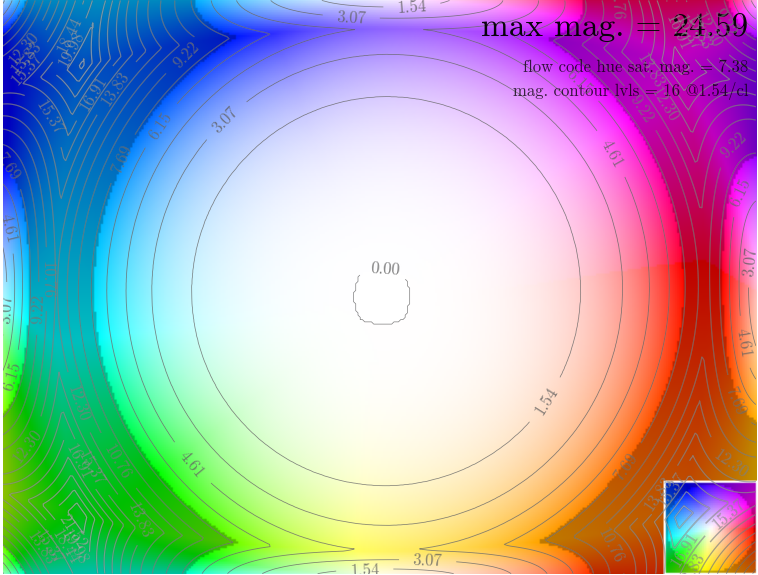} \end{tabular} &
        \rotatebox[origin=c]{-90}{\begin{adjustbox}{max width=0.19\textwidth} Not Normalized \end{adjustbox}}\\

        & &
        \begin{tabular}{@{}c@{}} \includegraphics[width=0.19\textwidth, cfbox=gray 0.1pt 0pt]{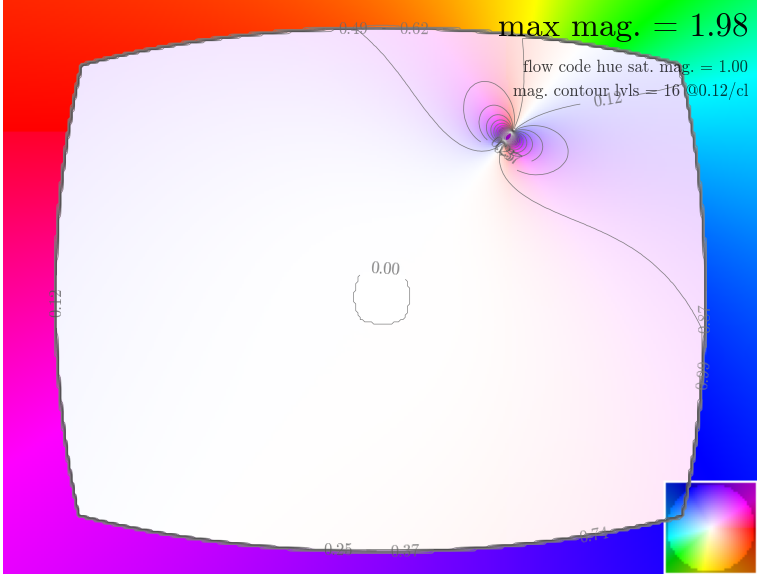} \end{tabular} &  
        \begin{tabular}{@{}c@{}} \includegraphics[width=0.19\textwidth, cfbox=gray 0.1pt 0pt]{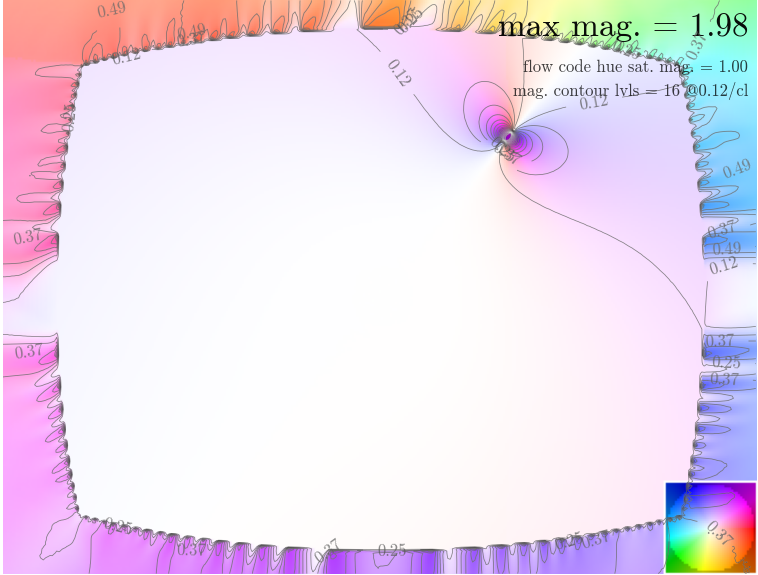} \end{tabular} &
        \begin{tabular}{@{}c@{}} \includegraphics[width=0.19\textwidth, cfbox=gray 0.1pt 0pt]{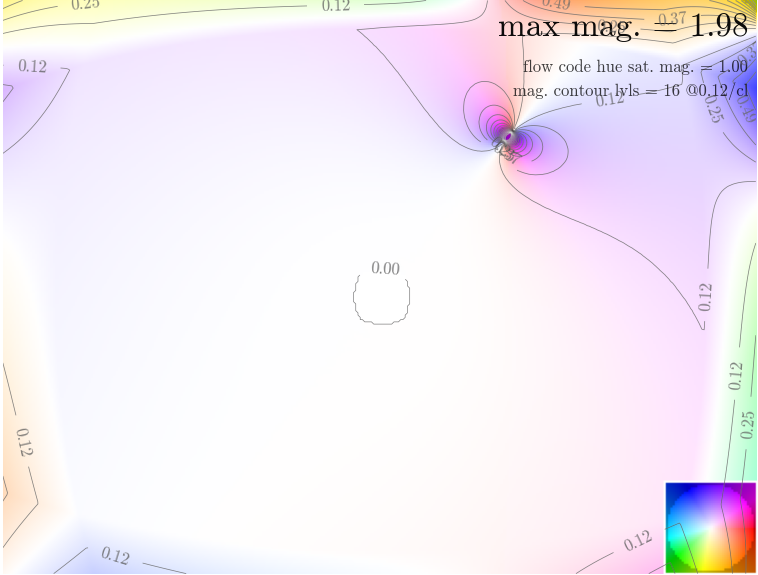} \end{tabular} &
        \begin{tabular}{@{}c@{}} \includegraphics[width=0.19\textwidth, cfbox=gray 0.1pt 0pt]{images/omap_extraction/normed/cv_rep_fill/lin_omap_diff.png} \end{tabular} &
        \rotatebox[origin=c]{-90}{\begin{adjustbox}{max width=0.19\textwidth} Normalized \end{adjustbox}}\\

        \midrule
        
        \multirow{4}{*}{\rotatebox[origin=c]{90}{\begin{adjustbox}{max width=0.6\textwidth}Angular Distorted Orientation Map Extraction \hspace{5ex} \end{adjustbox}}} & 
        \begin{tabular}{@{}c@{}} \includegraphics[width=0.19\textwidth, cfbox=gray 0.1pt 0pt]{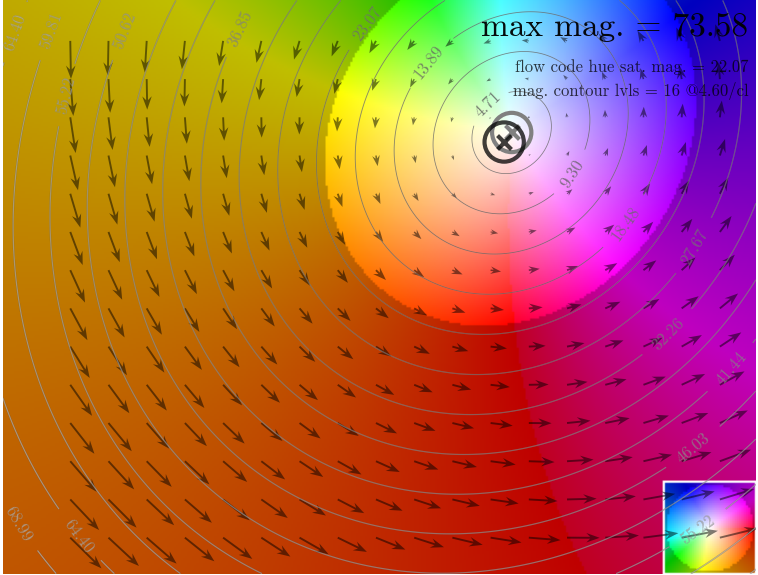} \end{tabular} &  
        \begin{tabular}{@{}c@{}} \includegraphics[width=0.19\textwidth, cfbox=gray 0.1pt 0pt]{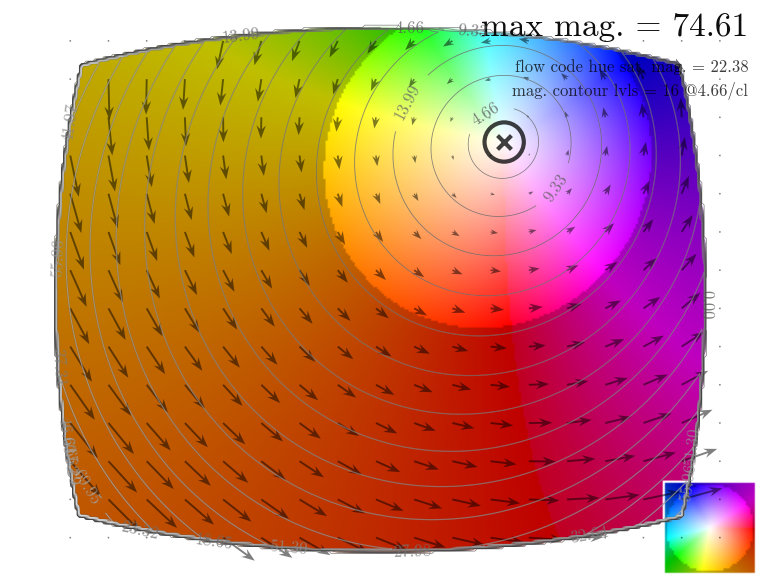} \end{tabular} &  
        \begin{tabular}{@{}c@{}} \includegraphics[width=0.19\textwidth, cfbox=gray 0.1pt 0pt]{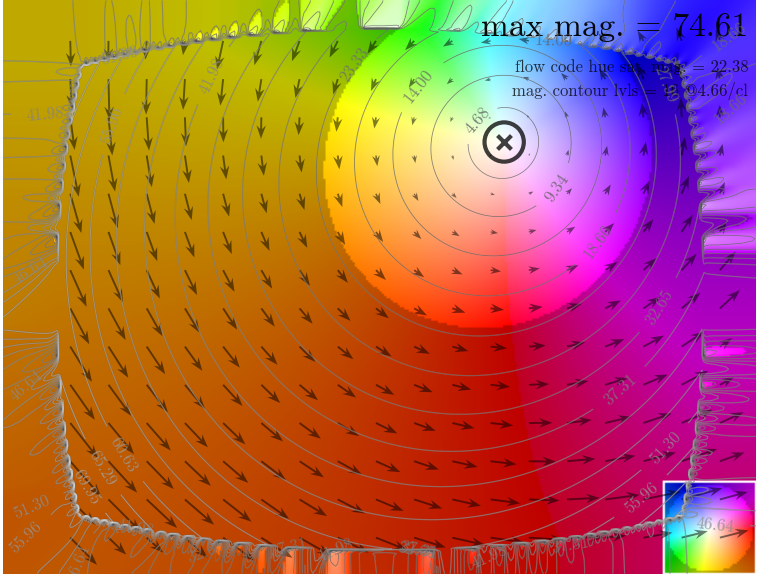} \end{tabular} &
        \begin{tabular}{@{}c@{}} \includegraphics[width=0.19\textwidth, cfbox=gray 0.1pt 0pt]{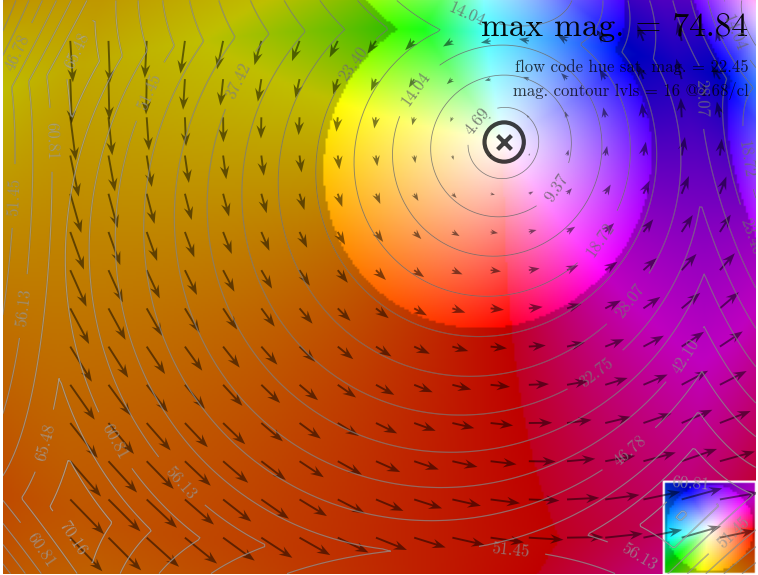} \end{tabular} &
        \begin{tabular}{@{}c@{}} \includegraphics[width=0.19\textwidth, cfbox=gray 0.1pt 0pt]{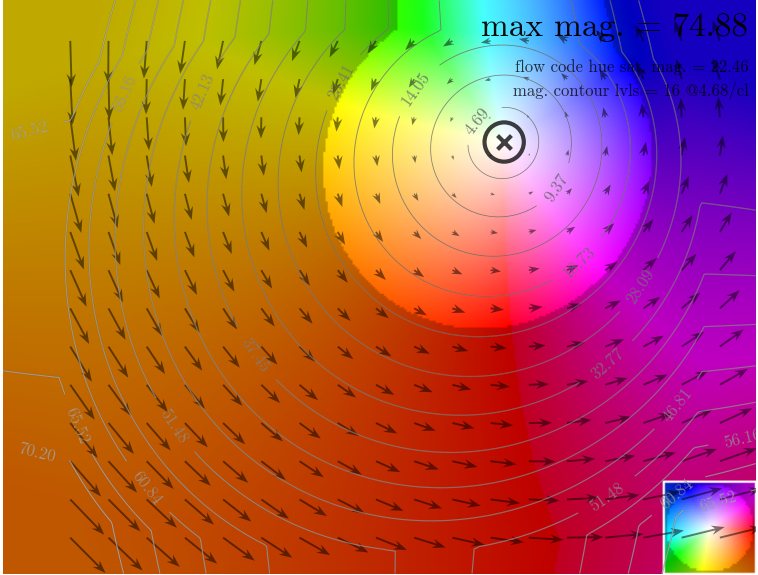} \end{tabular} &
        \rotatebox[origin=c]{-90}{\begin{adjustbox}{max width=0.19\textwidth} Not Normalized \end{adjustbox}}\\

        &
        \begin{tabular}{@{}c@{}} \includegraphics[width=0.19\textwidth, cfbox=gray 0.1pt 0pt]{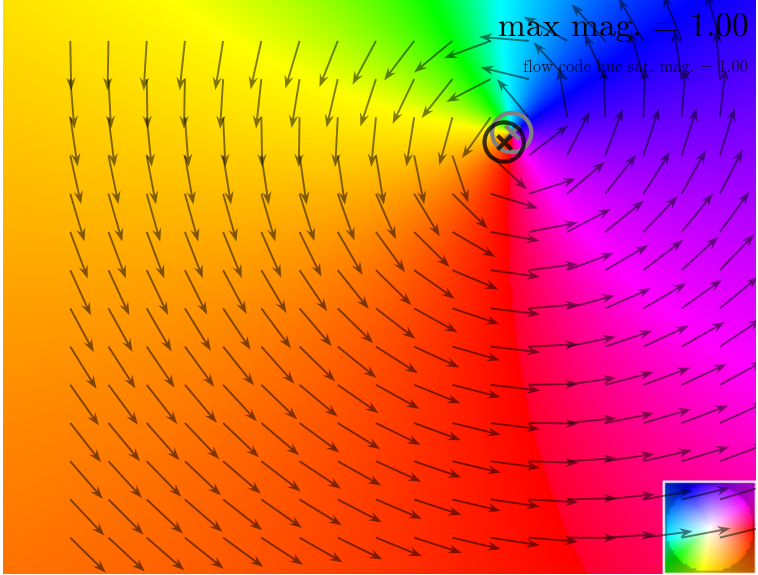} \end{tabular} &  
        \begin{tabular}{@{}c@{}} \includegraphics[width=0.19\textwidth, cfbox=gray 0.1pt 0pt]{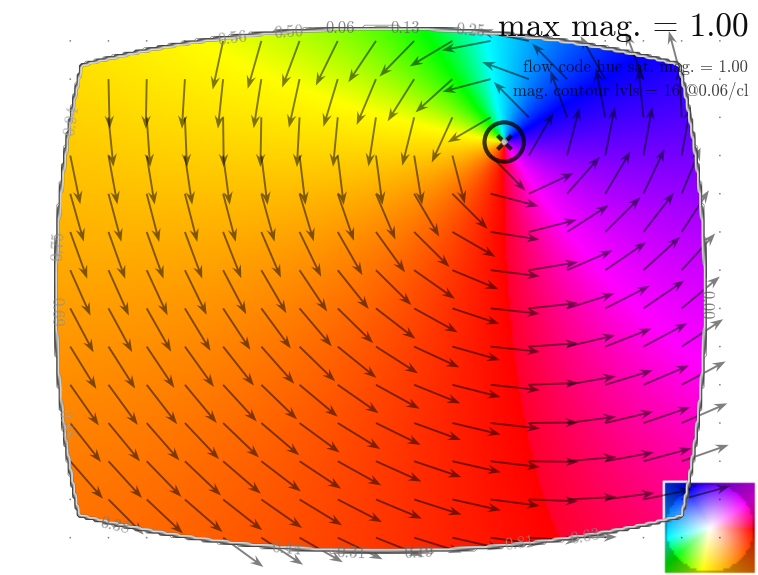} \end{tabular} &  
        \begin{tabular}{@{}c@{}} \includegraphics[width=0.19\textwidth, cfbox=gray 0.1pt 0pt]{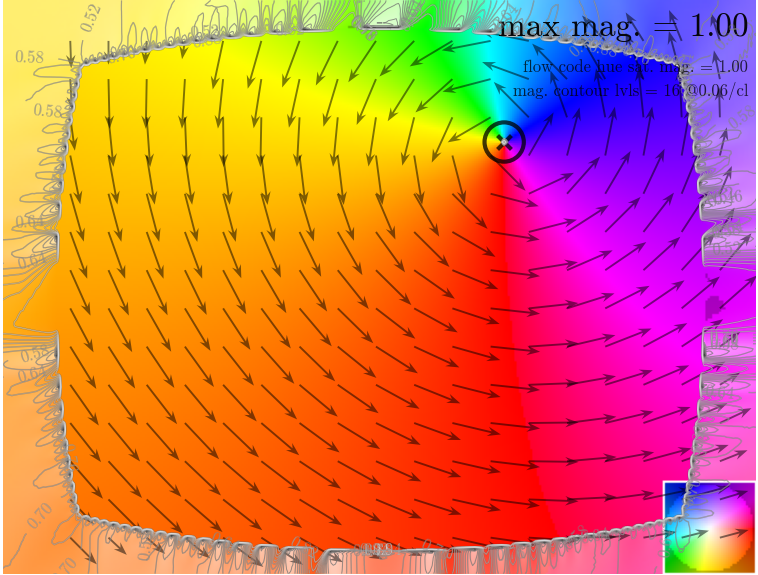} \end{tabular} &
        \begin{tabular}{@{}c@{}} \includegraphics[width=0.19\textwidth, cfbox=gray 0.1pt 0pt]{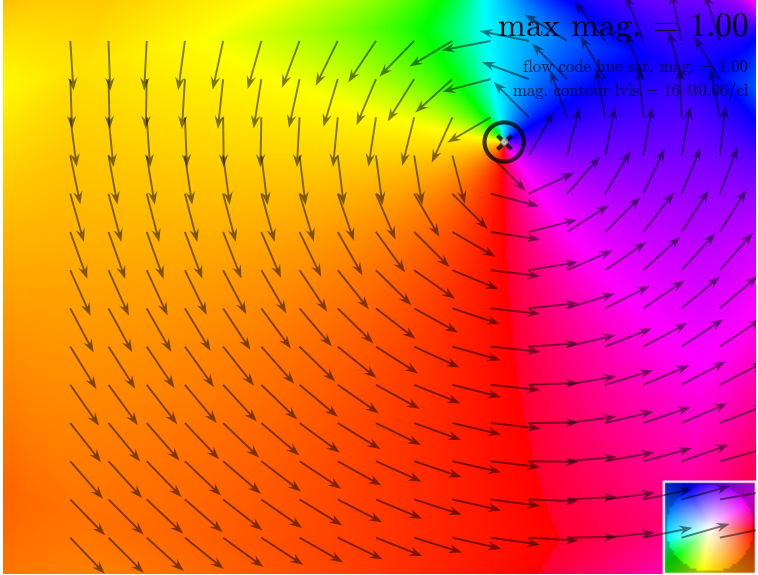} \end{tabular} &
        \begin{tabular}{@{}c@{}} \includegraphics[width=0.19\textwidth, cfbox=gray 0.1pt 0pt]{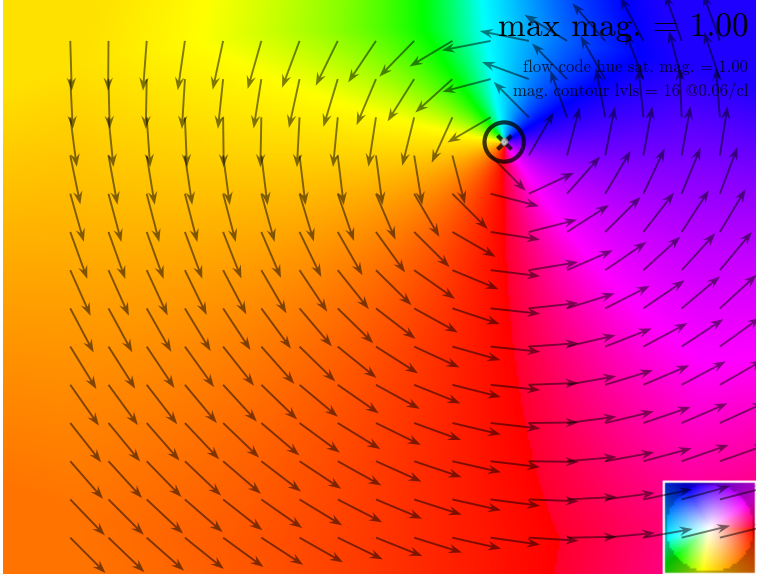} \end{tabular} & 
        \rotatebox[origin=c]{-90}{\begin{adjustbox}{max width=0.19\textwidth} Normalized \end{adjustbox}}\\
        
        &
        \multirow{2}{*}{\begin{tabular}{@{}c@{}} \includegraphics[width=0.18\textwidth, trim={0 2.5cm 0 0}, clip, cfbox=gray 0pt 0pt]{images/vels/vision_geom_ang.pdf} \end{tabular}} &  
        \begin{tabular}{@{}c@{}} \includegraphics[width=0.19\textwidth, cfbox=gray 0.1pt 0pt]{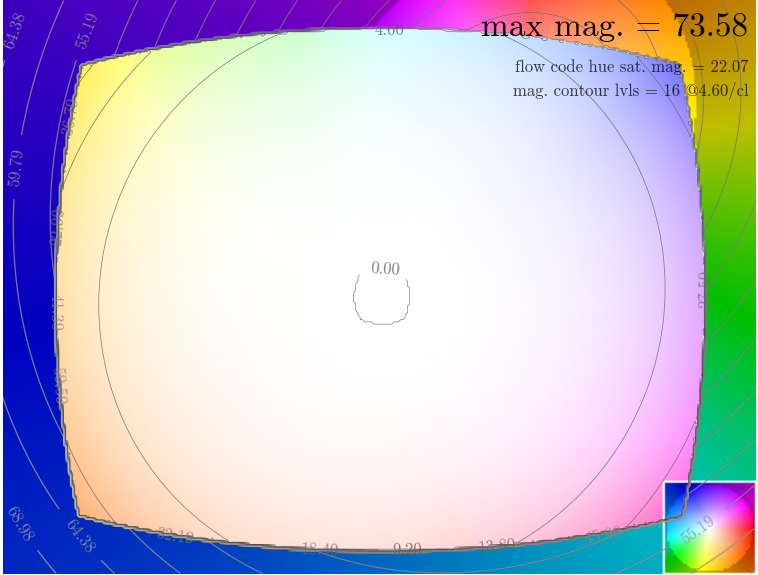} \end{tabular} &  
        \begin{tabular}{@{}c@{}} \includegraphics[width=0.19\textwidth, cfbox=gray 0.1pt 0pt]{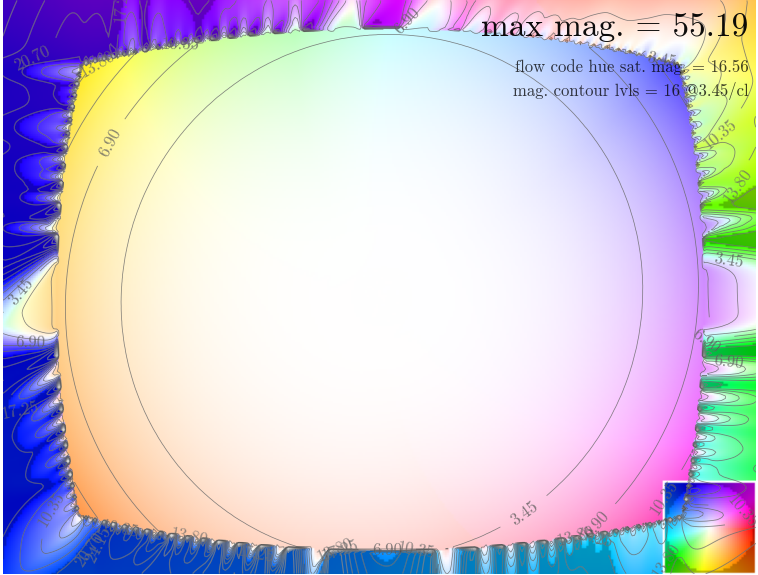} \end{tabular} &
        \begin{tabular}{@{}c@{}} \includegraphics[width=0.19\textwidth, cfbox=gray 0.1pt 0pt]{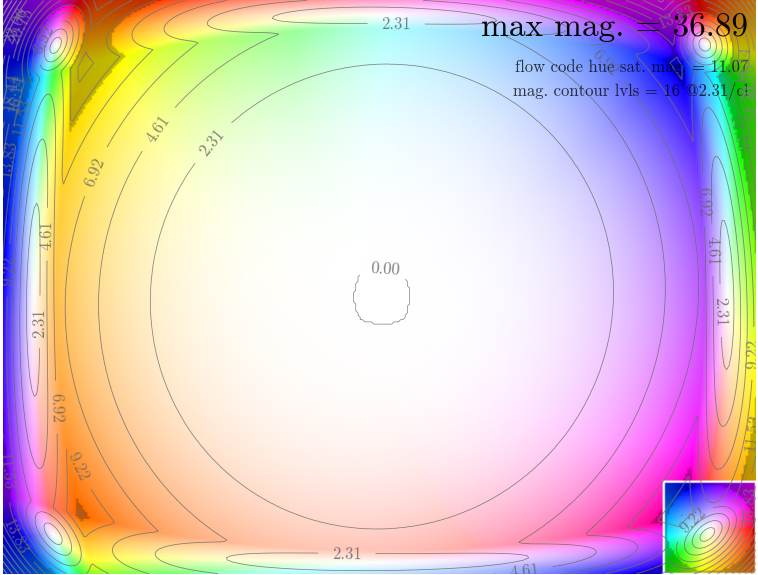} \end{tabular} &
        \begin{tabular}{@{}c@{}} \includegraphics[width=0.19\textwidth, cfbox=gray 0.1pt 0pt]{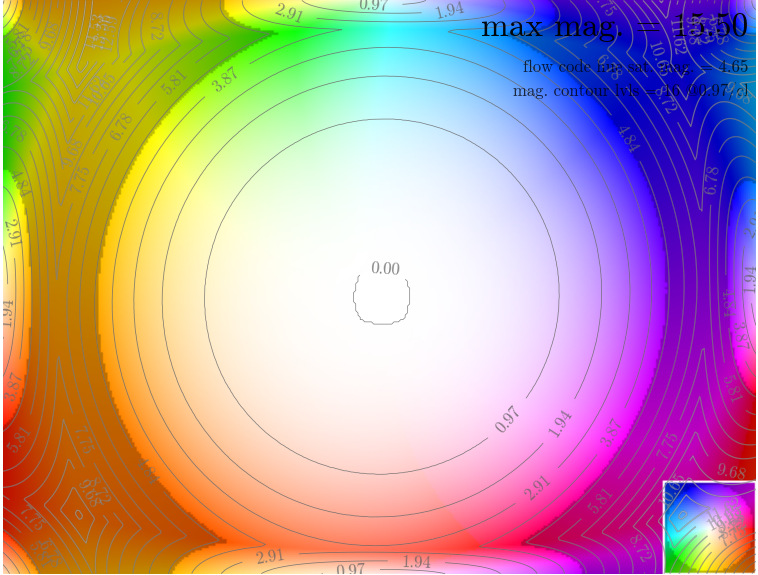} \end{tabular} &
        \rotatebox[origin=c]{-90}{\begin{adjustbox}{max width=0.19\textwidth} Not Normalized \end{adjustbox}}\\

        & &
        \begin{tabular}{@{}c@{}} \includegraphics[width=0.19\textwidth, cfbox=gray 0.1pt 0pt]{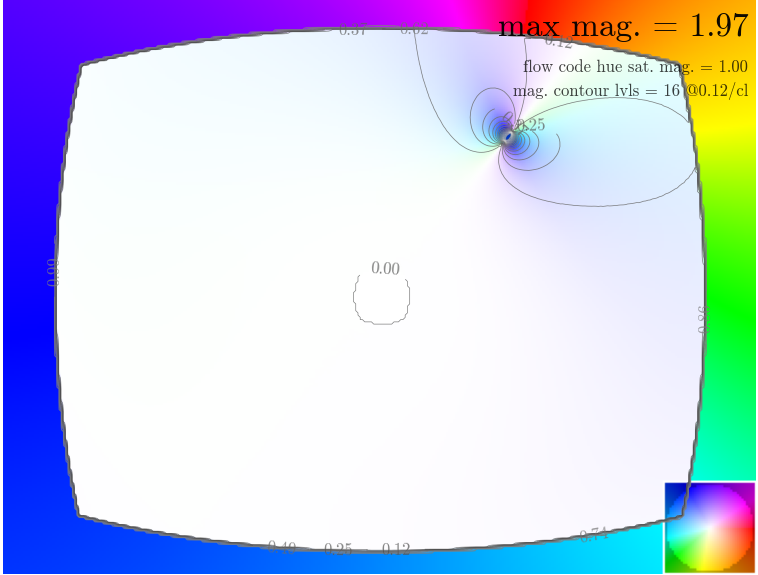} \end{tabular} &  
        \begin{tabular}{@{}c@{}} \includegraphics[width=0.19\textwidth, cfbox=gray 0.1pt 0pt]{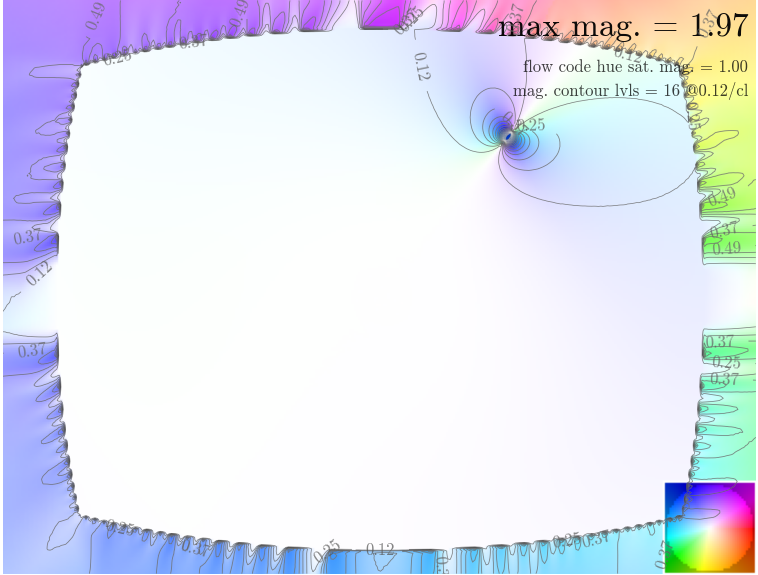} \end{tabular} &
        \begin{tabular}{@{}c@{}} \includegraphics[width=0.19\textwidth, cfbox=gray 0.1pt 0pt]{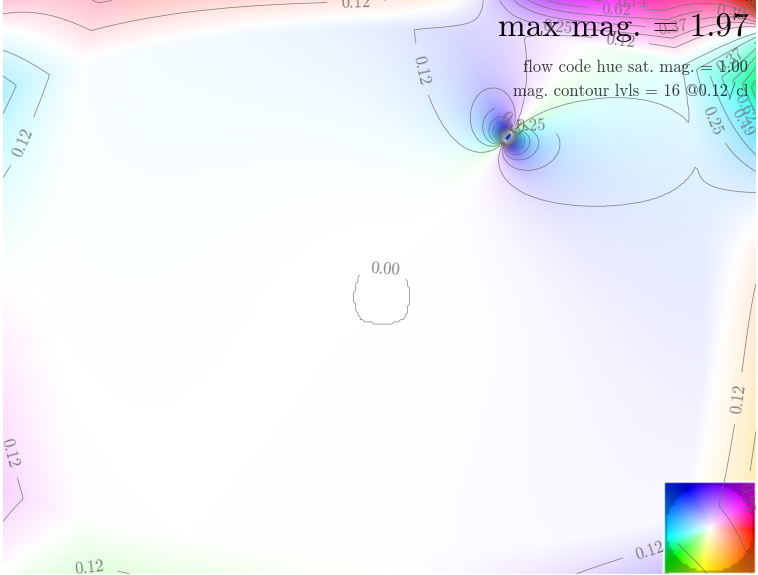} \end{tabular} &
        \begin{tabular}{@{}c@{}} \includegraphics[width=0.19\textwidth, cfbox=gray 0.1pt 0pt]{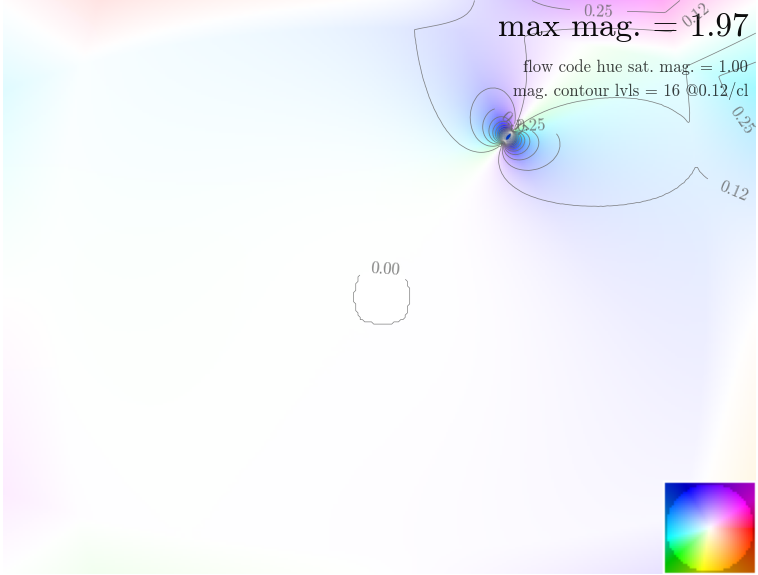} \end{tabular} &
        \rotatebox[origin=c]{-90}{\begin{adjustbox}{max width=0.19\textwidth} Normalized \end{adjustbox}}\\
    \end{tabular}
    \end{adjustbox}
    \caption{A layout of the orientation map extraction process for both linear
    (top section) and angular (bottom section) velocities. In each section (top
    and bottom), the top two rows show the orientation maps while the bottom
    two rows depict comparisons of the distorted orientation maps to the
    undistorted orientation maps (\ie,
    $\fourIdx{\text{dist}}{}{}{}{\mathcal{O}_{\cdot}} - \mathcal{O}_{\cdot}$).
    In both sections, rows 1 and 3 show the maps and comparisons, respectively,
    for the not-normalized case. Similarly, rows 2 and 4 show the maps and
    comparisons for the normalized case, respectively. Column 1 depicts the
    undistorted version, while columns 2-5 display the distorted versions.
    Column 2 demonstrates how the application of distortion creates empty
    regions in the map near the boundaries. In columns 3, 4, and 5, we
    highlight a variety of fill methods, namely Navier-Stokes, border reflect,
    and border replicate, respectively. In our evalutions, we utilize the
    border reflect method.}
    \label{fig:o_maps_extraction_supp}
\end{figure*}

Orientation maps are derived from the estimated 3D camera velocities and encode
pixel-wise directional information that serves as priors in the CM framework.
We project both the linear and angular velocity vector onto the image plane
using the intrinsic calibration of the camera. This yields the singularity
points with zero flow, which are then used to obtain the not-normalized
orientation maps as shown in rows 1 and 5 of \cref{fig:o_maps_extraction_supp}.
Subsequently, lens distortion is applied to obtain the distorted orientation
maps, which create empty regions around the boundaries. While filling can be
performed using various OpenCV libraries as illustrated in columns 3 and 4 of
\cref{fig:o_maps_extraction_supp}, we found the \texttt{BORDER\_REPLICATE}
strategy to work the most robustly for multiple data samples with varying
camera velocities. 

\section{Velocity Estimation from DSEC LiDAR Data}
\label{sec:velocity_esitmaion_from_dsec_lidar_data}
\begin{figure}[htb]
    \centering
    \adjincludegraphics[width=0.24\linewidth,trim={{0.02\width} {0.1\height} {0.18\width} {0.2\height}}, clip, cfbox=SILVER 0.1pt 0pt]{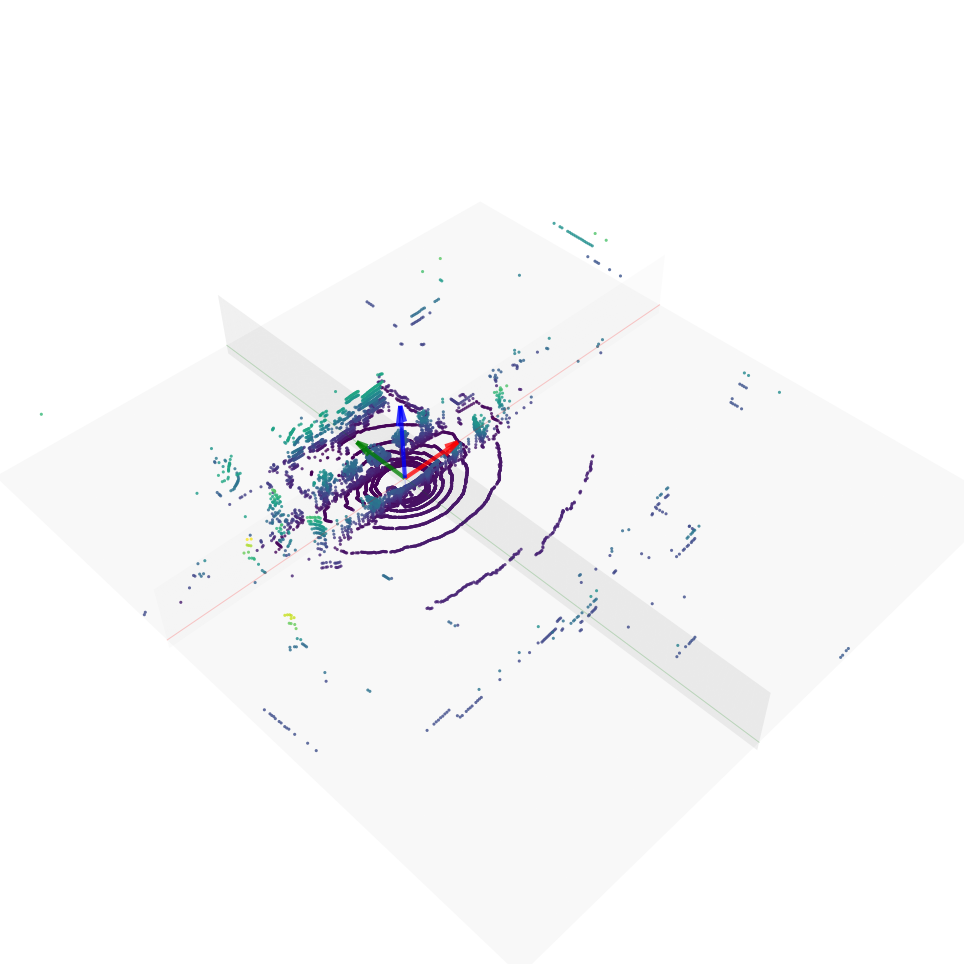}\hspace{0.1ex}%
    \adjincludegraphics[width=0.24\linewidth,trim={{0.02\width} {0.1\height} {0.18\width} {0.2\height}}, clip, cfbox=SILVER 0.1pt 0pt]{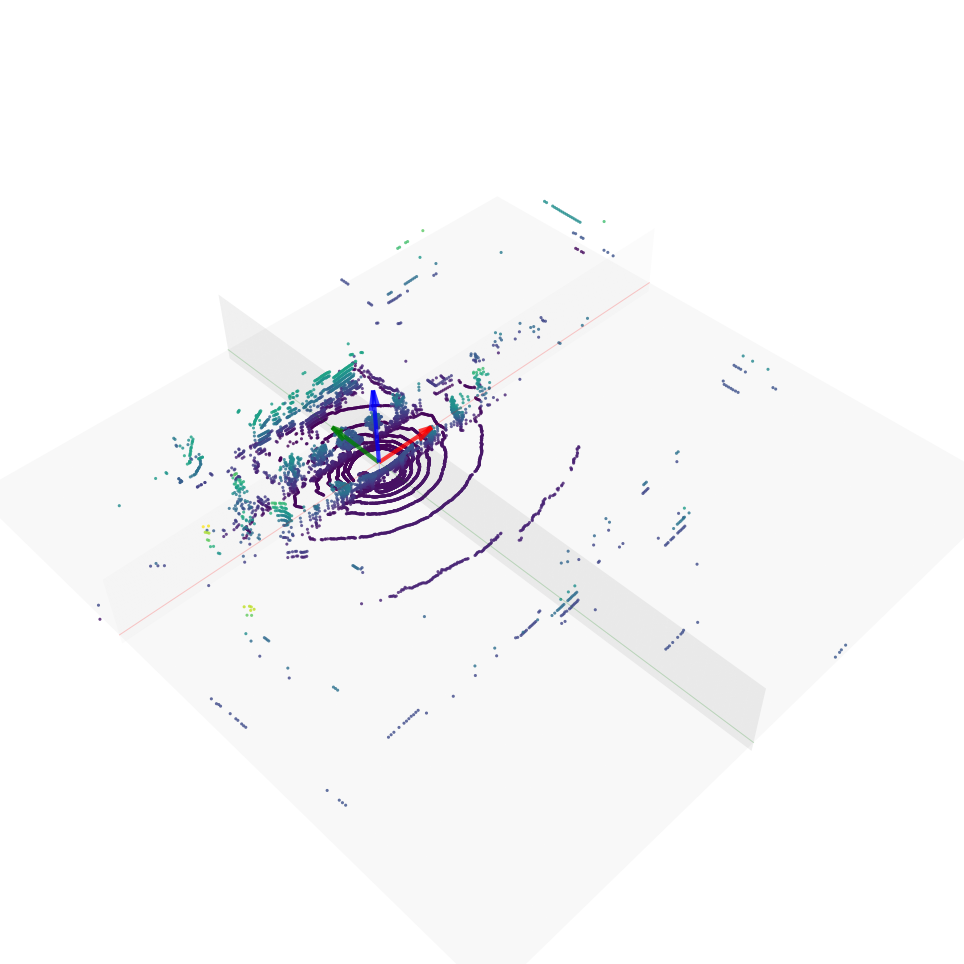}\hspace{0.1ex}%
    \adjincludegraphics[width=0.24\linewidth,trim={{0.02\width} {0.1\height} {0.18\width} {0.2\height}}, clip, cfbox=SILVER 0.1pt 0pt]{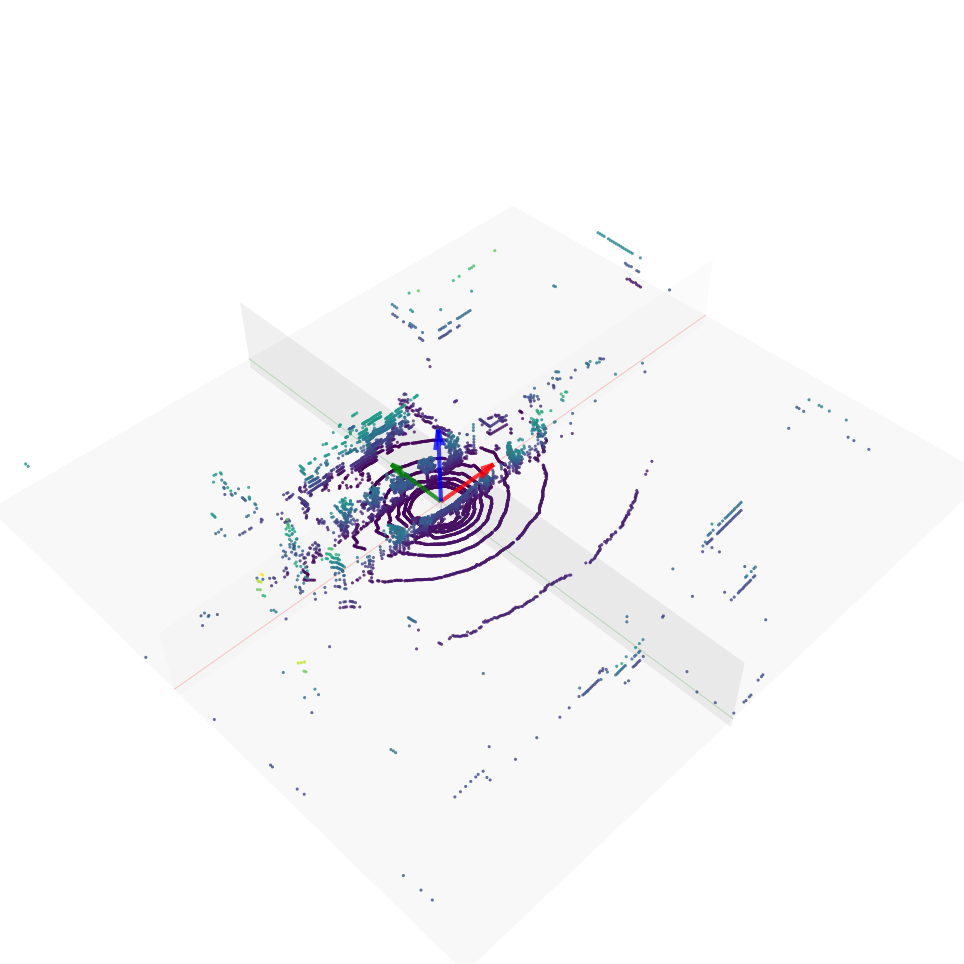}\hspace{0.1ex}%
    \adjincludegraphics[width=0.24\linewidth,trim={{0.02\width} {0.1\height} {0.18\width} {0.2\height}}, clip, cfbox=SILVER 0.1pt 0pt]{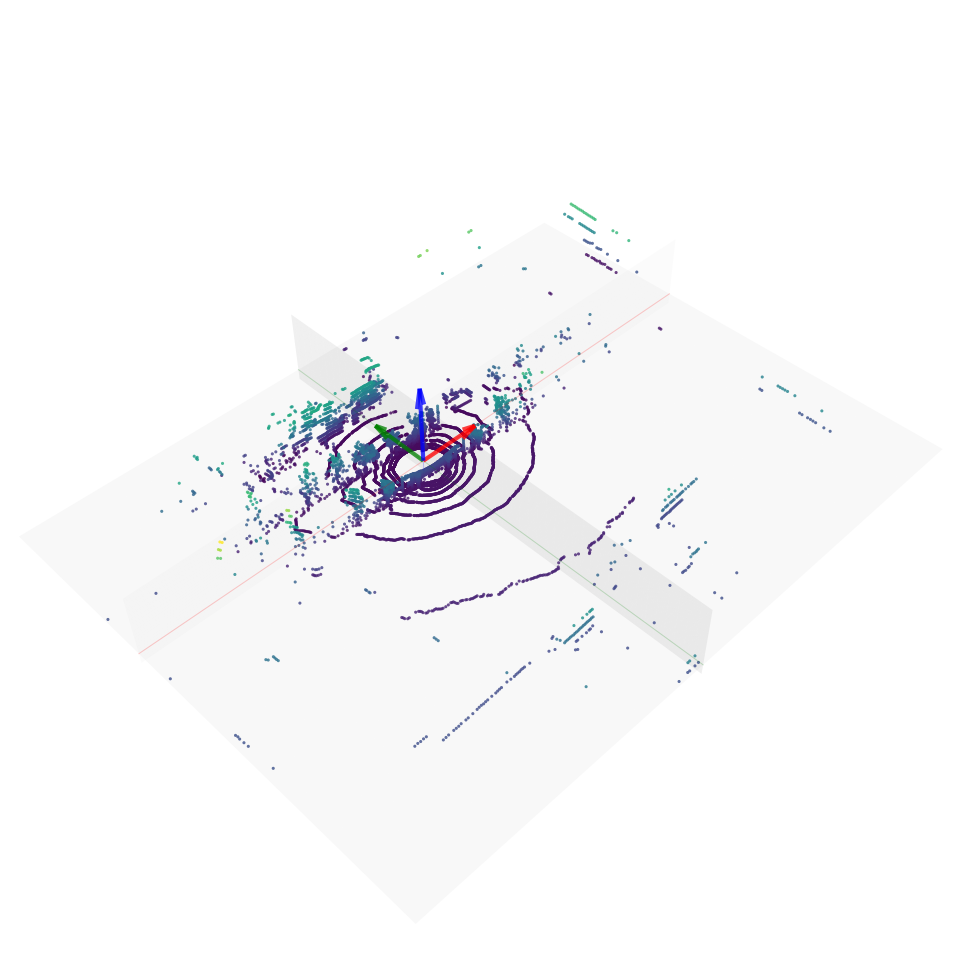}\\%
    
    \adjincludegraphics[width=0.24\linewidth,trim={{0.02\width} {0.1\height} {0.18\width} {0.2\height}}, clip, cfbox=SILVER 0.1pt 0pt]{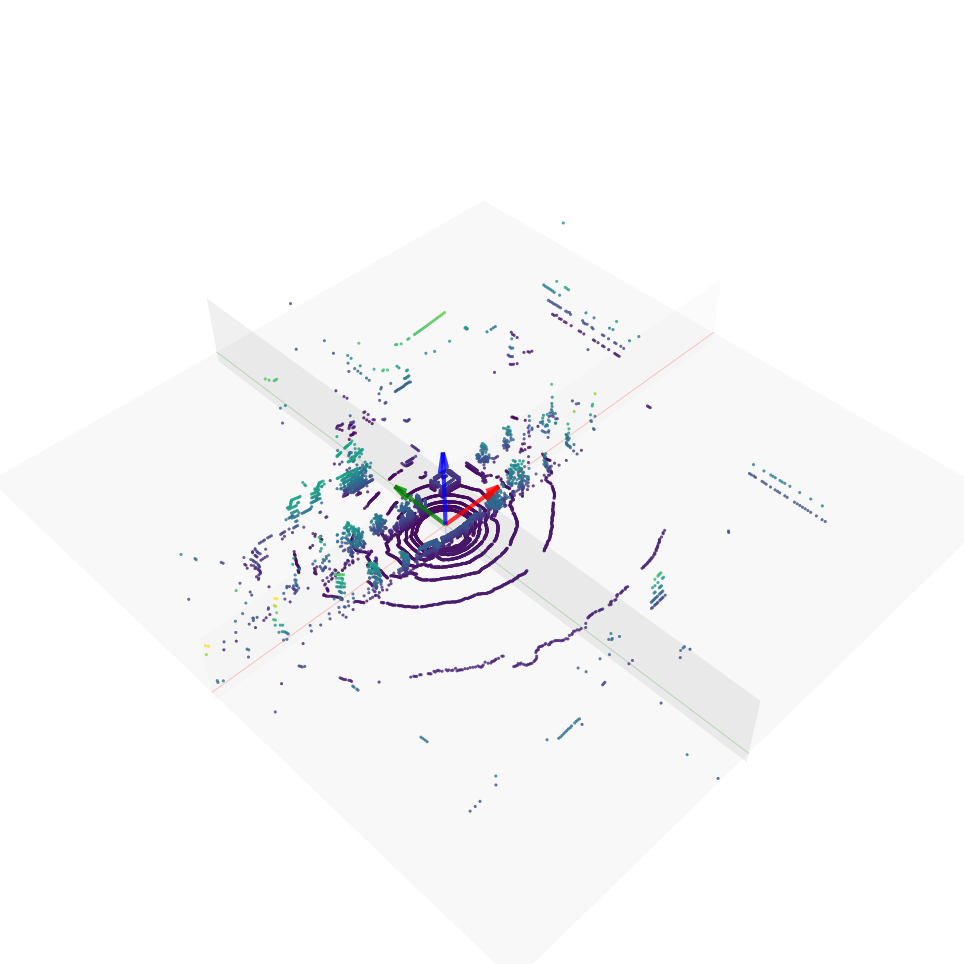}\hspace{0.1ex}%
    \adjincludegraphics[width=0.24\linewidth,trim={{0.02\width} {0.1\height} {0.18\width} {0.2\height}}, clip, cfbox=SILVER 0.1pt 0pt]{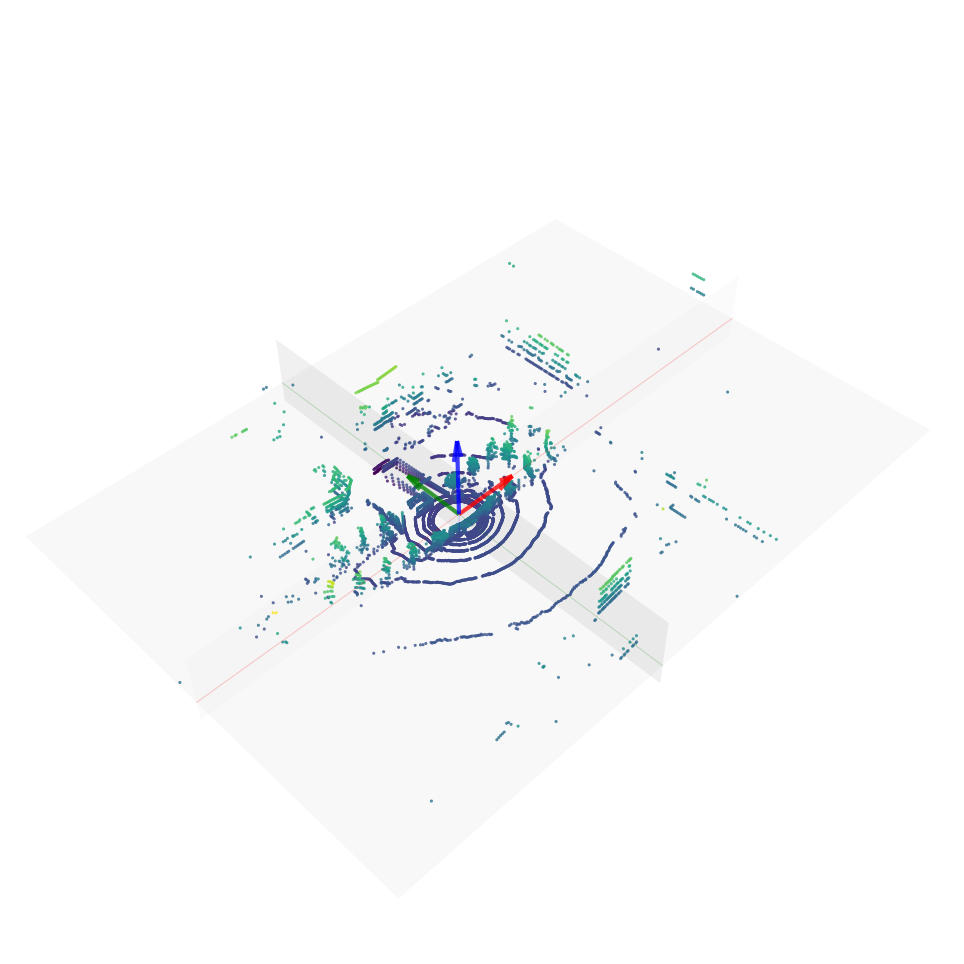}\hspace{0.1ex}%
    \adjincludegraphics[width=0.24\linewidth,trim={{0.02\width} {0.1\height} {0.18\width} {0.2\height}}, clip, cfbox=SILVER 0.1pt 0pt]{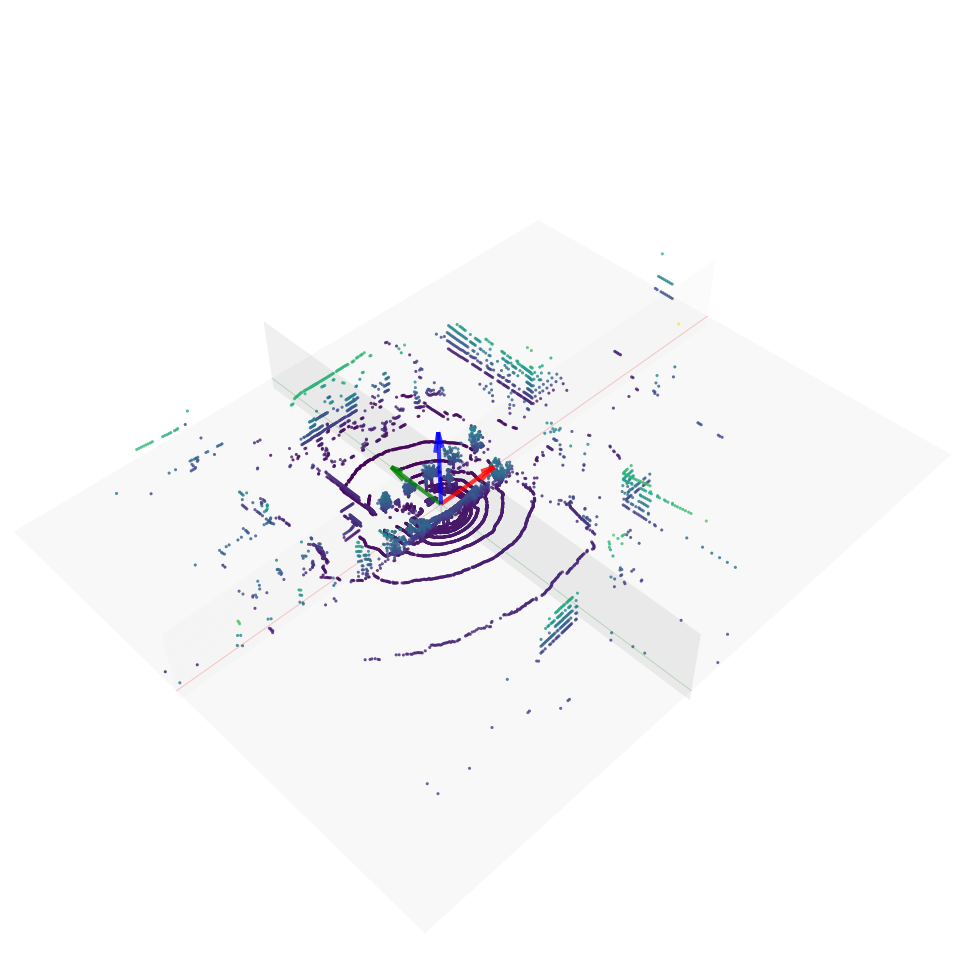}\hspace{0.1ex}%
    \adjincludegraphics[width=0.24\linewidth,trim={{0.02\width} {0.1\height} {0.18\width} {0.2\height}}, clip, cfbox=SILVER 0.1pt 0pt]{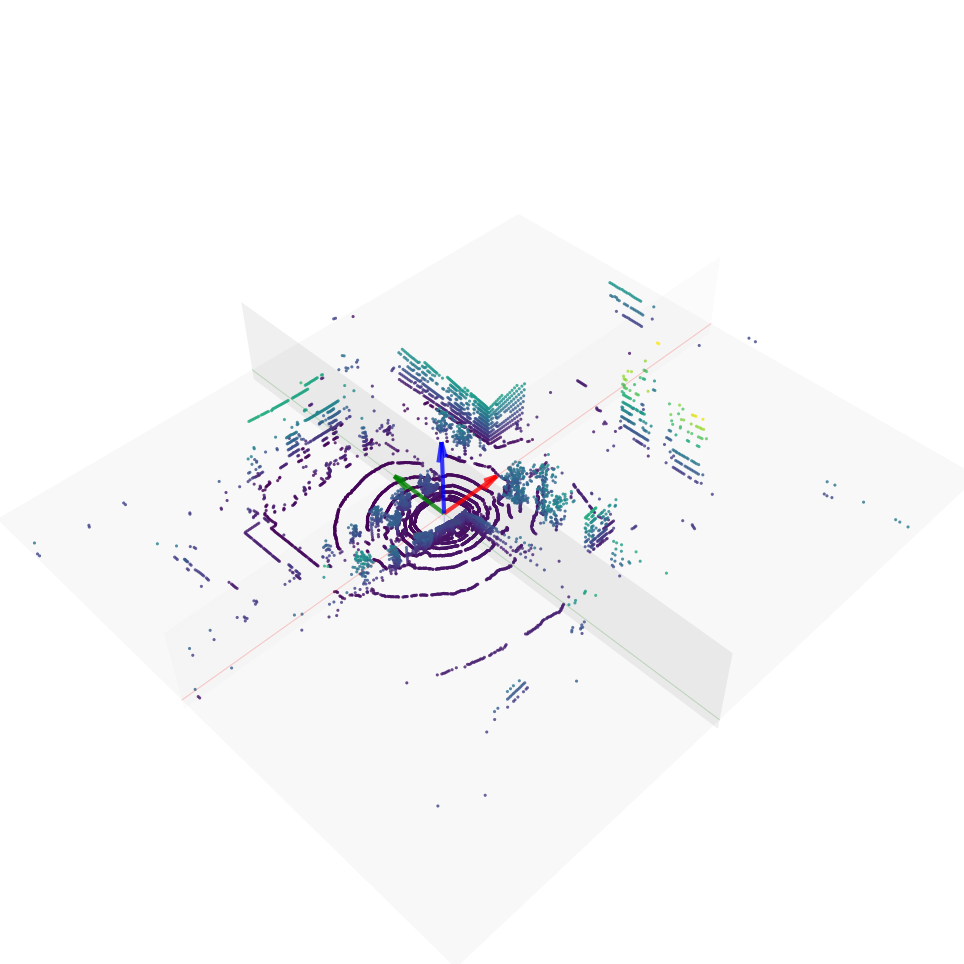}\\%
    
    \adjincludegraphics[width=0.24\linewidth,trim={{0.02\width} {0.1\height} {0.18\width} {0.2\height}}, clip, cfbox=SILVER 0.1pt 0pt]{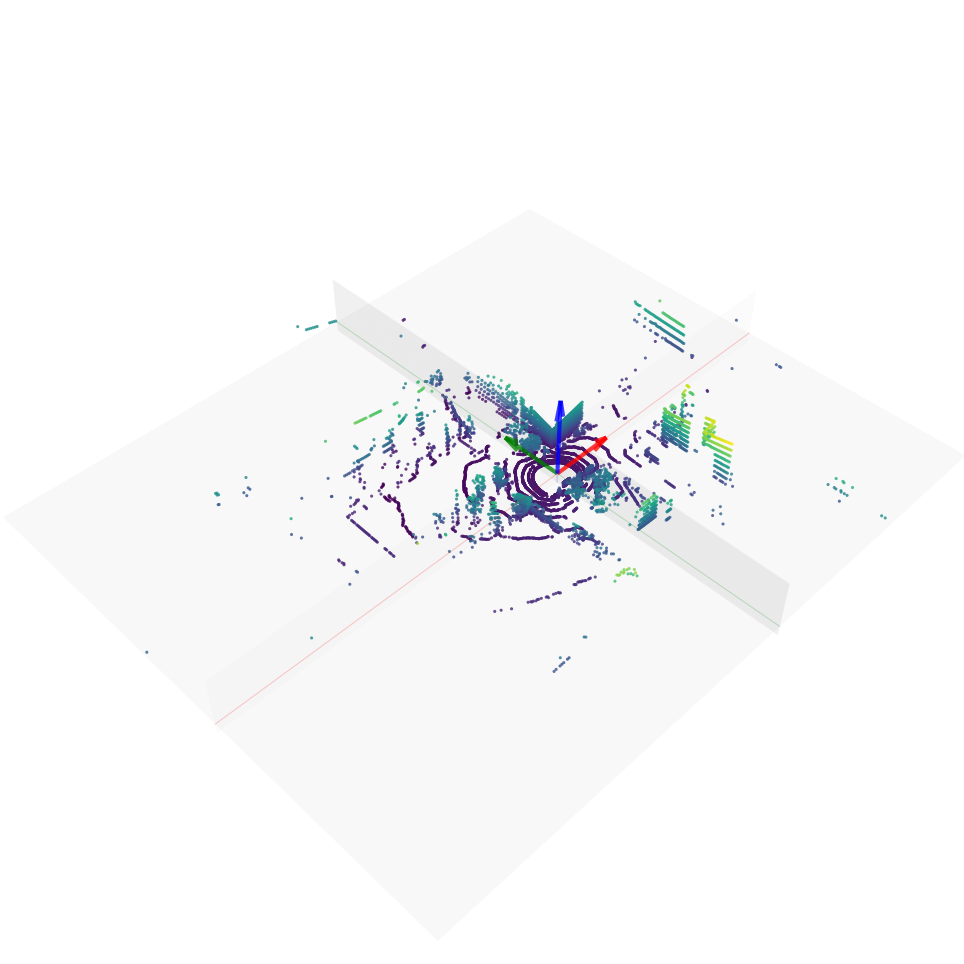}\hspace{0.1ex}%
    \adjincludegraphics[width=0.24\linewidth,trim={{0.02\width} {0.1\height} {0.18\width} {0.2\height}}, clip, cfbox=SILVER 0.1pt 0pt]{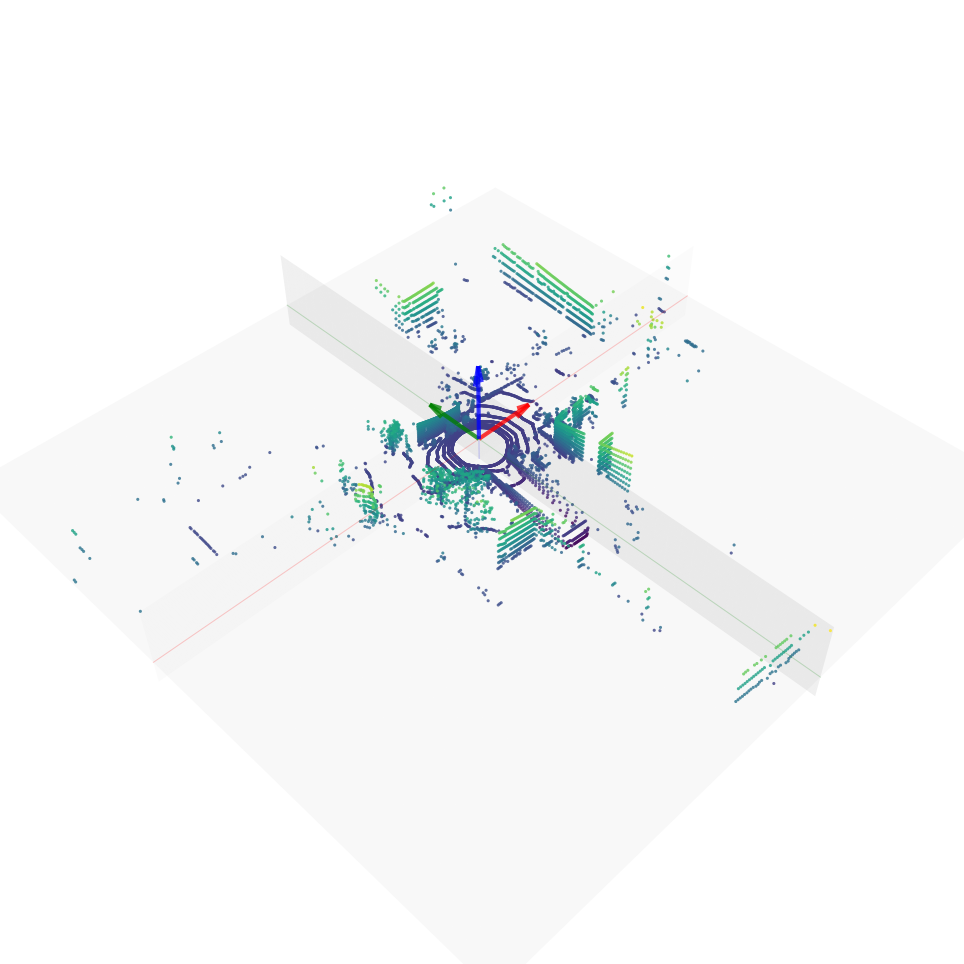}\hspace{0.1ex}%
    \adjincludegraphics[width=0.24\linewidth,trim={{0.02\width} {0.1\height} {0.18\width} {0.2\height}}, clip, cfbox=SILVER 0.1pt 0pt]{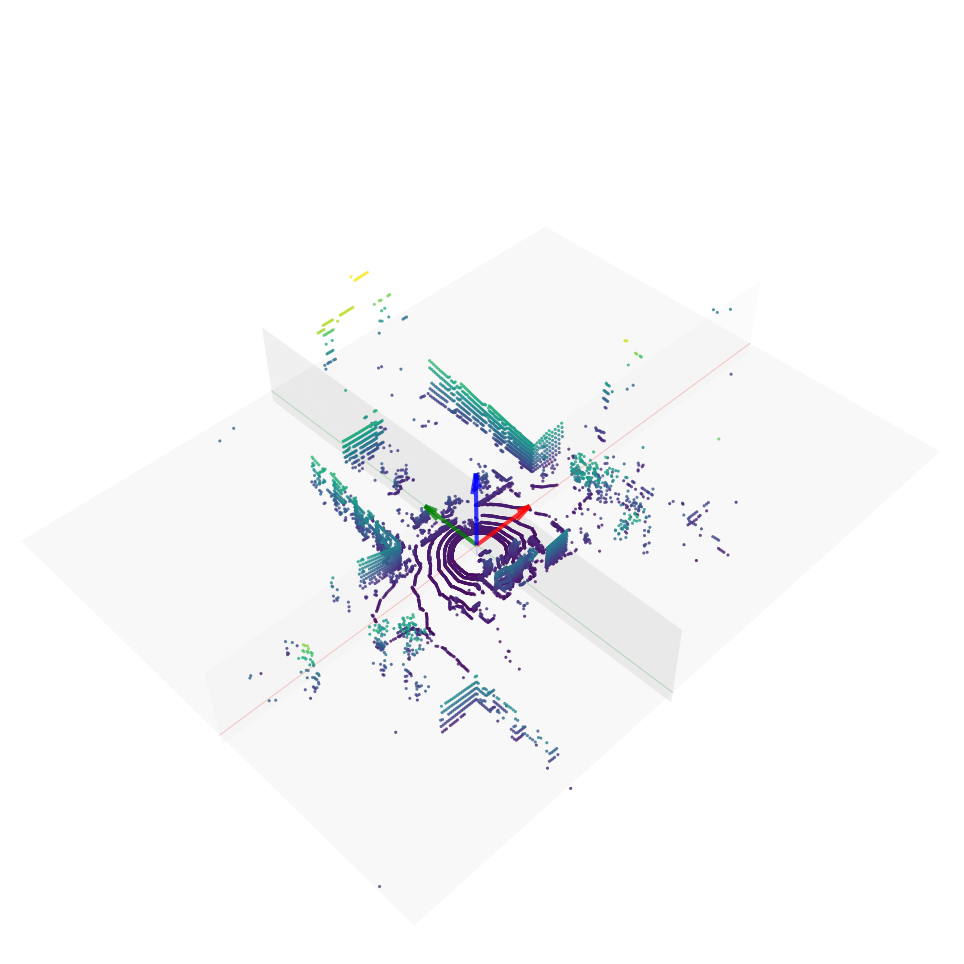}\hspace{0.1ex}%
    \adjincludegraphics[width=0.24\linewidth,trim={{0.02\width} {0.1\height} {0.18\width} {0.2\height}}, clip, cfbox=SILVER 0.1pt 0pt]{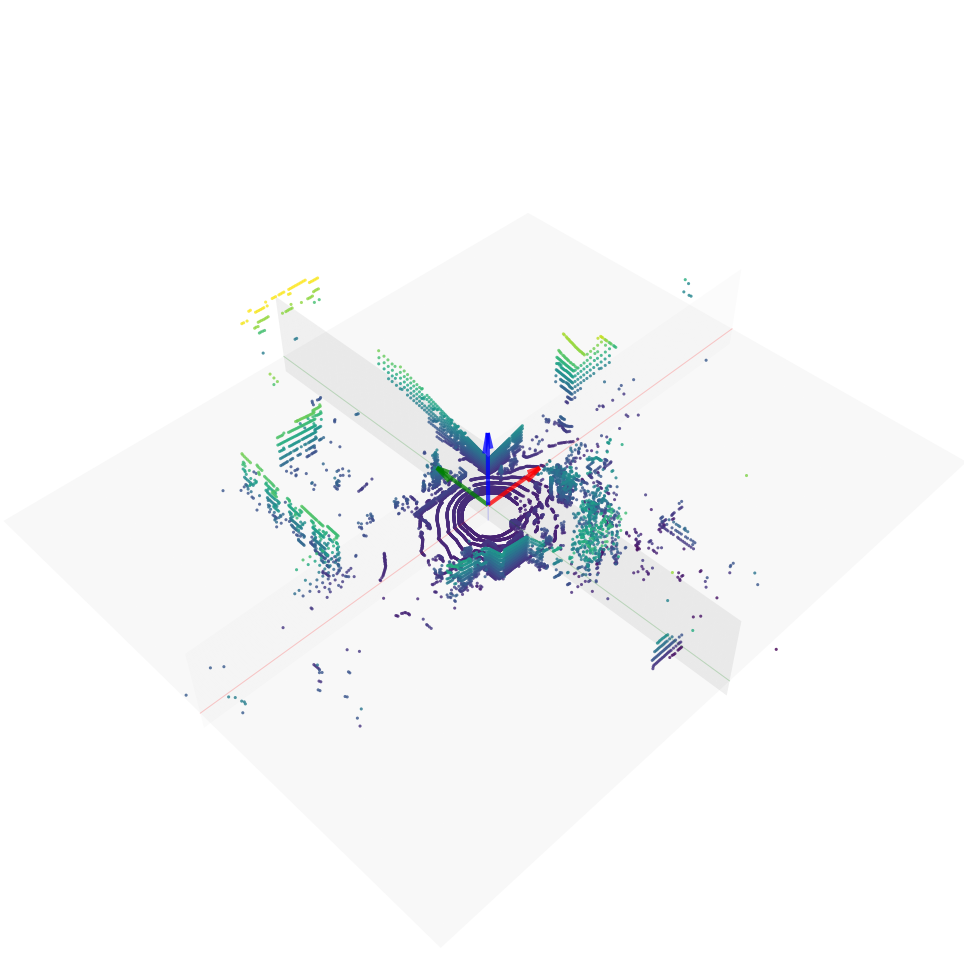}\\%
    
    \adjincludegraphics[width=0.24\linewidth,trim={{0.02\width} {0.1\height} {0.18\width} {0.2\height}}, clip, cfbox=SILVER 0.1pt 0pt]{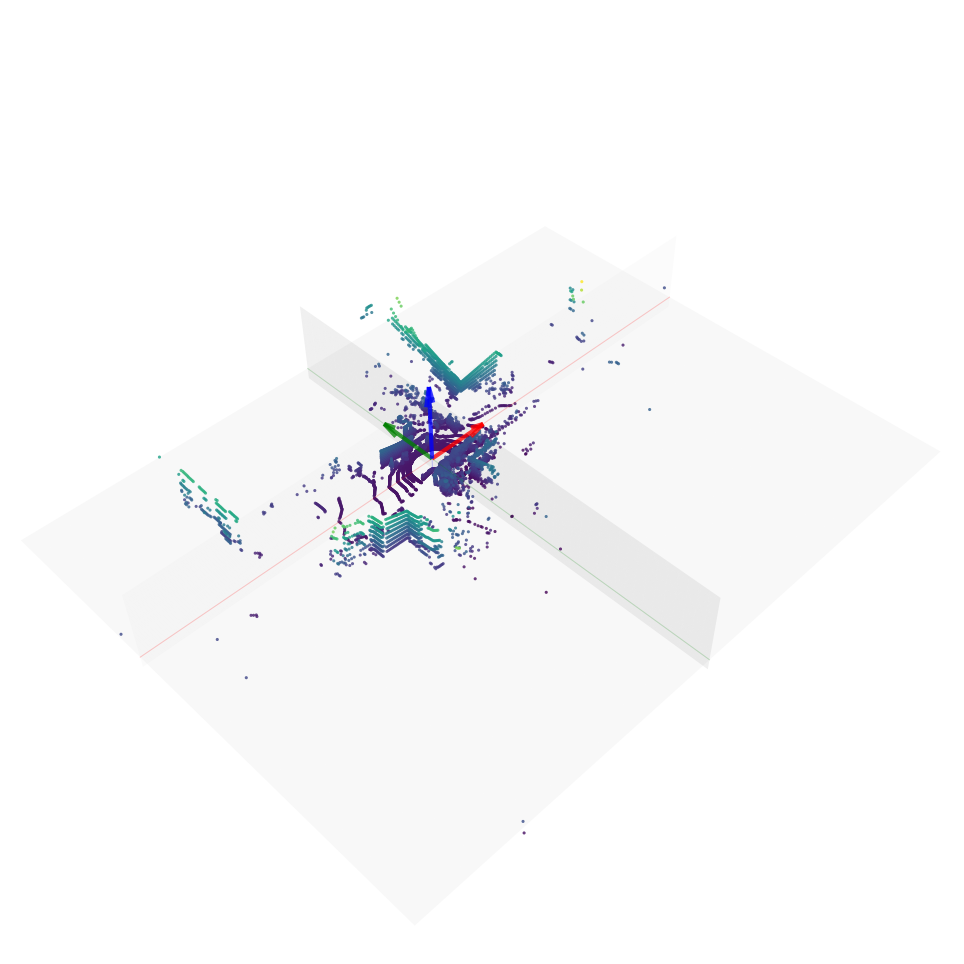}\hspace{0.1ex}%
    \adjincludegraphics[width=0.24\linewidth,trim={{0.02\width} {0.1\height} {0.18\width} {0.2\height}}, clip, cfbox=SILVER 0.1pt 0pt]{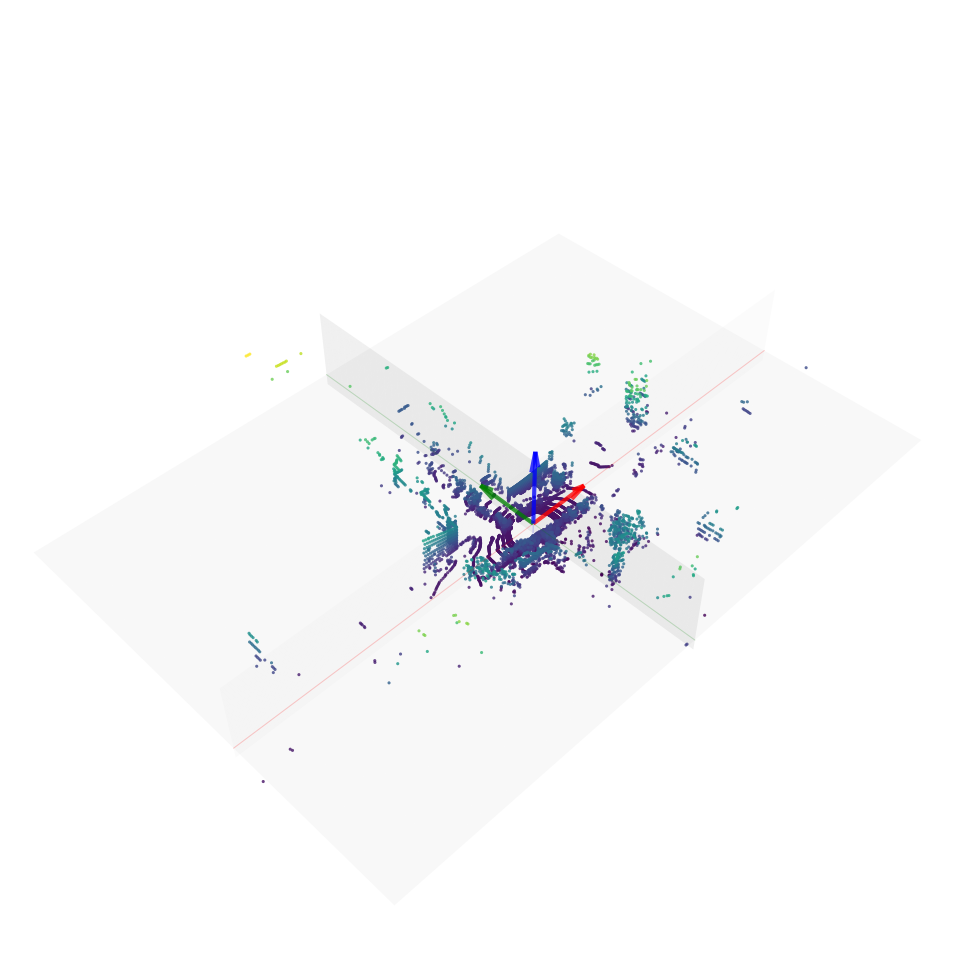}\hspace{0.1ex}%
    \adjincludegraphics[width=0.24\linewidth,trim={{0.02\width} {0.1\height} {0.18\width} {0.2\height}}, clip, cfbox=SILVER 0.1pt 0pt]{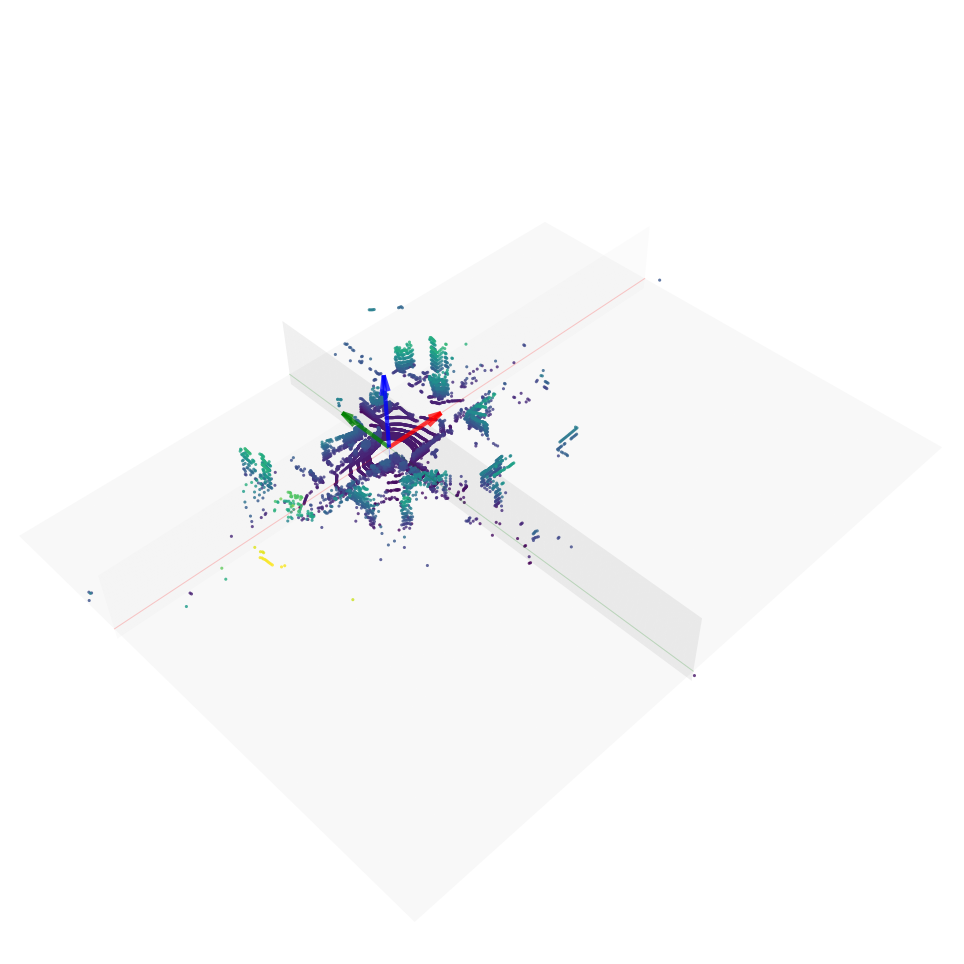}\hspace{0.1ex}%
    \adjincludegraphics[width=0.24\linewidth,trim={{0.02\width} {0.1\height} {0.18\width} {0.2\height}}, clip, cfbox=SILVER 0.1pt 0pt]{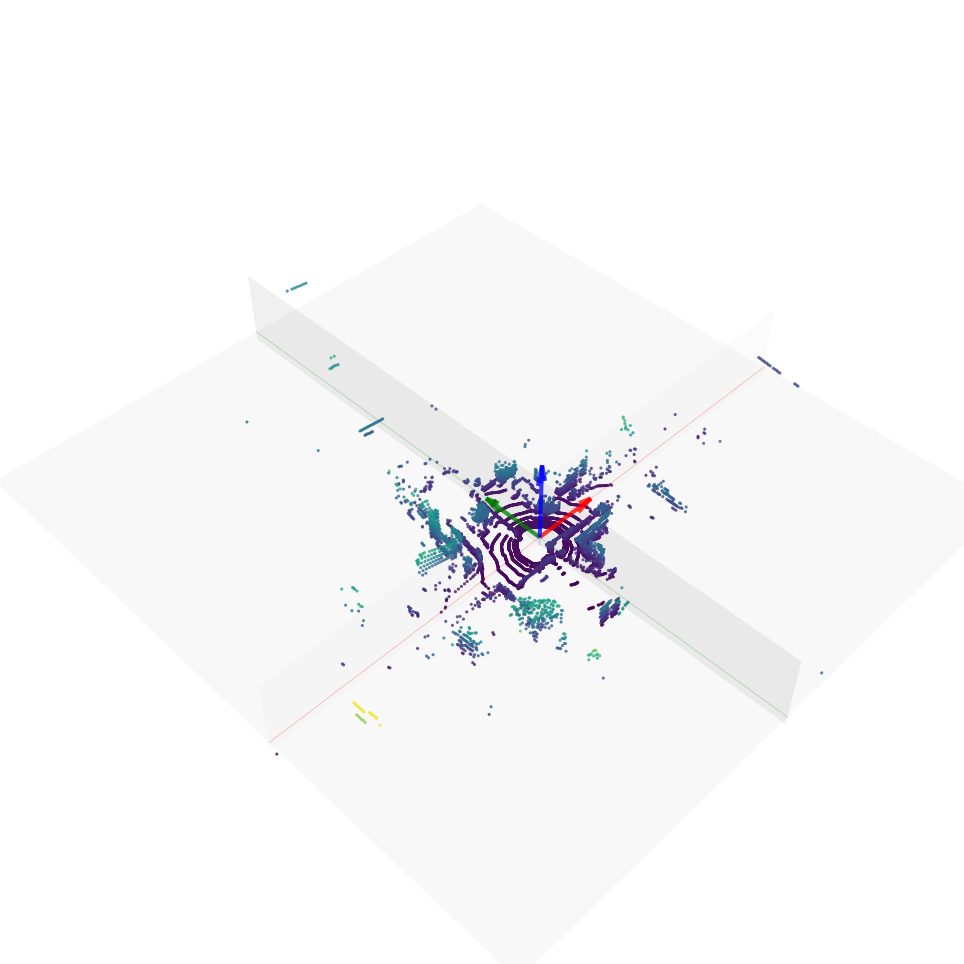}\\%
    
    \adjincludegraphics[width=0.24\linewidth,trim={{0.02\width} {0.1\height} {0.18\width} {0.2\height}}, clip, cfbox=SILVER 0.1pt 0pt]{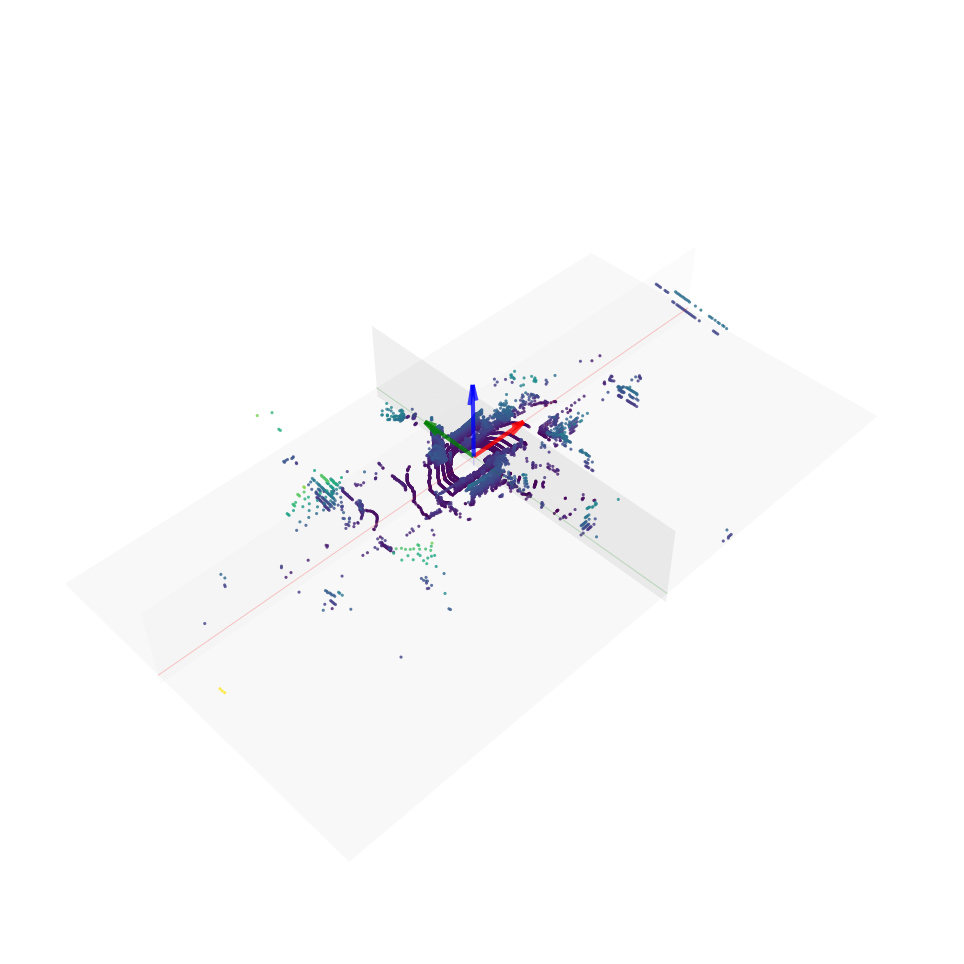}\hspace{0.1ex}%
    \adjincludegraphics[width=0.24\linewidth,trim={{0.02\width} {0.1\height} {0.18\width} {0.2\height}}, clip, cfbox=SILVER 0.1pt 0pt]{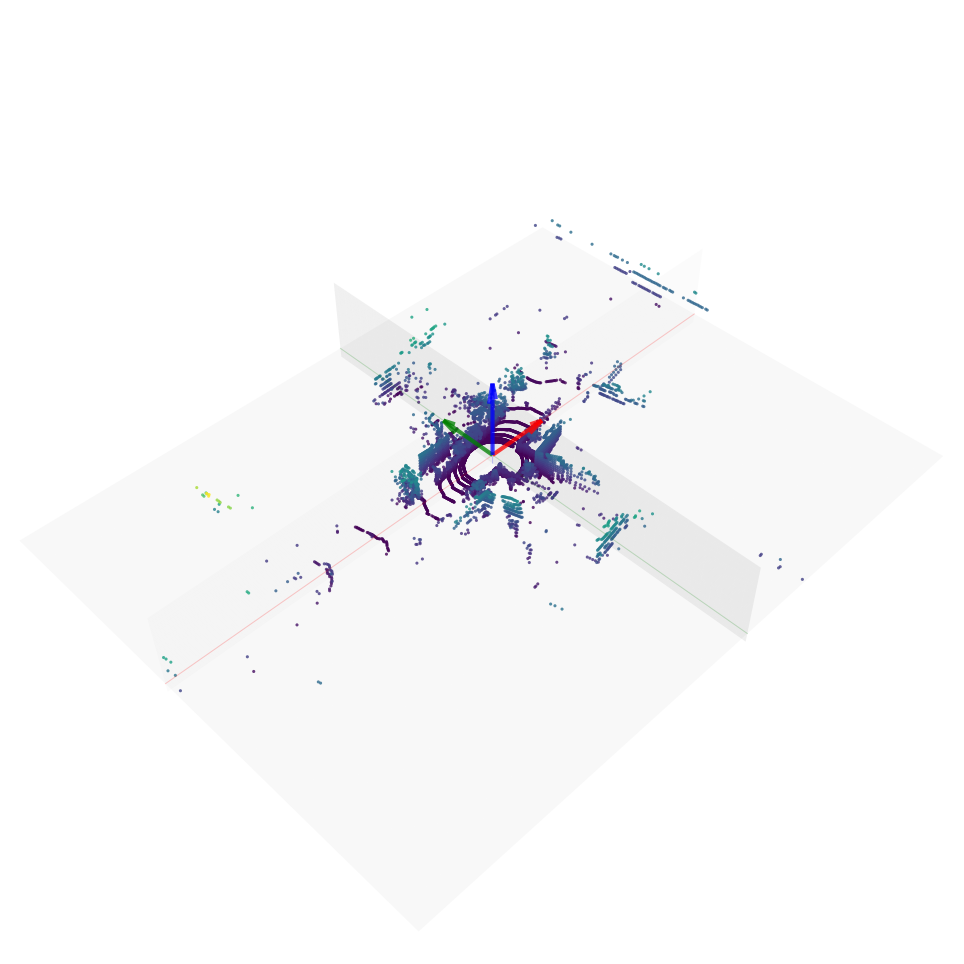}\hspace{0.1ex}%
    \adjincludegraphics[width=0.24\linewidth,trim={{0.02\width} {0.1\height} {0.18\width} {0.2\height}}, clip, cfbox=SILVER 0.1pt 0pt]{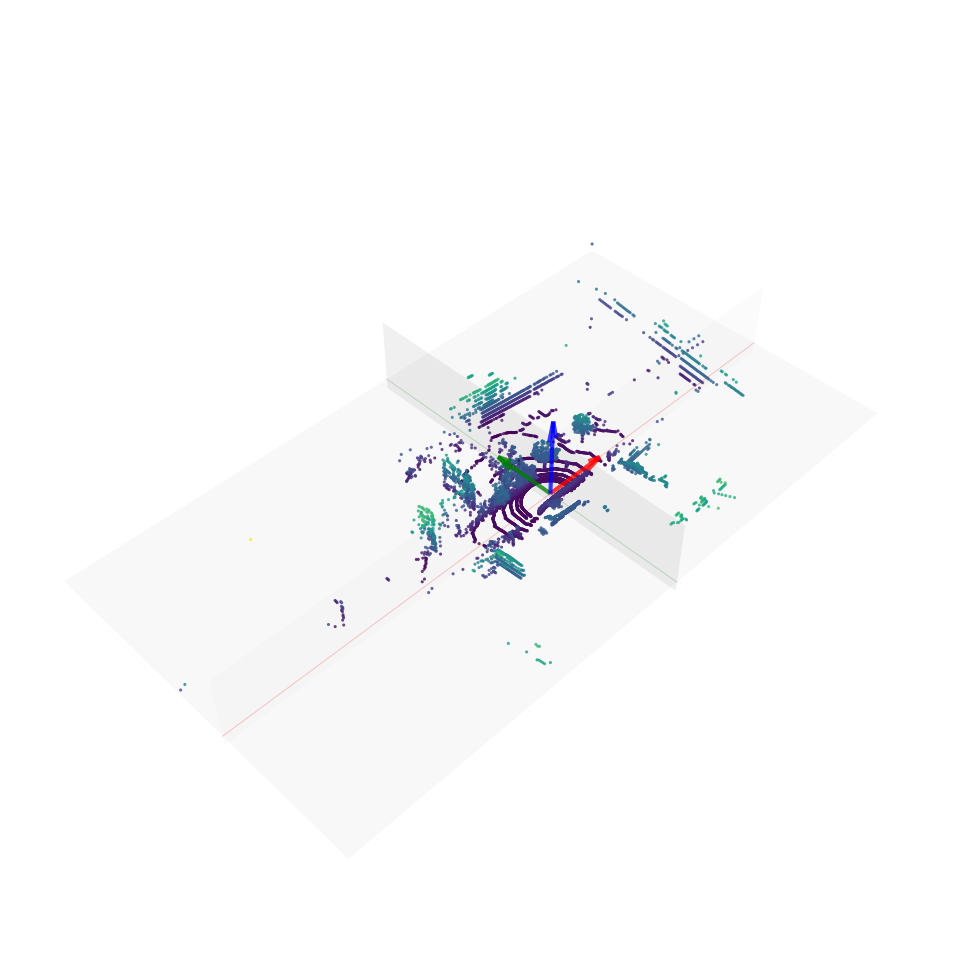}\hspace{0.1ex}%
    \adjincludegraphics[width=0.24\linewidth,trim={{0.02\width} {0.1\height} {0.18\width} {0.2\height}}, clip, cfbox=SILVER 0.1pt 0pt]{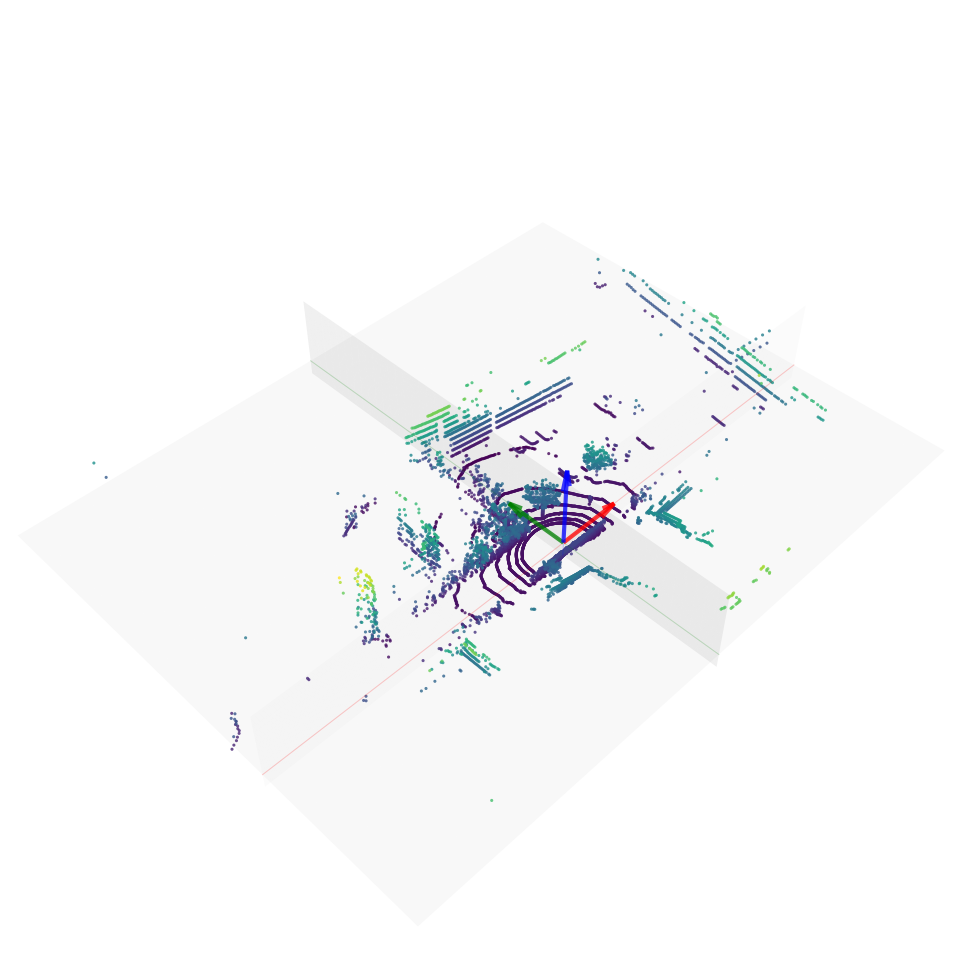}\\%
    
    \caption[DSEC LiDAR point cloud illustration]{Point clouds obtained from
    LiDAR scans for the DSEC sequence \texttt{thun\_00\_a}. For illustration
    purposes, the point clouds are sampled at regular indices where row 1 shows
    indices 0, 50, 100, 150, row 2 has indices 200, 250, 300, 350, row 3
    depicts indices 400, 450, 500, 550, row 4 contains indices 600, 650, 700,
    750, and row 5 displays indices 800, 850, 900, and 950 out of the total 976
    LiDAR scans for the sequence. Additionally, in each frame the LiDAR
    reference frame is drawn with the $x$-, $y$-, $z$-axis coded in red, green,
    and blue, respectively.}
    \label{fig:lidar_scans_thun00a}
\end{figure}

DSEC provides raw LiDAR scans (\cref{fig:lidar_scans_thun00a}) and IMU data,
but it does not include the processed ground-truth linear and angular
velocities of the camera. However, since our OPCM framework requires 3D camera
velocities to generate orientation priors, we estimate these velocities
directly from the LiDAR data using geometric registration techniques. Our
approach begins by converting each LiDAR scan into a structured 3D point cloud
in the LiDAR coordinate frame. Consecutive scans are paired and each scan is
rigidly aligned to the preceding scan using the iterative closest point
algorithm, which minimizes the Euclidean distance between corresponding points
in the two point clouds by estimating a rigid transformation $\mathbf{T} \in
\text{SE}(3)$ consisting of a rotation $\mathbf{R}$ and a translation
$\mathbf{t}$. 

The transformation $\mathbf{T}_{k \rightarrow k+1}$ between scans at timestamps
$t_k$ and $t_{k+1}$ encodes the rotation and translation over the time interval
$\Delta t = t_{k+1} - t_k$. The linear velocity is obtained by dividing by
$\Delta t$,
\begin{equation}
  \boldsymbol{v} = \frac{\mathbf{t}}{\Delta t}.
\end{equation}
The angular velocity is computed by first converting $\mathbf{R}$ to an
axis-angle representation, $\vartheta \hat{\upsilon}$, and then dividing by
$\Delta t$ to get 
\begin{equation}
  \boldsymbol{\omega} = \frac{\vartheta \hat{\upsilon}}{\Delta t}, 
\end{equation}
where $\vartheta$ is the rotation magnitude and $\hat{\upsilon}$ is the
rotation axis. Subsequently, both the linear and angular velocities,
$\boldsymbol{v}$ and $\boldsymbol{\omega}$, are transformed into the rectified
camera frame of the left event camera. Finally, we perform Kalman filtering to
obtain robust estimates of the camera velocities.

\section{Equivalence of Mean Squared Error, Negative Dot Product, and Negative
Cosine Similarity for Unit Orientation Vectors}
Let $\mathbf{u}, \mathbf{v} \in \mathbb{R}^d$ be unit vectors, i.e.,
$\Vert\mathbf{u}\Vert_2=\Vert\mathbf{v}\Vert_2=1$. We measure alignment via
their dot product $\mathbf{u}\cdot\mathbf{v}=\cos\theta$, where $\theta$ is the
angle between them. The squared Euclidean distance between $\mathbf{u}$ and
$\mathbf{v}$ expands as
\begin{equation}
  \Vert\mathbf{u}-\mathbf{v}\Vert_2^2 = \mathbf{u}\cdot\mathbf{u} + \mathbf{v}\cdot\mathbf{v} - 2\,\mathbf{u}\cdot\mathbf{v}.
\end{equation}
Since $\mathbf{u}$ and $\mathbf{v}$ are unit length, this reduces to
\begin{equation}
  \Vert\mathbf{u}-\mathbf{v}\Vert_2^2 = 1 + 1 - 2\,\mathbf{u}\cdot\mathbf{v} = 2 - 2\,\mathbf{u}\cdot\mathbf{v}.
\label{eq:square_euc_dist}
\end{equation}
Rearranging, we have
\begin{equation}
  \mathbf{u}\cdot\mathbf{v} = 1 - \tfrac{1}{2}\|\mathbf{u}-\mathbf{v}\|_2^2.    
\end{equation}
Thus, up to an affine transformation
\begin{equation}
  \underbrace{\Vert\mathbf{u}-\mathbf{v}\Vert_2^2}_{\text{MSE (per-sample)}} \;\propto\; 1 - \mathbf{u}\cdot\mathbf{v},
\end{equation}
and since \(\mathbf{u}\cdot\mathbf{v}=\cos\theta\) we have
\begin{equation}
  \Vert\mathbf{u}-\mathbf{v}\Vert_2^2 = 2(1-\cos\theta).
\end{equation}

Extending to a field of orientations (\eg, per-pixel unit vectors), let
$\{\mathbf{u}_i\}_{i=1}^N$ and $\{\mathbf{v}_i\}_{i=1}^N$ denote the
ground-truth orientation map and the estimated orientation field, respectively,
where $N=H\times W$ and $\Vert \mathbf{u}_i \Vert_2 = \Vert \mathbf{v}_i
\Vert_2 = 1$ for all $i$. The MSE per-pixel average is
\begin{align}
  \mathrm{MSE} &= \frac{1}{N}\sum_{i=1}^N \Vert\mathbf{u}_i-\mathbf{v}_i\Vert_2^2 \\
               &= \frac{1}{N}\sum_{i=1}^N \big(2 - 2\,\mathbf{u}_i\cdot\mathbf{v}_i\big) \\
               &= 2 - 2\left(\frac{1}{N}\sum_{i=1}^N \mathbf{u}_i\cdot\mathbf{v}_i\right).
\end{align}
Therefore, minimizing the MSE is equivalent to maximizing the average dot
product,
\begin{equation}
  \frac{1}{N}\sum_{i=1}^N \mathbf{u}_i\cdot\mathbf{v}_i,
\end{equation}
or equivalently minimizing the negative cosine similarity,
\begin{equation}
  \mathcal{L}_{\mathrm{cos}} = 1 - \frac{1}{N}\sum_{i=1}^N \mathbf{u}_i\cdot\mathbf{v}_i.
\end{equation}
Consequently, when the orientation vectors are normalized, the MSE, negative
dot product, and negative cosine similarity differ only by a positive scalar
\eqref{eq:square_euc_dist} and/or an additive constant. Thus, they have the
same minimizers and identical per-parameter optimization landscape up to
scaling and shifting. 

\end{document}